\documentclass[12pt]{article}
\usepackage{macros2e,din_a4,epsf,graphicx,%kay_labels,
newcomnd}

\begin{document}

%%%%%%%%%%%%%%%%%%%%%%%%%%%%%%%%%%%%%%%%%%%%%%%%%%%%%%%%%%%%%%%%%%%%%%%%
% Rewritten Macros for Diagrams, uses \graphicx and rescales pictures
% Copyright by Kay Wiese 1999
% Last Changes 21.1.1999
%%%%%%%%%%%%%%%%%%%%%%%%%%%%%%%%%%%%%%%%%%%%%%%%%%%%%%%%%%%%%%%%%%%%%%%%

\message{Reading macros for figures DynA...DynK}

%%%%%%%%%%%%%%%%%%%%%%%%%%%%%%%%%%%%%%%%%%%%%%%%%%%%%%%%%%%%%%%%%%%%%%%%
%Global variables
\def\bilderscale{0.32}
%-----------------^
%   Change this, if the font-size is reduced!
%
\newsavebox{\bilderbox}
\newlength{\bilderlength}
\newlength{\bilderhelp}
\newsavebox{\bilderone}
\newlength{\bilderonelength}
\newsavebox{\bildertwo}
\newlength{\bildertwolength}
%
%%%%%%%%%%%%%%%%%%%%%%%%%%%%%%%%%%%%%%%%%%%%%%%%%%%%%%%%%%%%%%%%%%%%%%%%

%%%%%%%%%%%%%%%%%%%%%%%%%%%%%%%%%%%%%%%%%%%%%%%%%%%%%%%%%%%%%%%%%%%%%%%%
%Auxiliary commands
\newcommand{\bild}[1]{\fboxsep0mm%
\sbox{\bilderbox}{%\fbox
{\includegraphics[scale=\bilderscale]{#1}}}%
\settowidth{\bilderlength}{\usebox{\bilderbox}}%
\parbox{\bilderlength}{\usebox{\bilderbox}}}
\newcommand{\savebild}[3]{\newsavebox{#2}%
\sbox{#2}{\bild{#3}}\newcommand{#1}{%
\ensuremath{\,\mathchoice{\usebox{#2}}%
{\settowidth{\bilderhelp}{\scalebox{0.7}{\usebox{#2}}}%
\parbox{\bilderhelp}{\scalebox{0.7}{\usebox{#2}}}}%
{\settowidth{\bilderhelp}{\scalebox{0.5}{\usebox{#2}}}%
\parbox{\bilderhelp}{\scalebox{0.5}{\usebox{#2}}}}%
{\settowidth{\bilderhelp}{\scalebox{0.35}{\usebox{#2}}}%
\parbox{\bilderhelp}{\scalebox{0.35}{\usebox{#2}}}}%
\,}}}%
\newcommand{\MOPE}[2]{{%
\mathchoice{%displaystyle
\sbox{\bilderone}{\ensuremath{\displaystyle#1}}%
\sbox{\bildertwo}{\ensuremath{\displaystyle#2}}%
\settowidth{\bilderonelength}{\rotatebox{90}{\ensuremath{\usebox{\bilderone}}}}%
\settowidth{\bilderonelength}{\rotatebox{90}{\ensuremath{\usebox{\bilderone}}}}%
\left(\!\usebox{\bilderone}\parbox{0mm}{\rule{0mm}{\bildertwolength}}\right.%
\hspace*{-0.5ex}\!\left|\!\parbox{0mm}{\rule{0mm}{\bilderonelength}}%
\usebox{\bildertwo}\!\right)}%
{%textstyle
\sbox{\bilderone}{\ensuremath{\textstyle#1}}%
\sbox{\bildertwo}{\ensuremath{\textstyle#2}}%
\settowidth{\bilderonelength}{\rotatebox{90}{\ensuremath{\usebox{\bilderone}}}}%
\settowidth{\bilderonelength}{\rotatebox{90}{\ensuremath{\usebox{\bilderone}}}}%
\left(\!\usebox{\bilderone}\parbox{0mm}{\rule{0mm}{\bildertwolength}}\right.%
\hspace*{-0.35ex}\!\left|\!\parbox{0mm}{\rule{0mm}{\bilderonelength}}%
\usebox{\bildertwo}\!\right)}%
{%scriptstyle
\sbox{\bilderone}{\ensuremath{\scriptstyle#1}}%
\sbox{\bildertwo}{\ensuremath{\scriptstyle#2}}%
\settowidth{\bilderonelength}{\rotatebox{90}{\ensuremath{\usebox{\bilderone}}}}%
\settowidth{\bilderonelength}{\rotatebox{90}{\ensuremath{\usebox{\bilderone}}}}%
\left(\!\usebox{\bilderone}\parbox{0mm}{\rule{0mm}{\bildertwolength}}\right.%
\hspace*{-0.1ex}\!\left|\!\parbox{0mm}{\rule{0mm}{\bilderonelength}}%
\usebox{\bildertwo}\!\right)}%
{%scriptscriptstyle
\sbox{\bilderone}{\ensuremath{\scriptscriptstyle#1}}%
\sbox{\bildertwo}{\ensuremath{\scriptscriptstyle#2}}%
\settowidth{\bilderonelength}{\rotatebox{90}{\ensuremath{\usebox{\bilderone}}}}%
\settowidth{\bilderonelength}{\rotatebox{90}{\ensuremath{\usebox{\bilderone}}}}%
\left(\!\usebox{\bilderone}\parbox{0mm}{\rule{0mm}{\bildertwolength}}\right.%
\hspace*{-0.1ex}\!\left|\!\parbox{0mm}{\rule{0mm}{\bilderonelength}}%
\usebox{\bildertwo}\!\right)}%
}}	
\newcommand{\DIAG}[2]{{%
\mathchoice{%displaystyle
\sbox{\bilderone}{\ensuremath{\displaystyle#1}}%
\sbox{\bildertwo}{\ensuremath{\displaystyle#2}}%
\settowidth{\bilderonelength}{\rotatebox{90}{\ensuremath{\usebox{\bilderone}}}}%
\settowidth{\bilderonelength}{\rotatebox{90}{\ensuremath{\usebox{\bilderone}}}}%
\left<\!\usebox{\bilderone}\parbox{0mm}{\rule{0mm}{\bildertwolength}}\right.%
\hspace*{-0.5ex}\!\left|\!\parbox{0mm}{\rule{0mm}{\bilderonelength}}%
\usebox{\bildertwo}\!\right>}%
{%textstyle
\sbox{\bilderone}{\ensuremath{\textstyle#1}}%
\sbox{\bildertwo}{\ensuremath{\textstyle#2}}%
\settowidth{\bilderonelength}{\rotatebox{90}{\ensuremath{\usebox{\bilderone}}}}%
\settowidth{\bilderonelength}{\rotatebox{90}{\ensuremath{\usebox{\bilderone}}}}%
\left<\!\usebox{\bilderone}\parbox{0mm}{\rule{0mm}{\bildertwolength}}\right.%
\hspace*{-0.35ex}\!\left|\!\parbox{0mm}{\rule{0mm}{\bilderonelength}}%
\usebox{\bildertwo}\!\right>}%
{%scriptstyle
\sbox{\bilderone}{\ensuremath{\scriptstyle#1}}%
\sbox{\bildertwo}{\ensuremath{\scriptstyle#2}}%
\settowidth{\bilderonelength}{\rotatebox{90}{\ensuremath{\usebox{\bilderone}}}}%
\settowidth{\bilderonelength}{\rotatebox{90}{\ensuremath{\usebox{\bilderone}}}}%
\left<\!\usebox{\bilderone}\parbox{0mm}{\rule{0mm}{\bildertwolength}}\right.%
\hspace*{-0.1ex}\!\left|\!\parbox{0mm}{\rule{0mm}{\bilderonelength}}%
\usebox{\bildertwo}\!\right>}%
{%scripscripttstyle
\sbox{\bilderone}{\ensuremath{\scriptscriptstyle#1}}%
\sbox{\bildertwo}{\ensuremath{\scriptscriptstyle#2}}%
\settowidth{\bilderonelength}{\rotatebox{90}{\ensuremath{\usebox{\bilderone}}}}%
\settowidth{\bilderonelength}{\rotatebox{90}{\ensuremath{\usebox{\bilderone}}}}%
\left<\!\usebox{\bilderone}\parbox{0mm}{\rule{0mm}{\bildertwolength}}\right.%
\hspace*{-0.1ex}\!\left|\!\parbox{0mm}{\rule{0mm}{\bilderonelength}}%
\usebox{\bildertwo}\!\right>}%
}}	
\newcommand{\reducedbildheightrule}[2]{{%
\mathchoice{\settowidth{\bilderlength}{\rotatebox{90}{\ensuremath{\displaystyle#1}}}%
\parbox{0mm}{\rule{0mm}{#2\bilderlength}}}%
{\settowidth{\bilderlength}{\rotatebox{90}{\ensuremath{\textstyle#1}}}%
\parbox{0mm}{\rule{0mm}{#2\bilderlength}}}%
{\settowidth{\bilderlength}{\rotatebox{90}{\ensuremath{\scriptstyle#1}}}%
\parbox{0mm}{\rule{0mm}{#2\bilderlength}}}%
{\settowidth{\bilderlength}{\rotatebox{90}{\ensuremath{\scriptscriptstyle#1}}}%
\parbox{0mm}{\rule{0mm}{#2\bilderlength}}}}}
\newcommand{\bildheightrule}[1]{\reducedbildheightrule{#1}{1}}
%

%%%%%%%%%%%%%%%%%%%%%%%%%%%%%%%%%%%%%%%%%%%%%%%%%%%%%%%%%%%%%%%%%%%%%%%%
%The Icons
%%%%%%%%%%%%%%%%%%%%%%%%%%%%%%%%%%%%%%%%%%%%%%%%%%%%%%%%%%%%%%%%%%%%%%%%
\savebild{\DynA}{\bildDynA}{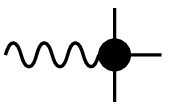}
\savebild{\DynB}{\bildDynB}{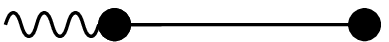}
\savebild{\DynC}{\bildDynC}{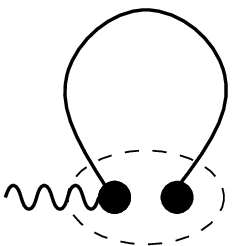}
\savebild{\DynD}{\bildDynD}{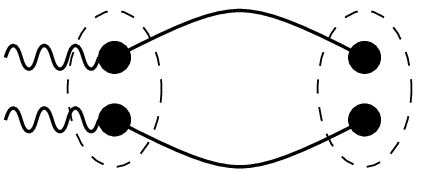}
\savebild{\DynE}{\bildDynE}{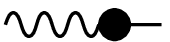}
\savebild{\DynF}{\bildDynF}{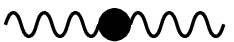}
\savebild{\DynG}{\bildDynG}{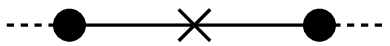}
\savebild{\DynH}{\bildDynH}{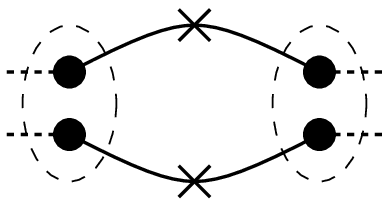}
\savebild{\DynI}{\bildDynI}{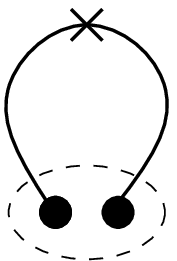}
\savebild{\DynJ}{\bildDynJ}{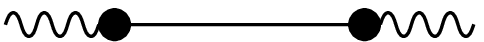}
\newcommand{\DynJL}{\stackrel{_\ind{L}}{\DynJ}}
\newcommand{\DynJT}{\stackrel{_\ind{T}}{\DynJ}}
\savebild{\DynJcross}{\bildDynJcross}{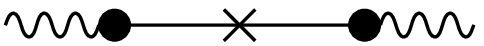}
\savebild{\DynK}{\bildDynK}{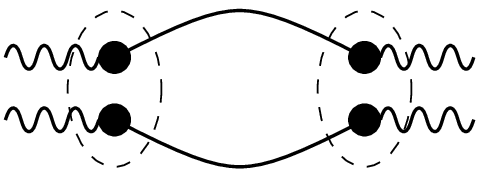}
\savebild{\DynKc}{\bildDynKc}{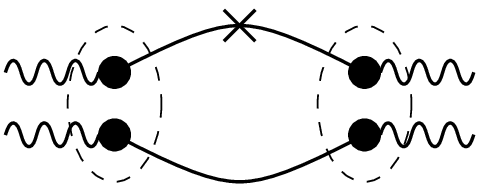}
\savebild{\DynKcc}{\bildDynKcc}{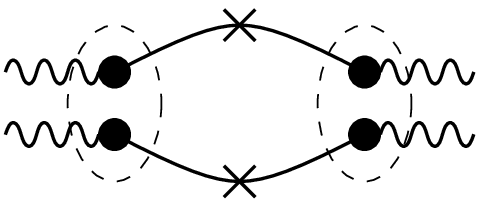}
\savebild{\DynL}{\bildDynL}{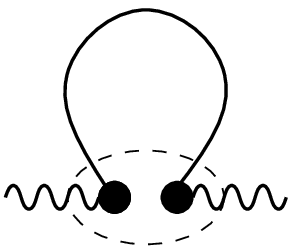}
\newcommand{\DynLL}{\DynL{\!\!\!\!\ind L}\,\,}
\newcommand{\DynLT}{\DynL{\!\!\!\!\ind T}\,\,}
\savebild{\DynO}{\bildDynO}{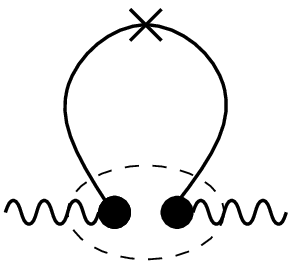}
\savebild{\DynP}{\bildDynP}{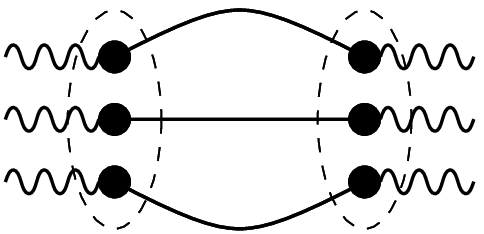}
\savebild{\DynPsub}{\bildDynPsub}{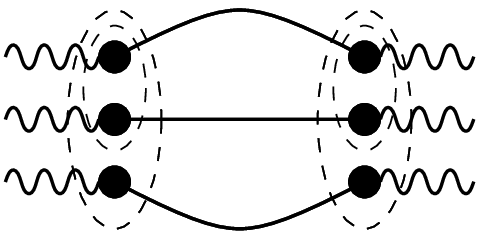}
\savebild{\DynQ}{\bildDynQ}{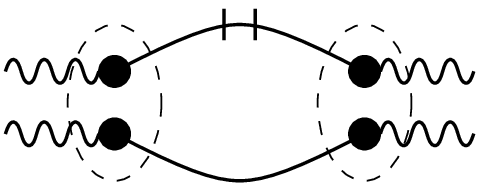}
\savebild{\DynR}{\bildDynR}{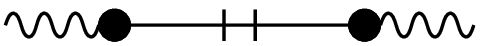}
\newcommand{\DynRL}{\stackrel{_\ind{L}}{\DynR}}
\newcommand{\DynRT}{\stackrel{_\ind{T}}{\DynR}}
%Macros uebertragen in new_bildermacros.sty bis hier 16.2.99
\savebild{\DynS}{\bildDynS}{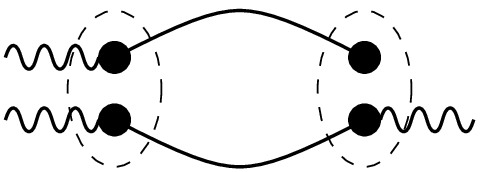}

\newcommand{\diagA}{\mathcal{A}}
\newcommand{\diagB}{\mathcal{B}}
\newcommand{\diagC}{\mathcal{C}}
%%%%%%%FIGURES%%%%%%%%%%%%%%%%%%
%
\newcommand{\fig}[2]{\epsfxsize=#1\epsfbox{#2}}

\begin{titlepage}

\noindent
\renewcommand{\thefootnote}{\fnsymbol{footnote}}
\parbox{1.85cm}{\epsfysize=1.85cm \epsfbox{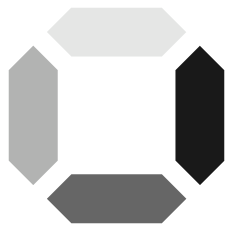}}
\hspace{0.5cm}
\parbox{0cm}{\epsfysize=1.6cm \epsfbox{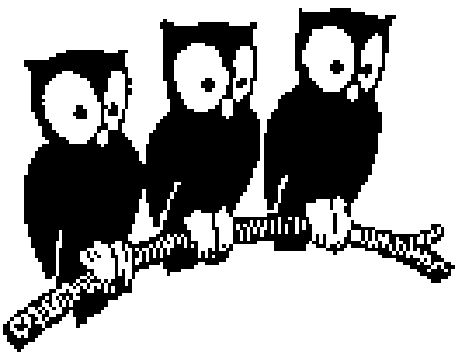}}\hfill%
\begin{minipage}{10cm}
\rightline{\small Preprint Uni GH Essen and LPTENS 98/35}%
\rightline{\small cond-mat/9808330v2}%
\rightline{\small \today}%
\end{minipage}
\vfill 
\centerline{\sffamily\bfseries\Large Polymers and manifolds in static random  
flows:}
\smallskip
\centerline{\sffamily\bfseries\Large a renormalization group study}
\vfill
\centerline{\bf\large Kay J\"org Wiese%
\footnote{Email: wiese@theo-phys.uni-essen.de}}
\smallskip
\centerline{\small Fachbereich Physik, Universit\"at GH Essen,  45117 Essen,
Germany}
\smallskip\smallskip
%\centerline{\bf and}
\smallskip\smallskip
\centerline{\bf\large Pierre Le Doussal%
\footnote{Email: ledou@physique.ens.fr}}\smallskip
\centerline{\small CNRS-Laboratoire de Physique Th\'eorique de l'Ecole
Normale Sup\'erieure,}
\centerline{\small 24 rue Lhomond, 75231 Paris
Cedex-France\thanks{LPTENS is a Unit\'e Propre du C.N.R.S.
associ\'ee \`a l'Ecole Normale Sup\'erieure et \`a l'Universit\'e Paris Sud}}

\vfill\vspace*{-0.5cm}
\begin{abstract}
We study the dynamics of a polymer or a $D$-dimensional elastic manifold
diffusing and convected in a non-potential static random flow
(the ``randomly driven polymer model''). 
We find that short-range (SR) disorder is relevant for $d \leq 4$ for
directed polymers (each monomer sees a different flow) and
for $d \leq 6$ for isotropic polymers (each monomer sees the
same flow) and more generally for $d<d_c(D)$ in the case of a manifold.
This leads to new large scale behavior, which we analyze
using field theoretical methods. We show that all divergences
can be absorbed in multilocal counter-terms which we compute
to one loop order. We obtain the non trivial
roughness $\zeta$, dynamical $z$ and transport exponents $\phi$
in a dimensional expansion. For directed polymers
we find $\zeta \approx 0.63$ ($d=3$), $\zeta \approx 0.8$
($d=2$) and for isotropic polymers $\zeta \approx 0.8$ ($d=3$).
In all cases  $z>2$ and the velocity versus applied force characteristics
is sublinear, i.e.\ at small forces $v(f) \sim f^\phi$
with $\phi > 1$. It indicates that this new state is glassy,
with dynamically generated barriers leading to trapping,
even by a divergenceless (transversal) flow. For random flows
with long-range (LR) correlations, we find continuously
varying exponents with the ratio $g_\ind L/g_\ind T$ of potential
to transversal disorder, and interesting crossover phenomena
between LR and SR behavior. For isotropic polymers new effects
(e.g.\ a sign change of $\zeta - \zeta_0$)
result from the competition between localization and stretching
by the flow. In contrast to purely potential 
disorder, where the  dynamics gets frozen, here the dynamical
exponent $z$ is not much larger than 2, making it easily
accessible by simulations.
The phenomenon of pinning by transversal disorder 
is further demonstrated using a two monomer ``dumbbell'' toy model.
\vspace{0.2cm}

 \end{abstract}\vspace*{-0.5cm}

\vfill

\centerline{\it Submitted for publication in Nuclear Physics B}

\vfill
\renewcommand{\thefootnote}{\arabic{footnote}}
\setcounter{footnote}{0}
\end{titlepage}
\setcounter{page}{2}
\tableofcontents
\newpage
\setcounter{page}{3}

\section{Introduction}
\begin{figure}[t]
%\vspace{-3mm}
\centerline{ \fig{0.8\textwidth}{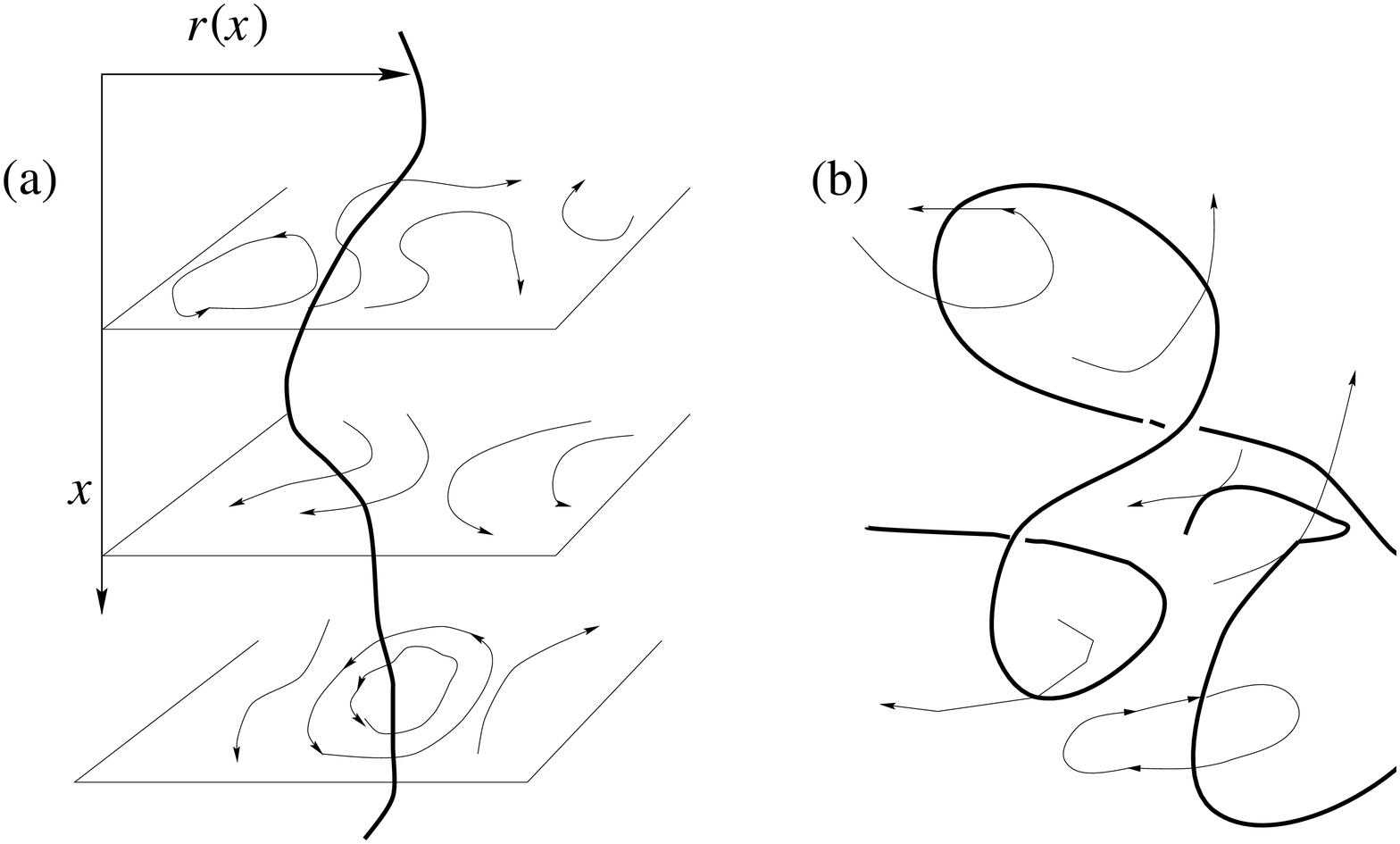} }
\caption{
{elastic manifolds (polymers $D=1$) in random flows:
(a) directed polymer (b) isotropic chain.}}
 \label{fig1}
\end{figure}%
\begin{figure}[t]
%\vspace{-3mm}
\centerline{ \fig{0.75\textwidth}{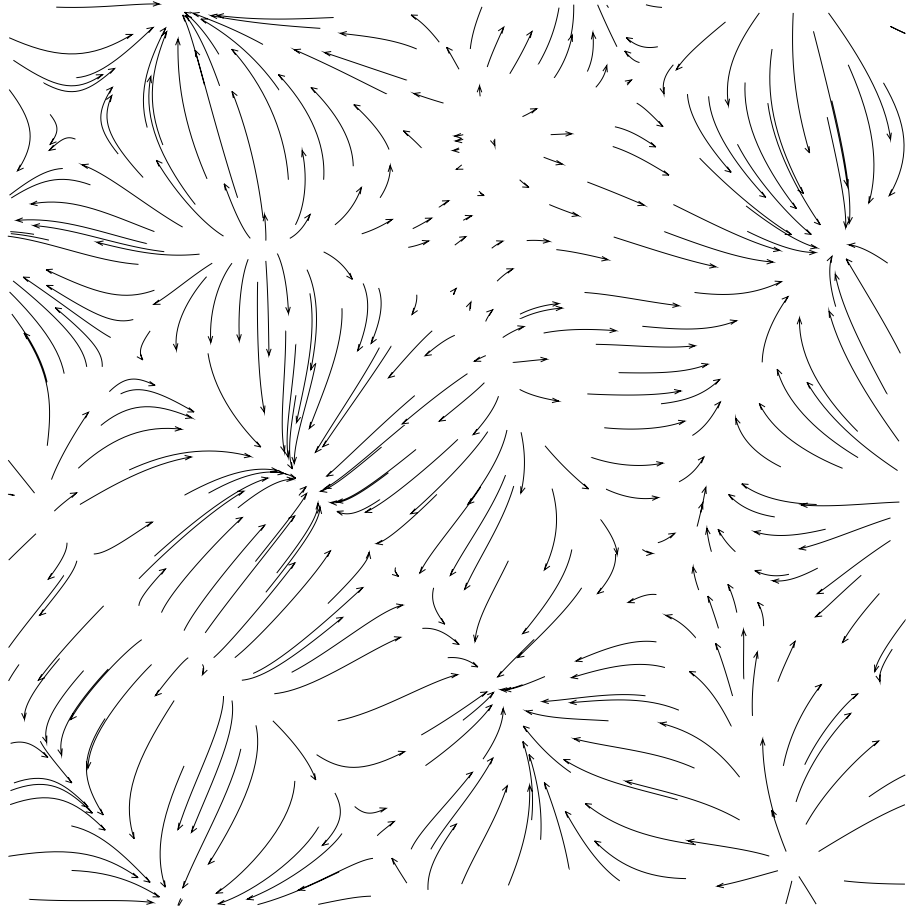} }
\caption{Longitudinal (potential) disorder, obtained from the disorder of 
Figure \protect\ref{f.isotrop} through application of the 
longitudinal projector.}
 \label{f.long}
\end{figure}%
\begin{figure}[t]
%\vspace{-3mm}
\centerline{ \fig{0.75\textwidth}{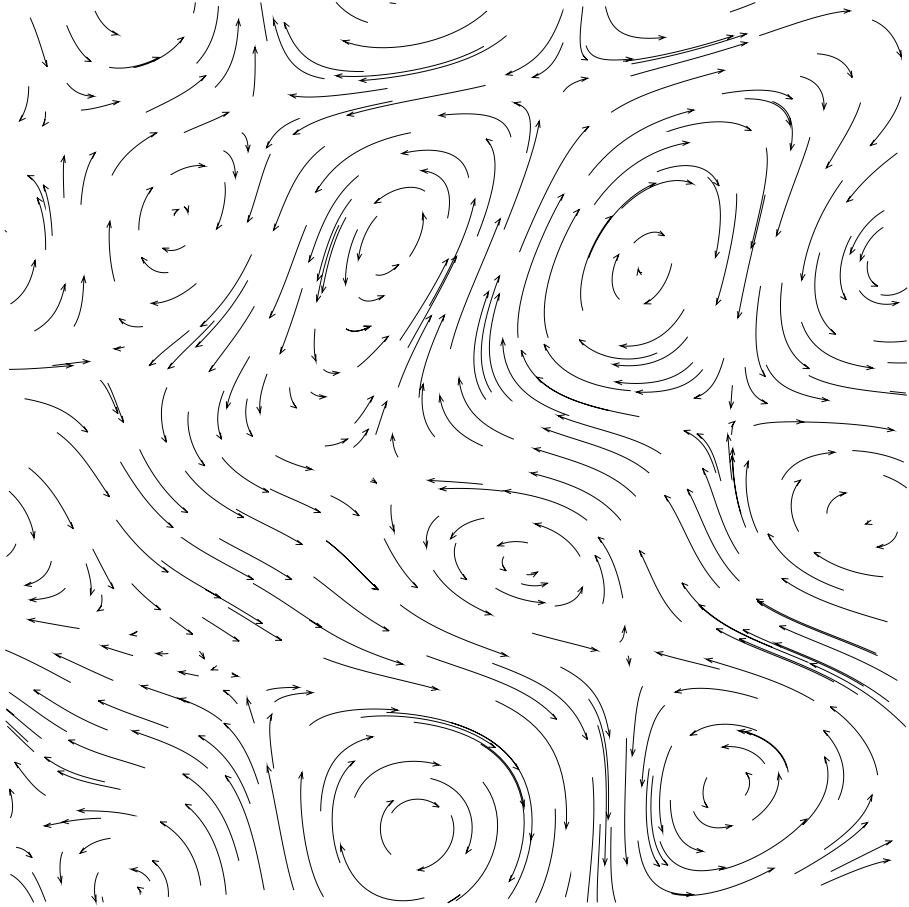} }
\caption{Transversal disorder, obtained from the disorder of 
Figure \protect\ref{f.isotrop} through application of the 
transversal projector.}
 \label{f.trans}
\end{figure}%
\begin{figure}[t]
%\vspace{-3mm}
\centerline{ \fig{0.75\textwidth}{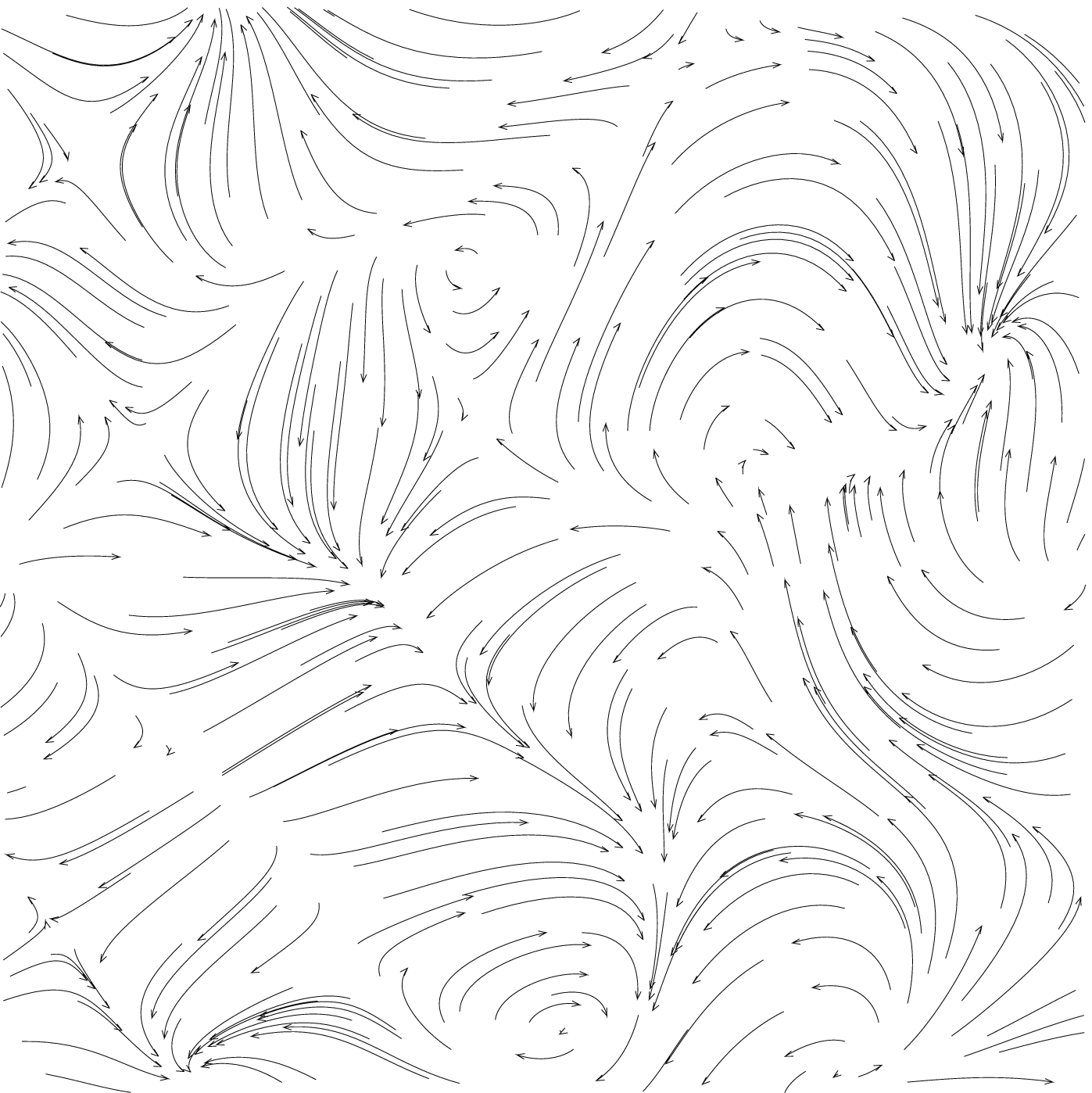} }
\caption{Isotropic disorder, obtained through superposition of 
$7\times7$ Fourier-modes. One observes purely transversal (``turbulent'')
regions and potential wells.}
 \label{f.isotrop}
\end{figure}

In this paper, we study the model of a polymer diffusing in a
static random flow and its extension to an
elastic manifold of arbitrary internal dimension $D$.
The velocity pattern of the flow $\vec v(r)$
is constant in time and leads to convection of the polymer
in addition to diffusion.
We are interested in the general case
of a {\it non-potential} flow: the extreme
example is the hydrodynamic divergenceless
flow, with $\nabla \cdot \vec v(r)=0$, but mixtures
of potential and divergenceless flows are also
considered. See figures \ref{f.long}, \ref{f.trans} and \ref{f.isotrop} for a visualization. The physical motivation to study this model,
which we call a ``randomly driven
polymer'', are the following:

First, it is a prototype of an out of equilibrium system
in the presence of quenched disorder which generalizes
the much studied problem of an elastic manifold
in {\it a random potential} \cite{Kardar97,%
NattermanEtAl1992,LeschhornEtAl1996,Fisher1985,%
DFisher86,BalentsFisher1993,MezardParisi1991,CugliandoloEtAl1996a,CugliandoloEtAl1996b,KinzelbachHorner1993a,KinzelbachHorner1993b},
i.e.\ for the present model the case of a purely {\it potential} flow.
This problem has been very popular in the last decade
due to its numerous physical realizations
such as vortex lines in superconductors, interfaces in random magnets and
charge density waves \cite{vortex_review,FukuyamaLee1978,GiamarchiBookYoung,LemerleEtAl1998}.
Equally fascinating, albeit more formal, relations
exist to growth processes
and potential fluid dynamics, via the Cole-Hopf transformation onto the
Burgers and Kardar-Parisi-Zhang equations \cite{KPZ}. In addition, it is
one of the simplest models which
exhibits glassy properties, typically ultra slow dynamics,
anomalously small response to external
perturbations and pinning 
\cite{NattermanEtAl1992,vortex_review,NarayanFisher93},
which are also
present in more complex systems, such as spin glasses \cite{VincentHammannOcio1992,BinderYoung1986,Sherrington1998}
or random field systems \cite{NattermannBookYoung,Anderson87}. However
little is known about what of the
glassy properties remains in the presence of {\it non-potential}
disorder. This question is important for numerous
physical systems such as
driven systems in the presence of quenched disorder
\cite{Krug1995,BalentsFisher1995,%
GiamarchiLeDoussal1996,SchmittmannBassler96}
or domain growth in the presence of shear
\cite{Onuki86}. The model studied in this paper
is simple enough to
allow for some answers to this question.

The second motivation concerns real systems of
elastic objects convected by hydrodynamic flows.
Recently, experiments on polymers in
spatially varying stationary flows, such as
elongational flows have been performed \cite{PerkinsSmithChu1997,DeGennes1997,LongViovyAjdari1996a,LongViovyAjdari1996b} and
these experiments can be extended to longer objects
and more complex flows. Of course
the model studied here is a toy model in the sense that it
disregards hydrodynamic forces (the back reaction of the
polymer on the flow)
and additional time dependent components
of the flow. The main interest
is to illustrate in a simple way some of the fascinating
properties arising from the competition between the
convecting flow and the elasticity of the structure. We
believe that some of these properties, for instance
the  existence of preferred
regions in the flow, will be present in more
realistic situations.
Experimentally, inhomogeneities in the
polymer distribution
and non-linearities in the response of the drift velocity $v(f)$
to an uniform applied external force $f$
are worth being investigated \cite{BonnEtAl1993}.

In the case of nonpotential disorder new effects
arise from the competition between
on one hand disorder and elasticity -- which tend to
create pinned or frozen states -- and on the
other hand the energy which is constantly pumped
into the system and which tends to destroy the glassy
properties. Very few systematic studies of these effects
exist at present. One extensive work was performed
in the context of driven lattices
\cite{GiamarchiLeDoussal1996,LeDoussalGiamarchi1997}:
it was shown that some glassy features,
such as pinning in the displacement
transverse to the motion, barriers and strongly
non linear transverse velocity
force-characteristics, survive in the moving state.
Some glassy features were also found to survive
in  infinite range random spin systems with non relaxational
dynamics
\cite{CugliandoloKurchanLeDoussalPeliti97}.
Another example, studied long ago
\cite{Luck1983,Fisher1984,KravtsovLernerYudson1985,Peliti1984,FisherFriedanQiuShenker1985,BouchaudComtetGeorgesLeDoussal87,%
BouchaudComtetGeorgesLeDoussal88,%
HonkonenKarjalainen1988,Stepanow92}
and recently solved analytically
in the limit of $d \to \infty$ for long-range
(LR) disorder \cite{LeDoussalCugliandoloPeliti97},
is the problem of a single particle diffusing
in a quenched random flow, the $D=0$ limit of the present model.
This problem was  found to exhibit some glassy features,
also seen in recent simulations of a related spin
model \cite{CugliandoloKurchanLeDoussalPeliti97}. In the present
paper we generalize the single particle study to
an elastic manifold, which is expected to lead to new
physics due to the crucial role of
internal elasticity in generating barriers.
It also generalizes the studies of the dynamics of self-avoiding
manifolds \cite{Wiese97a,Wiese97b} to a quenched disorder situation,
though self-avoidance is found to be perturbatively less
relevant in most of the regimes studied here.

In this paper we work directly in finite dimension
using dynamical renormalization group techniques,
and establish the existence of new universal fixed points
for short-range disorder. We also treat the case of
long-range disorder, which leads to a line of fixed
points. In all cases we compute explicitly the non trivial roughness exponent,
the dynamical exponent and the velocity force characteristics at small
applied force $f$. We study two cases: either the polymer is directed
(each monomer sees a different flow) or it is isotropic
(all monomers are in the same flow) as depicted on \Fig{fig1}.
A short account of our results has already
appeared in \cite{LeDoussalWiese97a}, which was presented using a
Wilson RG technique as in related studies
\cite{LeDoussalGiamarchi1997,CarpentierLeDoussal1997}.
The aim of the present paper is to give a detailed account
of our results using a
field theoretical derivation.
This derivation, based on the powerful multilocal operator
product expansion (MOPE)
techniques developed in  the context of self-avoiding manifolds  
\cite{DDG3,DDG4,WieseDavid96b,WieseDavid95,DavidWiese96a,Wiese97a,Wiese97b},  
deserves a detailed presentation, as it
is  generally useful to treat problems
involving manifolds. In addition, we give here a detailed
treatment of the crossover between long-range and 
short-range disorder, as well as an in depth discussion
of the results.

This article is organized as follows: In section
\ref{Model and physical observables} we define the model, the
physical observables and present a simple analysis based on
dimensional and Flory arguments.
Section \ref{FFTRG} is devoted to the field theoretic treatment
of the model. We start by defining the generating  functional
(section \ref{Definition of the field-theoretic model}), and
then show that divergences are
short-range (section \ref{General considerations about renormalization}).
This allows to identify local (section \ref{local div})
and bilocal (section \ref{DO ren}) counter-terms, summarized in
section \ref{List MOPE}. Their divergent parts
are calculated in sections \ref{The residues in the directed case}
and \ref{residues isotropic}, followed by a discussion of 
the particle limit in section \ref{particle limit}.
This yields the RG-equations
in section \ref{The flow-equations}
and \ref{Other RG-functions and critical exponents}.

The section which contains a presentation and a
discussion of the results, and can be read independently
is section \ref{Results and discussion}.
There we analyze the RG equations,
report the results of our calculation
and specify them to the four cases of interest
(directed and isotropic disorder, short-range and long-range correlations).
The crossover from long-range to short-range correlated disorder is discussed
in section \ref{Crossover}.
We compare with previous results for Hartree limit ($d=\infty$)
and particle case ($D=0$) and discuss the crossover 
from short-range to long-range correlated disorder. This is followed
by an analysis of the effect of
self-avoidance in section \ref{Inclusion of SA}.
The section terminates with the study of
a toy model, involving 2 particles
in a (directed) random flow which illustrates
how barriers are generated dynamically.
Conclusions and perspectives
are presented in section \ref{Conclusion}.
Some technicalities are relegated
to appendices \ref{appendixfree}, \ref{Extrapolations}
and \ref{finite cor to DO}.

\section{Model and physical observables}
\label{Model and physical observables}

\subsection{The model}

We consider a manifold
of internal dimension $D$ parameterized by a $d$-component
field $r_{\alpha}(x)$.
The polymer corresponds to $D=1$
($x$ labels the monomers), and a single particle
to $D=0$. Membranes have $D=2$.
We study the Langevin dynamics (for the dimensional analysis see
section \ref{Dimensional analysis})
\be \label{depart}
\dot r_\alpha(x,t) = \lambda \left( \Delta r_\alpha(x,t) +\eta_\alpha(x,t)
+ F_\alpha[r(x,t),x] \right) \ ,
\ee
where the term proportional to $\Delta r_\alpha(x,t)$ is the 
derivative of the internal (entropic) elasticity (``stiffness'').
The Gaussian thermal noise $\eta_{\alpha}(x, t)$ satisfies
\be \label{thermal}
\langle \eta_{\alpha}(x, t) \eta_{\beta}(x',t')  \rangle=
\frac2{\lambda} \delta_{\alpha \beta} \delta(t-t') \delta^D(x-x')\ .
\ee
Angular brackets denote thermal averages and
overbars disorder averages.

$F_{\alpha}[r,x]$ is a Gaussian quenched random force field
of correlations:
\bea   \label{corr}
\overline{F_{\alpha}[r,x]  F_{\beta}[r',x']}
= \Delta_{\alpha \beta}(r-r') h(x-x')
\ .
\eea

We study  two main cases of interest illustrated in Fig.\ \ref{fig1}. 
If the manifold is {\it directed} (e.g.\ a polymer oriented by an
external field), then
\be
h(x-x')=\delta^D(x-x') \ .
\ee
If the manifold is {\it isotropic}, (e.g.\ a Gaussian chain in
a static flow), the force field does not depend
on the internal coordinate
\be
F_{\alpha}[r,x] = F_{\alpha}[r]
\ee
and
\be
h(x-x')=1 \ .
\ee
Note that the isotropic polymer (or more generally the manifold)
in a static flow can also be seen as a directed polymer in
a flow completely correlated (i.e.\ constant) along 
its internal dimension.

We consider a
statistically rotationally invariant force field
with both a potential (L)
(``electric'') and a divergence-free (T)
part (``magnetic'') which
 depends only on the distance $|r-r'|$.
Both parts  contribute separately to the correlator:
\be
\Delta_{\alpha \beta}(r)  = -
\partial_{r_\alpha}  \partial_{r_\beta}
\Delta_\ind L (|r|) -  (
\delta_{\alpha \beta} \sum_{\gamma} \partial^2_{r_\gamma}
 -  \partial_{r_\alpha}  \partial_{r_\beta} )
\Delta_\ind T (|r|)
\ .
\ee
In Fourier space we use
\be
\Delta_{\alpha \beta}(r)=\int_k \Delta_{\alpha \beta}(k) e^{ik\cdot r}
\ee
where we specify the normalization of the $k$-integral below.
There are several  cases of interest for the correlation
of the random force.

\medskip
\leftline{\bf Model SRF: force with short-range correlations}
The {\em force} correlator scales like
a $\delta$-distribution  $\Delta_{\alpha \beta}(r-r') \sim \delta^d(r-r')$,
with however a non-trivial index structure. In Fourier space the correlator
reads at small $k$
\be
\Delta_{\alpha \beta}(k) =
{g}_{{\mbox{\scr L}}} P^{{\mbox{\scr L}}}_{\alpha \beta}(k)
+ {g}_{\mbox{\scr T}} P^{\mbox{\scr T}}_{\alpha \beta}(k) \ ,
\ee
where $P_{\alpha\beta}^{\mbox{\scr L}}(k)=k_{\alpha}k_{\beta}/k^2$
is the longitudinal and $P_{\alpha\beta}^{\mbox{\scr T}}(k)=\delta_{\alpha  
\beta} -P_{\alpha
\beta}^{\mbox{\scr L}}(k)$ the transverse projector.
As we shall show below, this is the generic situation,
since it is relevant for all short-ranged correlations, even those, for which 
one starts with a force correlator which is formally shorter ranged than
the $\delta$-distribution (e.g.\ decaying faster than $1/r^d$).

\medskip
\leftline{\bf Model SRP: potentials with short-range correlations}

In this case, the {\em potential} correlator scales like
a $\delta$-distribution $\Delta_{\ind L,\ind T}(r-r') \sim \delta^d(r-r')$.
The  corresponding force-correlator reads in Fourier space at small $k$
\be
\Delta_{\alpha \beta}(k) =
\tilde{g}_{\mbox{\scr L}} k_{\alpha} k_{\beta}
+ \tilde{g}_{\mbox{\scr T}} ( \delta_{\alpha \beta} k^2 - k_{\alpha} k_{\beta} )
\equiv k^2 \left[ \tilde g_\ind L P^\ind L_{\A\B}(k) +  \tilde g_\ind T  
P^\ind T_{\A\B}(k)\right]
\ .
\ee
Except for $\tilde g_\ind T=0$, which is preserved by a fluctuation  
dissipation theorem
\cite{vortex_review,DFisher86,Ebert96},
this model is unstable and will renormalize to
model SRF discussed above (except in the transversal case $g_\ind L=0$ for the particle 
$D=0$).

\medskip
\leftline{\bf Model LR: forces with long  range correlations}

In this model, correlations at large scale obey
a power law $\Delta_{\alpha \beta}(r-r') \sim |r-r'|^{-a}$, with $a<d$.
The Fourier transforms of the correlators
with respect to $r$ behave, for small values of their argument $k$, as

\be
\Delta_{\alpha \beta}(k) \sim |k|^{a-d} (
{g}_{{\mbox{\scr L}}} P^{\mbox{\scr L}}_{\alpha \beta}(k)
+ {g}_{\mbox{\scr T}} P^{\mbox{\scr T}}_{\alpha \beta}(k) )
\ .
\ee

Let us also note that in order to simplify calculations, we normalize the
integral over $k$  such that
\be \label{norm int k}
 \int_k |k|^{a-d}\rme^{-C k^2} = C^{-a/2} \ .
\ee

\subsection{Physical observables}
We study the time translational invariant
steady state. It can be constructed by taking a finite size
sample $L$ and considering the large time limit first, before
taking the system size to $\infty$. We will not address here
the subtle issue of whether, or in which case, these limits commute
(for a partial discussion in the case $g_\ind T>0$ see e.g.\ 
\cite{LeDoussalCugliandoloPeliti97}). The
radius of gyration (or roughness) exponent $\zeta$
is then defined by
\be \label{def zeta}
\overline{\langle (r(x,t)-r(x',t))^2 \rangle}
\sim |x-x'|^{2 \zeta}\ ,
\ee
and  the single monomer
diffusion exponent  $\nu$ by
\be \label{def nu}
\overline{\langle (r(x,t)- r(x,t'))^2 \rangle}
\sim |t-t'|^{2 \nu} \ .
\ee
We assume
scaling behavior
\be \label{def z}
\overline{\langle (r(x,t)-r(0,0))^2 \rangle}
\sim |x|^{2 \zeta}\, \overline{B}(t/x^z)
\ee
with $\zeta = z \nu$.

We also consider the anomalous dimension of the elasticity of the membrane,
$\beta$ (for the precise definition see \Eq{exp beta def}), which evaluates
to 0 in the directed case.

The drift velocity under
a small additional applied force $f$ in \Eq{depart} is
\be
v \sim \overline{\langle r(x,t) \rangle}/t \ .
\ee
We find
\be
	v \sim f^{\phi}
\ee
at small $f$, with
\be
\phi=\frac{z-\zeta}{2-\zeta+\beta} >1 \ ,
\ee
indicating trapping of  the polymer
by the flow.

\subsection{Dimensional analysis}
\label{Dimensional analysis}
\begin{figure}[tb]%
%\vspace{-3mm}
\centerline{ \fig{0.7\textwidth}{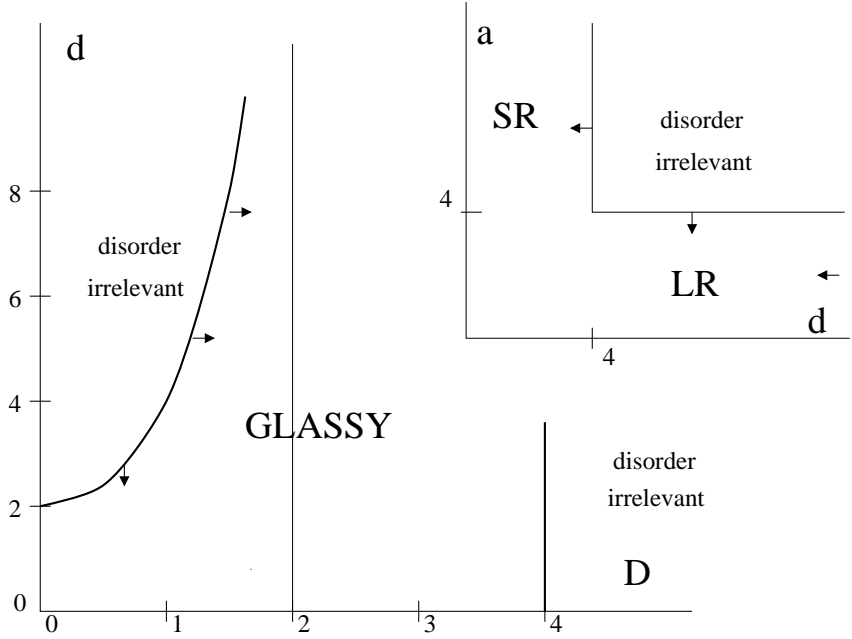} }
\caption{
Regions in the ($d,D$) plane where SR disorder is relevant
for a directed manifold. Inset: Regions in the ($a,d$) plane
where SR or LR disorder is relevant
for the directed polymer $D=1$.}
\label{fig-correl}
\end{figure}%
Power counting at the Gaussian fixed point
without disorder yields (denoting e.g.\ $\left[r\right]_x$ the dimension 
of $r$ in units of $x$)
\be
\left[r\right]_x= \zeta_0=\frac{2-D}2 \ ,\qquad  \left[r\right]_t=\nu_0 = \frac{2-D}4\ , \qquad \tilde \zeta_0=  \left[\eta\right]_x=\left[\tilde r\right]_x=-\frac{2+D}2
\ ,\qquad \beta=0 
\ee
for $D<2$. The dimensions of the couplings are 
\bea 
\E^{\ind{dir}}(D,d) &=& -\left[g_\ind L\right]_x^{\ind{dir}}
=-\left[g_\ind T\right]_x^{\ind{dir}} = 2-\zeta_0 a \\
\E^{\ind{iso}}(D,d) &=&  -\left[g_\ind L\right]_x^{\ind{iso}}
=-\left[g_\ind T\right]_x^{\ind{iso}} = 2+D - \zeta_0 a
\eea
in the directed and isotropic case respectively. Thus  
short range disorder (formally $a:=d$) 
is relevant for positive $\E$, i.e.\ when (see \Fig{fig-correl})
$d<d_c^{\ind{dir}}(D)$ for directed manifolds and
$d<d_c^{\ind{iso}}(D)$ for isotropic manifolds, with:
\be
d^{\ind{dir}}_c(D) = \frac4{2-D}\ , \qquad d^{\ind{iso}}_c(D) =  
\frac{4+2 D}{2-D} 
\ .
\label{criticaldim}
\ee
Thus we find that short-range disorder is
relevant for directed polymers when
$d<4$ and for isotropic polymers for $d<6$.
This is interesting because it means that 
in physical dimensions ($d=2$, $d=3$) polymers
should be affected by quenched random flows while for particles
short-range disorder is irrelevant for $d=3$ and only marginal
in $d=2$. Disorder can also be relevant if the interaction is
long enough ranged: disorder is relevant for 
$a<a_c(D) = d_c(D)$ for LR disorder.

A qualitative idea of the effect of disorder
is obtained by using Flory-like estimates.
Balancing in  \Eq{depart} friction or elasticity with
disorder \footnote{The idea of the Flory-estimate is to
suppose that all terms in the equation of motion scale in
the same way, and by this to obtain the critical exponents.
For directed manifolds in short-range correlated forces, 
one sets $\partial_t r \sim \Delta r \sim F(r,x)$,
replaces $\partial_t$ by $1/t$, $\Delta$ by $1/x^2$
and from \Eq{corr} $F(r,x)$ by $r^{-d/2}x^{-D/2}$ to obtain
 $r/t \sim r/x^2 \sim r^{-d/2}x^{-D/2} $, and the exponents
 stated in the main text. For 
 isotropic manifolds in short-range correlated forces,
the same argument reads $r/t \sim r/x^2 \sim r^{-d/2} $.} yields
\be \label{flory}
\zeta^{\ind{dir}}_{\mbox{\scr F}} = 2 \nu^{\mbox{\scr  
dir}}_{\mbox{\scr F}} = \frac{4-D}{2+d}
~,\qquad\zeta^{\ind{iso}}_{\mbox{\scr F}} = 2 \nu^{\mbox{\scr  
iso}}_{\mbox{\scr F}} = \frac4{2+d}
\ee
for directed and isotropic manifolds, respectively.
Results for LR disorder are obtained upon replacing $d \to a$. 
Note that this argument is too rough to 
distinguish between the potential or divergenceless nature
of the flow. It suggests however that 
the manifold will be stretched by disorder.
As we will see below, the precise calculation will indeed
confirm that the manifolds are stretched in most (but not all!)
of the cases, and will yield the deviations away
from Flory.

The expected ``phase diagram'' in the $d,D$ plane 
is illustrated in Fig. \ref{fig-correl}
for directed manifolds. The new glassy regime extends
from the line $d=d_c(D)$ to the upper critical (internal)
dimension $D=D_c=4$ beyond which disorder is perturbatively
irrelevant. For isotropic manifolds the diagram has the same
topology except that the upper critical internal dimension
is pushed to $D_c = + \infty$.
We know of three ways to investigate perturbatively
the nature of the glassy regime in Fig. \ref{fig-correl}.
First one can start from large $d$ (any $D$), which was
done in \cite{LeDoussalCugliandoloPeliti97}. The second
approach is to start from $D=4$ using a functional RG
method \cite{LeDoussalWieseProgress}. The last approach 
is to start from the line $d=d_c(D)$ which is the aim of
this paper.

\section{Field theoretic treatment of the renormalization group equations}
\label{FFTRG}

\subsection{Definition of the field-theoretic model}
\label{Definition of the field-theoretic model}
We start from the equation of motion (\ref{depart})
and convert it into an effective field theory following
 \cite{Janssen92}. The effective dynamical action
in prepoint (It\^o) discretization reads
\be
J\left[ r , \tilde r \right] = \int_{x,t} \tilde r_\A(x,t)
\left( \dot r_\alpha(x,t) + \lambda \left( -\Delta r_\alpha(x,t) +\eta_\alpha(x,t)
+ F_\alpha[r(x,t),x] \right) \right)  \ .
\ee
The noise $\eta(x,t)$ as well as the disorder force $F(r)$ are
Gaussian, see Eqs.~\eq{thermal} and \eq{corr}, and can  thus be integrated  
over. This gives the dynamic
functional
\bea
J\left[ r, \tilde r \right] &=& \int_{x, t} \tilde r(x, t) \left(
\dot r(x, t) -\lambda \Delta r(x,t) \right) -\lambda \tilde r(x, t)^2 \nn\\
&&\qquad -\half \lambda^2\int_{x,y, t, t'}
 \tilde r_{\alpha}(x, t) \Delta_{\alpha \beta} (r(x, t)-r(y, t')) h(x-y)
\tilde r_\beta(y, t')
\label{model}
\ .
\eea
Here and in the following, contraction over indices is implied, where ever
confusion is impossible.
In order to simplify notations, we will in the following suppress
the factor of $\lambda$, i.e.\ set
\be
 \lambda  t \to t \ .
\ee
This is not problematic, as $\lambda$ always appears
with time. At the end of the calculations one has to replace
$t$ by $\lambda t$ which is necessary in order to get the renormalization
factors correct.
Note also that the factor $\half$ disappears, if we use the time-ordered vertex.

The disorder correlator is
\be
\Delta_{\alpha \beta} (r(x,t)-r(y,t)) = \int_k \rme^{ik (r(x,t)-r(y,t))}
\left(g_\ind L P^\ind L_{\A\B}(k) + g_\ind T P^\ind T_{\A\B}(k) \right)  
|k|^{a-d} h(x-y) \ .
\ee
The function $h(x)$ may be specified to the directed case
with
\be
h(x-x') = \delta^D(x-x')
\ee
or to the isotropic case
\be
h(x-x') =1 \ .
\ee
Eventually, one may also want to study
an interpolating case, where
\be
h(x-x') = \delta(x_\perp-x'_\perp) \ .
\ee
In order to simplify notations, we introduce the following  graphical symbols
\bea
\DynA&=& \tilde r(x,t) (-\Delta_x) r(x,t)\nn\\
\DynE&=& \tilde r(x,t) \dot r(x,t)\nn\\
\DynF&=& \tilde r(x,t)^2\nn\\
_\A \DynJ_\B&=& \int_k \tilde r_\A(x,t)
	\rme^{ik (r(x,t)-r(y,t'))}\tilde r_\B(y,t') |k|^{a-d} h(x-y) \\
\DynJ&=& \int_k \tilde r(x,t)
	\rme^{ik (r(x,t)-r(y,t'))}\tilde r(y,t') |k|^{a-d}h(x-y) \nn\\
\DynJL&=&  \int_k P^\ind L_{\A\B}(k)\tilde r_\A(x,t)
	\rme^{ik (r(x,t)-r(y,t'))}\tilde r_\B(y,t') |k|^{a-d}h(x-y) \nn\\
\DynJT&=&  \int_k P^\ind T_{\A\B}(k)\tilde r_\A(x,t)
	\rme^{ik (r(x,t)-r(y,t'))}\tilde r_\B(y,t') |k|^{a-d}h(x-y)\nn
\ .
\eea
The action takes the symbolic form
\be
J\left[ r,\tilde r \right] = \int_{x, t} \DynE + \DynA - \DynF -\half   
\int_{x,y, t, t'}
\left(\DynJL g_\ind L + \DynJT g_\ind T \right)
\ .
\ee

The free theory ($g_\ind L=g_\ind T=0$) is easily investigated to obtain
the free response $R(x,t)$ and correlation-function $C(x,t)$:
\bea
R(x,t) &=& \frac1d \left< r(x,t) \tilde r(0,0) \right>_0 \\
C(x,t) &=& \frac1d \left< \half (r(x,t) -r(0,0))^2  \right>_0 \ ,
\eea
where $\left<\ldots \right>_0$ denotes averages in the free theory.
Explicit analytical expressions are given in appendix \ref{appendixfree}.
Here it is only important to note that for $0<D<2$ both functions
are well defined in the continuum. $C(x,t)$
vanishes for $x=t=0$, and
is monotonically increasing in both arguments.  They are non-trivially
related by
the fluctuation-dissipation theorem
\be \label{FDT}
 \Theta(t) \frac{\partial}{\partial t} C(x,t) =  R(x,t) \ .
\ee

\subsection{General considerations about renormalization}
\label{General considerations about renormalization}
Before embarking upon calculations, let us follow \cite{DDG4,Wiese97a} to
show that divergences are
always short-range. This is necessary for the renormalization program to
work. To this aim write down the perturbation-expansion,
the interaction vertex being
\be \label{e:DynJdef}
\DynJ :=  \int_k \tilde r(x,t) \,  \rme^{i k (r(x,t)-r(y,t'))}
\,\tilde r(y,t') \ ,
\ee
where for the sake of transparency, we have neglected the index structure.
 The reader will easily be
able to generalize the analysis given below to the model defined
in \Eq{model}.

The perturbative expansion of an observable $\cal O$ can then
be written as
\be \label{3.24}
\left< { \cal O} \right> = \mbox{Norm}
\sum_n \frac {g^n} { n !}
\int_{\{x_i,t_i\}} \left< {\cal O} \DynJ^n \right>_0 \ ,
\ee
where $\left<\ldots\right>$ denotes the expectation value with respect
to the free, non-interacting theory, and the normalization Norm has to be  
chosen so that $\left<1 \right>=1$.
The integral is taken over all arguments of the
interaction vertex.
We claim that divergences only occur at short distances and short
times. To prove this look at a typical expectation value with
fixed $x_i$ and $t_i$
\bea
\left<{\cal O}\DynJ^n \right>_0 &=& \sum_\alpha \int_{\{k_i\}}  
f_\alpha(x_l-x_m,t_l-t_m,k_l,k_m)
\rme^{-\sum_{i,j} Q_{ij} k_i k_j  } \ ,
\label{14}
\eea
where each contribution consists of a function $f_\alpha$, which is a  
polynomial product
of correlation and response functions and $k$'s and an exponential factor, with
\be
Q_{ij}=-C(x_i-x_j,t_i-t_j) \ .
\ee
As $f_\alpha$ is a regular function of the distances, divergences in the
$k$-integral in \Eq{14} can only occur 
 if   $Q_{ij}$ is not a positive form.
We will show that $Q_{ij}$ is a positive form for all $k_i$ which
satisfy the constraint
\be \label{neutral}
\sum_i k_i = 0 \ .
\ee
This constraint always holds, see \Eq{e:DynJdef}.
For equal times it is just the statement that the
Coulomb energy of a globally neutral assembly of charges is
positive by identifying $C$ with the Coulomb-propagator and
$k_i$ with the charges.
In the dynamic case, write
\bea
Q_{ij} &=& \int \frac{\rmd^D\!p}{(2\pi)^D} \int\frac{\rmd  
\omega}{2\pi}\,\frac2{\omega^2 +\left( p^2\right)^2}
\left( \rme^{ip(x_i-x_j) + i\omega (t_i-t_j)} -1 \right)
\ .
\eea
The exponential in \Eq{14} then reads
\bea
\sum_{i,j} k_i k_j Q_{ij} &=&
\int \frac{\rmd^D\!p}{(2\pi)^D}
\int \frac{\rmd \omega}{2\pi}\,
\frac2{\omega^2 +\left( p^2\right)^2}
\sum_{ij} k_i k_j
\left( \rme^{ip(x_i-x_j)+i\omega (t_i-t_j)} -1 \right)
\nn\\
&=&
\int \frac{\rmd^D\!p}{(2\pi)^D}
\int \frac{\rmd \omega}{2\pi}\,
\frac2{\omega^2 +\left( p^2\right)^2}
\left|\sum_{i} k_i\, \rme^{ip x_i+\omega t_i}\right|^2
\ .
\eea
To get the second line, \Eq{neutral} has been used.
Note that again due to \Eq{neutral},
the integral is ultraviolet convergent and thus
 positive. It vanishes if and only if the charge-density, regarded as a  
function of space {\em and}\/ time, vanishes.
This is possible if and only if endpoints of the dipoles (which form
the interaction) are at the same point in space {\em and} time.

Thus when integrating in \Eq{3.24} over $x_i$ and $t_i$, no
 divergence can occur at finite distances. To renormalize the
theory, only short distance divergences have to be
removed by adding appropriate counter-terms.
Thay are analyzed via a multilocal operator product expansion
(MOPE) \cite{DDG3,DDG4}. For a detailed
discussion of the MOPE and examples see \cite{DDG3,DDG4,%
WieseDavid96b,DavidWiese96a,WieseHabil}. 
The main result of this analysis is, like in scalar field theory,
that {\em in any order of perturbation theory, only a finite number
of counter-terms is necessary and that these
 are exactly those terms, which are
already present in the original theory.} The theory is thus 
multiplicatively renormalizable. 
To actually evaluate the counter-terms,  a multilocal
operator product expansion is used, which is described in 
the next sections. The general idea is always to 
study the behavior of $n$ disorder-vertices on approaching their end-points. 
The resulting term is expanded into a product of a (multi-local) operator,
times a MOPE-coefficient, which is a homogenous function of the distances
involved. Power-counting then indicates that as long as the operator is 
relevant or marginal 
in the RG-sense, the integral over the distances possesses an 
UV-divergence, and thus requires a counter-term  in the Hamiltonian. 
On the other hand, when the operator is irrelevant in the RG-sense,
then the integral over the MOPE-coefficient is globally UV-convergent,
and no counter-term is needed. This procedure is constructed such, that  possible sub-divergences are taken care of recursively as 
in the BPHZ-scheme%of perturbative renormalizability
. A formal 
proof of this statement is given in the context of a simpler 
model in \cite{DDG4}, see also \cite{WieseHabil}.

\subsection{Divergences associated with local operators}
\label{local div}
We now analyze our specific model, using the techniques of the multilocal
operator product expansion (MOPE) \cite{DDG3,DDG4}.
The first class of divergences stems from configurations
where the two end-points of the interaction vertex are approached.
The  interaction vertex is:
\bea \label{a:1}
	\DynJL g_\ind L + \DynJT g_\ind T&=&\int_k \,\tilde  
r_\A (x,t)
 \rme^{ik\left[r(x,t)-r(y,t')\right]} \tilde  
r_\B(y,t') \times\nn\\
&& \hspace{1.8cm}\times \,(P^\ind L_{\A\B}(k) g_\ind L + P^\ind  
T_{\A\B}(k)g_\ind T)\,
 |k|^{a-d}h(x-y)
\ .\qquad\quad
\eea
For short-range forces set $a=d$.

In order to extract the divergences for small $x-y$ and $t-t'$,
the first possibility is not to contract any response-field.
We then start by  normal-ordering the r.h.s.\ of \Eq{a:1}. 
As usual in statistical field theory, we define the normal-ordered
operator $\,:\!{\cal O}\!:\,$ starting from the operator $\cal O$ by
\be
	\,:\!{\cal O}\!:\, = {\cal O} -\mbox{all self-contractions}
\ee
where the self-contractions are taken with respect to the free
theory. The idea is that in order to calculate the insertion of
$\,:\!{\cal O}\!:\,$ into some expectation value only contractions of
$\,:\!{\cal O}\!:\,$ with the other operators in this expectation value
have to be taken. This implies that the operator
$\,:\!{\cal O}_1(x,t){\cal O}_2(0,0)\!:\,$
is always free of divergences for $(x,t) \to (0,0)$.

Using the identity
\be
	\rme^{i k r(x,t) }\, \rme^{-ik r(y,t')} =
\,:\!\rme^{i k r(x,t) } \rme^{-i k r(y,t') }\!:\,\rme^{-k^2 C(x-y,t-t')}
\ .
\ee
one can expand the normal-ordered
vertex-operators on the r.h.s.\ for small $x-y$ and $t-t'$.
The leading contribution is
\be
\mbox{\bf 1}\, \rme^{-k^2 C(x-y,t-t')} \ ,
\ee
yielding the first term in  the short-distance
expansion of \Eq{a:1} (for the normalization of the $k$-integral cf.\  
\Eq{norm int k}):
\bea
\lefteqn{\int_k \rme^{-k^2 C(x-y,t-t')} \,:\!\tilde r_\A (x,t)
 \tilde r_\B(y,t')\!: \,(P^\ind L_{\A\B}(k) g_\ind L + P^\ind  
T_{\A\B}(k)g_\ind T) |k|^{a-d}
h(x-y)} \nn\\
 &=& \tilde r\left({x+y \over 2},{t+t' \over 2}\right)^{\!2}
\!\left(\! g_\ind T\left(1-\frac 1 d\right) + g_\ind L \frac1d \right)  
\!C(x-y,t-t')^{-a/2}h(x-y)
+\mbox{subleading terms} \nn\\
\label{loc cont 1}
&=& \DynF \left( g_\ind T\left(1-\frac 1 d\right) + g_\ind L \frac1d \right)  
C(x-y,t-t')^{-a/2}
h(x-y)+\mbox{subleading terms} \ .
\eea
By going from the first to the second line, two things have been used:
First, since the integral over $k$ is rotationally invariant, only
terms with $\alpha=\beta$ survive, with a geometric factor
of $1/d$ for the transversal and  $(1-1/d)$ for the 
longitudinal projector. Second, since $\,:\!\tilde r(x,t) \tilde r(y,t')\!:\,$
is {\em normal-ordered}, i.e.\ all contractions between the two fields $\tilde r$ have been eliminated and no short-distance singularities remain,
it can be written as 
\bea
:\!\tilde r(x,t) \tilde r(y,t')\!:\ &=& \ :\!\tilde r\!\left(\frac{x+y}2,\frac{t+t'}2
\right)^{\!2}\!\!:\, + 
:\!\tilde r\!\left(\frac{x+y}2,\frac{t+t'}2
\right)\left[2(x-y)\nabla\right] \tilde r\!\left(\frac{x+y}2,\frac{t+t'}2
\right) \!\!:\nn\\
 &&\qquad+ 
:\!\tilde r\!\left(\frac{x+y}2,\frac{t+t'}2
\right)\left[2(t-t')\frac{2\partial}{\partial t}\right] \tilde r\!\left(\frac{x+y}2,\frac{t+t'}2
\right) \!\!:\nn\\
 &&\qquad+\, {\cal O}\left((x-y)^2\right) + {\cal O}\left((t-t')^2\right) +
 {\cal O}\left(x-y\right) {\cal O}\left(t-t'\right)
\eea
From this expansion, only the first term has to be retained and will
give a correction to the marginal operator $\DynF=\tilde r^2$ in the action. 
The following terms are {\em irrelvant} (in the RG-sense)
 with respect to this dominant
term and can therefore be neglected. Dropping the explicit normal ordering
of $\tilde r^2$ which is a pure number, and zero by analytic continuation,
leads to \Eq{loc cont 1}.

Note that when expanding the product of two operators as sum of normal
ordered ones, this sum runs over all possible contractions. 
The second contribution to  the normal-ordered product of \Eq{a:1}
is therefore 
obtained upon contracting one response field. Due to causality, this must 
be the field with the smaller time-argument. For simplicity, let us
take $t>0$ and put $y=t'=0$. By the same procedure as above, we
 obtain the following contribution:
\be \label{A6}
\int_k \,:\!\tilde r_\A (x,t)
\rme^{i k (r(x,t)-  r(0,0)) }\!:\, R(x,t) (ik)_\B
\rme^{-k^2 C(x,t)}
\,(P^\ind L_{\A\B}(k) g_\ind L + P^\ind T_{\A\B}(k)g_\ind T) |k|^{a-d}h(x-y)
\ .
\ee
The next step is to expand $\,:\!\rme^{i k (r(x,t)-  r(0,0)) }\!:\,$
about $(x,t)$. (It is important to expand about $(x,t)$
as otherwise  $\tilde r(x,t)$ had to be expanded, too.)
This expansion is
\be \label{A7}
:\!\rme^{i k (r(x,t)-  r(0,0)) }\!:\, = \mbox{\bf 1} + (ik)_\G
	\left( t\, \dot r_\G(x,t) \!+\! (x\nabla) r_\G(x,t) \!-\! \half  
(x\nabla)^2 r_\G(x,t)\right)
 	+\mbox{subleading terms} \ .
\ee
Upon inserting \Eq{A7} into \Eq{A6} and integration over $k$,
only terms even in $k$ survive. We can also neglect the term linear in
$x$, which is odd under space reflection. The remaining terms are
\bea
\lefteqn{\int_k :\!\tilde r_\A (x,t)
\left( t \dot r_\G(x,t) - \half (x\nabla)^2 r_\G(x,t)\right)\!\!:\,
(ik)_\G (ik)_\B R(x,t)\rme^{-k^2 C(x,t)}\times}\nn\\
&& \qquad \qquad \times \,(P^\ind L_{\A\B}(k) g_\ind L + P^\ind  
T_{\A\B}(k)g_\ind T)|k|^{a-d} h(x) \nn\\
&=& - R(x,t) \frac {a }{2d} C(x,t)^{-a/2-1}\, h(x)
 \left( t \DynE + \frac{x^2}{2D}\DynA \right)g_\ind L
\ . \hspace{2cm}
\eea
For the contribution proportional to $\DynA$, we have retained from the
tensor operator $\tilde r(x,t) (x\nabla)^2 r(x,t)$ only
the diagonal contribution $\frac1D \tilde r(x,t) x^2 (\Delta)r(x,t)$,
which is sufficient at 1-loop order. For the subtleties associated
with  the insertion of this operator at the 2-loop level
cf.\ \cite{WieseDavid96b}.
Using the FDT, \Eq{FDT}, this can still be simplified to
\be \label{loc cont 2}
\frac{\partial }{\partial t} C(x,t)^{-a/2} \left( \frac t d \DynE +  
\frac{x^2}{2dD}\DynA \right)g_\ind L \ .
\ee
\Eqs{loc cont 1} and \eq{loc cont 2} contain all possible divergent terms in
the short-distance expansion of \Eq{a:1} and all terms which have to be
taken into account in 1-loop order.
Notably, due to causality, no term independent of $\tilde r$ appears.

\subsection{The renormalization of the disorder (divergences associated with
bilocal operators)} \label{DO ren}
\begin{figure}[t]
\setlength{\unitlength}{1mm}
\hfill
\begin{picture}(50,50)
%\put(0,0){\framebox(50,50) { } }
\put(0,25){\parbox{35mm}{\epsfxsize=35mm\epsfbox{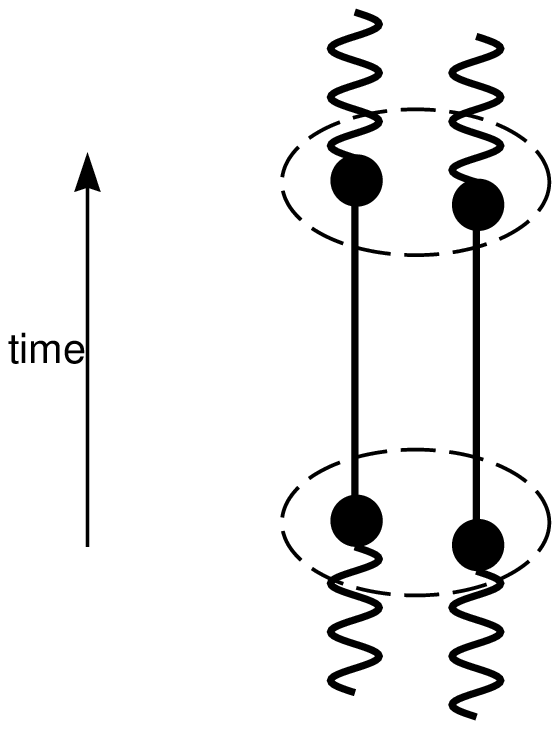}}}
\put(16,15){$\scriptstyle 0$}
\put(16,36){$\scriptstyle t$}
\put(36,13){$\scriptstyle -\tau$}
\put(36,34){$\scriptstyle t-\sigma$}
\end{picture}
\hfill
\begin{picture}(50,50)
%\put(0,0){\framebox(50,50) { } }
\put(0,25){\parbox{35mm}{\epsfxsize=35mm\epsfbox{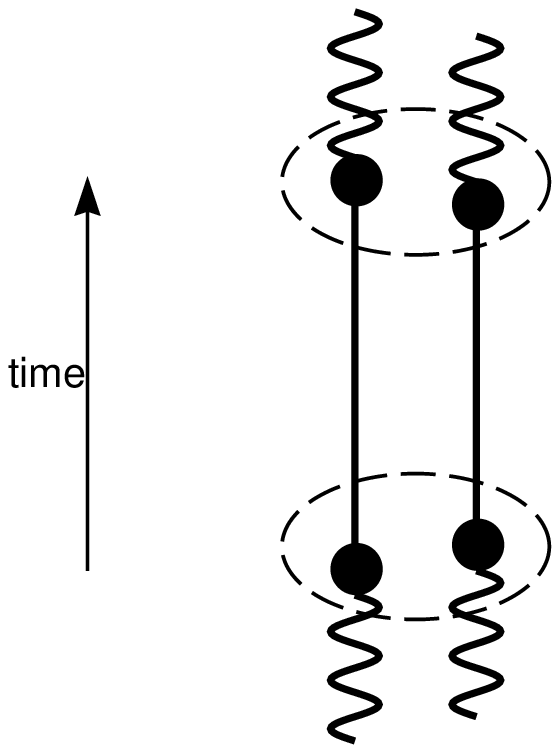}}}
\put(16,14){$\scriptstyle 0$}
\put(16,38){$\scriptstyle t$}
\put(36,15){$\scriptstyle \tau$}
\put(36,36){$\scriptstyle t-\sigma$}
\end{picture}
\hfill\hspace{0cm}
\caption{The diagrams $D_1$ (left) and $D_2$ (right).}
\label{disren}
\end{figure}

There are also UV-divergent configurations associated to bilocal
operators, which  renormalize the disorder, and which are
depicted on \Fig{disren}. A dashed line indicates that
the end-points of the interaction vertices are approached in space and time.
Up to permutations of the two
interaction vertices, there are two possibilities to order their
end-points in time, namely $D_1$ and $D_2$ on \Fig{disren}.
We first calculate $D_ 1$, starting from
\bea
&&\hspace{-2mm}\tilde r_\A (y,t) \int_k \left( g_\ind T P^\ind T_{\A\B}(k) +  
g_\ind L P^\ind L_{\A\B}(k)
\right)|k|^{a-d}h(x-y) \rme^{ik(r(y,t)-r(x,0))} \tilde r_\B(x,0) \times \\
&& \hspace{0.6cm} \times \,\tilde r_\G (y',t-\sigma) \int_p \left( g_\ind T  
P^\ind T_{\G\D}(p) + g_\ind L P^\ind L_{\G\D}(p)
\right)|p|^{a-d}h(x'-y') \rme^{ik(r(y',t-\sigma)-r(x',-\tau))} \tilde  
r_\D(x',-\tau)
\nn \ .
\eea
For small $x-x'$, $y-y'$, $\tau$ and $\sigma$, with $\tau, \sigma>0$,
there is one contribution for the renormalization of the interaction
vertex. First, due to causality,  $\tilde r_\G (y',t-\sigma)$ and
$\tilde r_\D(x',-\tau)$  have to be contracted with a correlator field
in order to obtain two response
fields at the end. Then the short-distance expansion for
nearby vertex-operators reads:
\bea
\rme^{ik r(y,t)} \rme^{ip r(y',t-\sigma)} &\approx&
\rme^{i(k+p) r(y,t)}  \,\rme^{kp C(y-y',\sigma)}  
\nn\\\rme^{-ik r(x,0) }\rme^{-ip r(x',-\tau)}  &\approx&
\rme^{-i(k+p) r(x,0) }  \rme^{kp C(x-x',\tau)}\ .
\eea%
This yields up to subleading terms:
\bea
\lefteqn{\int_k\int_p
\tilde r_\A (y,t) \tilde r_\B(x,0)
\rme^{i(k+p)(r(y,t)-r(x,0))}\left( g_\ind T P^\ind T_{\A\B}(k) + g_\ind L  
P^\ind L_{\A\B}(k)
\right)
\left( g_\ind T P^\ind T_{\G\D}(p) + g_\ind L P^\ind L_{\G\D}(p)
\right)} \nn\\
&& (ik)_\G (-ik)_\D
R(y-y',\sigma) R(x-x',\tau) \rme^{kp (C(y-y',\sigma) + C(x-x',\tau) )}
|k|^{a-d}|p|^{a-d}
h(x-y)h(x'-y')\ .\nn\\
&&
\eea
In the next step, first $k$ and second $p$ are shifted:
\be
	k \longrightarrow k-p\ , \qquad p \longrightarrow p+\frac k 2
\ .
\ee
The result is
\bea
&&\!\!\int_k
\tilde r_\A (y,t) \tilde r_\B(x,0)
\rme^{ik(r(y,t)-r(x,0))}
\left[ \int_p
\left( g_\ind T P^\ind T_{\A\B}(p-k/2) + g_\ind L P^\ind L_{\A\B}(p-k/2)
\right)\right. \nn\\
& & \qquad \quad
\left( g_\ind T P^\ind T_{\G\D}(p+k/2) + g_\ind L P^\ind L_{\G\D}(p+k/2)
\right) (k/2-p)_\G (k/2-p)_\D |p+k/2|^{a-d}|p-k/2|^{a-d}\nn\\
&& \qquad\quad \left.
R(y-y',\sigma) R(x-x',\tau) \rme^{(k^2/4-p^2) (C(y-y',\sigma) + C(x-x',\tau) )}
h(x-y)h(x'-y')\right] \ . \hspace{1cm}
\label{3.33}
\eea
To compute the correction proportional to the disorder vertex, the expression 
in the rectangular brackets is  expanded for small $k$.
As the integral has
a well-defined limit for $k\to0$, convergent for $d>2$, neither a term  of the 
form $k_\A k_\B/k^2$, nor a term of the form $|k|^{a-d}$ with
$a<d$ can be generated.
We conclude that in the LR-case only SR-disorder is
generated. In the remainder of this sections, we therefore focus
on short-range correlated forces setting $a=d$. The leading term of the  
above expansion is then (the $p$-integral being defined in \Eq{norm int k})
\bea
\lefteqn{_\A \DynJ_\B \int_p \left( g_\ind T \left(\delta_{\A\B} -p^{-2}p_\A p_\B \right)
+g_\ind L p^{-2}p_\A p_\B  \right)
g_\ind L p^2 \rme^{-p^2\left( C(y-y',\sigma)+C(x-x',\tau) \right)} \times} \nn\\
&& \hspace{2.6cm} \times R(y-y',\sigma) R(x-x',\tau)h(x'-y')\nn\\
&& \qquad = \DynJ\left(  g_\ind T \left(1-\frac1d \right)+g_\ind L \frac1d    
\right)
	 g_\ind L\times \nn\\
&& \hspace{2.6cm} \times\frac d2 \left( C(y-y',\sigma)+C(x-x',\tau) \right)^{-d/2-1}
 \dot C(y-y',\sigma) \dot C(x-x',\tau) h(x'-y')\nn \\
&& \qquad = \DynJ\left(  g_\ind T \left(1-\frac1d \right)+g_\ind L \frac1d    
\right)
	 g_\ind L\times \nn\\
&& \hspace{2.6cm} \times
\frac 2 {d-2} \frac \partial {\partial \tau}
\frac \partial {\partial \sigma}
\left( C(y-y',\sigma)+C(x-x',\tau) \right)^{-d/2+1} h((x-y)-(x'-y')) \ .
\nn\\ &&
\eea
In the last formula we have rearranged the product of $h$-functions,
using the fact that they are always $\delta$-distributions for the cases of  
interest.
Note also that we have used the (perturbative) FDT, \Eq{FDT}, for the
first transformation.

%Let us also mention that the same analyticity for small $k$ prevents
%corrections to disorder when its leading term in Fourier
%representation starts like
%\be
%\int_k |k|^{-a} \rme^{ik(r-r')} \qquad\mbox{with} \qquad a>0 \ .
%\ee
%As a consequense, the disorder vertex is not renormalized in
%the long-range case, and the renormalization of the disorder is
%completely driven by the renormalization of the field $r$.

The second possible way to do the contraction, see  $D_2$ on figure \ref{disren}, 
is performed similarly. The leading term is
\bea
\DynJ\ \frac{-1}d g_\ind L^2 %\times \nn\\
%&& \hspace{3cm} \times
\frac 2 {d-2} \frac \partial {\partial \tau}
\frac \partial {\partial \sigma}
\left( C(y-y',\sigma)+C(x-x',\tau) \right)^{-d/2+1} h((x-y)-(x'-y')) \ . \nn\\
&&
\eea

Note that there are two other possible contractions,
which can be obtained from $D_1$ and $D_2$ by replacing
$\tau$ and $\sigma$ with $-\tau$ and $-\sigma$ respectively.
Together they add up to
\bea
&&\DynJ\  g_\ind L g_\ind T %\times \nn\\
%&& \hspace{3cm} \times
\frac 4 {d-2} \left( 1-\frac1d \right)\frac \partial {\partial \tau}
\frac \partial {\partial \sigma}
\left( C(y-y',\sigma)+C(x-x',\tau) \right)^{-d/2+1} \times \nn\\
&& \qquad \qquad\times h((x-y)-(x'-y'))
\ .
\label{DO diagram}
\eea
This result is remarkable in several respects: First, the contribution
to the renormalization of disorder from the disorder--disorder
contraction is {\em isotropic}. We will see that this stabilizes the
isotropic fixed point. Second, there is no divergent contribution
in the purely transversal or purely longitudinal case at 1-loop order.
(Note however that there are finite contributions in the
transversal case, see appendix \ref{finite cor to DO}, such that 
isotropic disorder will be generated at 2-loop order.) 

\subsection{Beyond the leading order}
The MOPE-coefficients evaluated in the last two subsections
are sufficient to derive the RG-equations at leading order. 
Subleading orders are analyzed similarly, with the result that
the only diverging contributions, and thus those which have to 
be renormalized, are proportional to the terms already present
in the action. An explicit example is the 
second order term from the contraction of three disorder-vertices
towards a single one as
\be
\DynP \ \longrightarrow \ \DynJ \ .
\ee

Another question is, whether the fixed point as given 
in leading order, is stable towards perturbations w.r.t.\ subleading 
terms, i.e.\ those which are irrelevant at 
the trivial (Gaussian) fixed point. This is a difficult
question, and cannot be studied in its generality here. 
However, we shall show later in section 
\ref{Inclusion of SA} that the physically 
important perturbation by  self-avoidance is indeed irrelevant
at the 1-loop fixed point to be described below.

\subsection{List of the MOPE-coefficients}
\label{List MOPE}
We have identified the 1-loop divergences, and shall now
compute explicitly the counter-terms needed to render the
action finite to 1-loop order. To summarize, we first give a list of
the corrections to each term in the original action in the
form of the MOPE-coefficients for
the diagonal case ($g_\ind L=g_\ind T$). Then the MOPE-coefficients
for the  transversal and longitudinal disorder are expressed
with its help. To simplify notations,
the fluctuation-dissipation theorem \eq{FDT} has been used.
Concerning the notation: $( A | B )$ means the contribution from the  
contraction $A$
proportional to $B$, and is expressed in terms of the relative
distances. From \Eq{loc cont 1} we obtain
\bea \label{MC1}
\MOPE \DynL  \DynF  &=& C(x,t)^{-a/2}h(x) \\
\label{MC2}
\MOPE \DynLL  \DynF  &=& \frac 1 d\MOPE \DynL   
\DynF  \\
\label{MC3}
\MOPE \DynLT  \DynF  &=& \left( 1- \frac 1 d\right)\MOPE  
\DynL  \DynF 
\ .
\eea
The other local divergences are determined from  \Eq{loc cont 2}:
\bea \label{MC4}
\MOPE \DynL  \DynE  &=& \frac 1 d\, t\frac{\partial}{\partial t}
C(x,t)^{-a/2} h(x)\\
\label{MC5}
\MOPE\DynLL  \DynE  &=& \MOPE \DynL  \DynE  \\
\label{MC6}
\MOPE \DynLT  \DynE  &=& 0
\eea
and
\bea \label{MC7}
\MOPE \DynL  \DynA  &=& \frac 1 {2dD} \, x^2\,
\frac{\partial}{\partial t}
C(x,t)^{-a/2} h(x)\\
\label{MC8}
\MOPE \DynLL  \DynA  &=& \MOPE \DynL  \DynA  \\
\label{MC9}
\MOPE \DynLT  \DynA  &=& 0
\ .
\eea
Diagrams proportional to $\DynA$, renormalizing the stiffness,  will of
course vanish in the directed case.

The  bilocal divergences, proportional to the
disorder vertex, are
\bea\label{MC10}
\MOPE \DynK  \DynJ  &=& \frac 4{d-2} \left(1-\frac1d\right)
\frac{\partial}{\partial \tau}\frac{\partial}{\partial \sigma}
\left(C(x,\tau)+C(y,\sigma)\right)^{-d/2+1} h(x-y)\  \ \ \ \ \ \nn\\
\label{MC11}\\
\MOPE {\DynK  {_\ind L \atop ^\ind T}} \DynJ  &=&
\half \MOPE \DynK  \DynJ \\
\label{MC12}
\MOPE {\DynK  {_\ind L \atop ^\ind L}} \DynJ  &=& 0 \\
\label{MC13}
\MOPE {\DynK  {_\ind T \atop ^\ind T}} \DynJ  &=& 0 \ .
\eea
Remember that in the LR-case, this class of  diagrams vanishes.

\subsection{The residues in the directed case}
\label{The residues in the directed case}
Let us start by analyzing the directed case, i.e.~$h(x)=\delta^D(x)$.
Using the MOPE coefficients given in the last subsection, we can
calculate the diverging part of the
diagrams as pole in
\be
	\E= 2 - \zeta_0 a \ ,
\ee
where
\be
 \zeta_0= \frac{2-D} 2 \ ,
\ee
and setting $a=d$ for SR.
For $\E>0$, the integrals are UV-convergent and we put
an IR-cutoff $L$ in position space and $L^2$ in time.
Note that reintroducing $\lambda$ this would yield a
``physical'' time cut-off $\lambda L^2$.
The precise procedure is irrelevant at 1-loop order,
cf.\ \cite{WieseDavid95}. Note that
\be
h(x) =\delta^D(x)
\ee
always eliminates one space integration.

Start with an explicit example. Let us define
a Feynman-diagram as the integral of the MOPE-coefficient over
all distances in space and time, bounded
by $L$ and $L^2$ respectively (for details, see \cite{WieseDavid96b}). For  
the MOPE-coefficient $\MOPE \DynL  \DynF $
this reads
\bea
\DIAG \DynL  \DynF_{\!\!L}
&:=& \int_{t<L^2} \MOPE \DynL  \DynF  \nn\\
&=& \int_0^{L^2} \rmd t \,  C(0,t)^{-a/2} \nn\\
&=& P \int_0^{L^2} \rmd t \,   t^{-\zeta_0 a/2} \nn\\
&=&  P \frac2\E L^{\E}
\eea
where
\be \label{P}
P= C(0,t=1)^{-a/2}= \left( \frac{(2-D) (4\pi)^{D/2}}2\right)^{a/2}
\ .
\ee
We introduce the following abbreviation for the pole-term, which has
to be subtracted in the minimal subtraction scheme
\be \label{P'}
\diagA^\ind{dir}:=\DIAG \DynL  \DynF_{\!\!\E} = 2 P
\ .
\ee
The divergent contribution to the friction coefficient is
obtained from
\bea
\DIAG \DynL  \DynE_{\!\!L}
&=& \int_{t<L^2} \MOPE \DynL  \DynE  \nn\\
&=& \int_0^{L^2} \rmd t \, \frac td \frac{\partial}{\partial t} C(0,t)^{-a/2}\nn\\
&=& -\int_0^{L^2} \rmd t \, \frac 1d  C(0,t)^{-a/2} + {\cal O}(\E^0)\nn\\
&=&  - \frac 1 d P \int_0^{L^2} \rmd t \,   t^{-\zeta_0 a/2} + {\cal O}(\E^0)\nn\\
&=&  - \frac 1 d P \frac2\E L^{\E} + {\cal O}(\E^0)\ ,
\eea
with $P$ introduced in \Eq{P}.
The pole-term, which has
to be subtracted in the minimal subtraction scheme, is
\bea
\DIAG \DynL  \DynE_{\!\!\E} &=& -\frac 2 d
 P
\ .
\eea
Remarkably
\be \label{rel1}
\DIAG \DynL  \DynF_{\!\!\E}
=-d \DIAG \DynL  \DynE_{\!\!\E} \ .
\ee
This relation can be viewed as a consequence of the (perturbative)
FDT, and is responsible for the non-renormalization of the temperature
in the longitudinal case, where a FDT holds in the full theory.

Since the disorder-vertex is local in $x$-space (``statistical
tilt invariance''), there is
no  contribution
proportional to $\tilde r (-\Delta) r$:
\be \label{tilt-cons}
\DIAG \DynL  \DynA_{\!\!\E} = 0
\ .
\ee
The renormalization of the disorder interaction is (for $a=d$ only)
 obtained from
\bea
\lefteqn{\DIAG \DynK  \DynJ_{\!\! L}=} \nn\\
&=&\int_0^L \rmd^{D}\! x \int_0^{L^2} \rmd \tau \int_0^{L^2} \rmd \sigma\,
\frac 4{d-2} \left(1-\frac1d\right)
\frac{\partial}{\partial \tau}\frac{\partial}{\partial \sigma}
\left(C(x,\tau)+C(x,\sigma)\right)^{-d/2+1}\nn\\
&=& \frac 4{d-2} \left(1-\frac1d\right)
S_D\int_0^L \frac{\rmd x}x x^D\,\left(2C(x,0)\right)^{-d/2+1} +  
\mbox{finite terms}\ ,
\eea
resulting in
\be \label{Q}
\diagB^\ind{dir}:=\DIAG \DynK  \DynJ_{\!\E} =
\frac 4{d-2} \left(1-\frac1d\right) Q \ ,
\ee
where
\be \label{Q'}
Q = \left(  \frac2{(2-D)S_D}  \right)^{1-d/2} S_D
\ .
\ee
with $S_D = 2 \pi^{D/2}/\Gamma[D/2]$.
This terminates the evaluation of the Feynman-diagrams in the directed
case.

\subsection{The residues in the isotropic (non-directed) case}
\label{residues isotropic}
\begin{figure}[t] 
\centerline{%\fboxsep0mm\fbox%
{\unitlength1mm \begin{picture}(110,95)
\put(0,80){$\tilde I(D)$}
\put(75,0){$D$}
\put(10,5){\epsfxsize=10cm \epsfbox{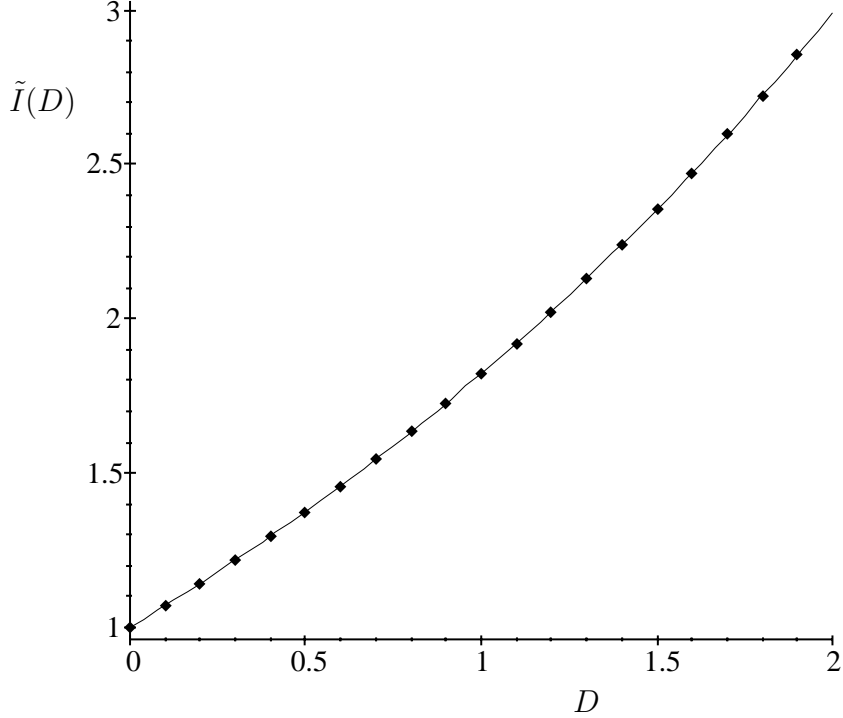}}
\end{picture}}}
%\vspace{0.3cm}
\caption{The function $\tilde I(D)$. The dots have been obtained numerically,
the interpolating line is a polynomial fit of fifth degree, later used in
the extrapolations.}
\label{tilde I(D)}
\end{figure}
In the isotropic, i.e.\ non-directed case with
\be
h(x)=1\ ,
\ee
the dimensional regularization parameter is
\be
	\E= 2 +D - \zeta_0 a \ .
\ee
We  now parallel the calculation of diagrams from the last subsection.
The term proportional to $\DynF$ (see \Eq{MC1}) is
\bea
\DIAG \DynL  \DynF_{\!\! L} &=& \int_0^L \rmd^D\!x  
\int_0^{L^2} \rmd t
\,\MOPE \DynL  \DynF  \nn\\
&=& S_D\int_0^L \frac{\rmd x}x x^D \int_0^{L^2} \rmd t \, C(x,t)^{-a/2}\nn\\
&=& S_D\frac{L^\E}{\E} \int_0^{\infty}\rmd t \, C(1,t)^{-a/2} +\mbox{finite}\ ,
\eea
where in addition $a$ has be set to $a_c(D)$, defined by $\E(D,a_c)=0$.
Here $a_c = 2 (2+D)/(2-D)$.
The residue is
\bea
\diagA^\ind{iso}:=\DIAG \DynL  \DynF_{\!\!\E} &=&
{S_D} \int_0^\infty \rmd t\, C(1,t)^{-a_c(D)/2} \nn\\
&=:& I(D) \ .
\label{def I(D)}
\eea
where the expression for the free correlator (obtained e.g.\ via
time integration of the free response function) reads
\bea
 C(1,t) &=& \frac{\Gamma\left(\frac{D}{2}\right)}{2 (2-D) \pi^{\frac{D}{2}}} +
\frac{1}{4 \pi^{\frac{D}{2}}} 
\Gamma\left(-1+\frac{D}{2}, \frac{1}{4 t}\right)  \nn\\
&=& \frac{1}{2 (2-D) \pi^{\frac{D}{2}}}
\left[ \Gamma\left(\frac{D}{2}\right) + (4 t)^{\frac{2-D}{2}} 
e^{- \frac{1}{4 t}} -
\Gamma\left(\frac{D}{2},\frac{1}{4 t}\right) \right] \ ,
\eea
where we have abbreviated the incomplete $\Gamma$-function 
as
\be
\Gamma(a,b)=\int_b^\infty \rmd t\, t^{a-1}\rme^{-t} \ .
\ee

We could not find a closed analytical form for the integral
$I(D)$. 
Using the approximation for the correlator
\be \label{approx correl}
	C(1,t) \approx \frac 2 {(2-D) (4\pi)^{D/2}}
\left( |t| +\frac 2 D\right)^{\zeta_0} \ ,
\ee
which works astonishingly well numerically, we obtain
for the integral $I(D)$
\bea
I(D)&=& S_D \left( \frac{(2-D)(4 \pi)^{D/2}}2 \right)^{a/2}
\left( \frac 2 D \right)^{1-D/2} \tilde I(D) \nn\\
\label{I(D)}
&=& S_D \left( S_D (2-D) \right)^{a/2} \times \left( 2^{D-2} \Gamma\left(
\frac D 2 \right) \right)^{a/2} 
\left( \frac 2D \right)^{1-D/2} \tilde  I(D) \ .
\eea
where $a$ must be set to $a=a_c(D)$.
If the  approximation in \Eq{approx correl} was exact,
 $\tilde I(D)$ would equal 1. This is not the case, and
the value for $\tilde I(D)$ obtained from
numerical integration is plotted on figure \ref{tilde I(D)}.
It is a slowly varying function though, well behaved
at $D=0$ and $D=2$. 
Note also that in \Eq{I(D)} a factor of $S_D \left( S_D (2-D) \right)^{a/2}$
is factorized
which is common to all other diagrams and will therefore cancel at the end.

Proceeding equivalently, the diagram correcting the friction coefficient
is calculated as
\bea
\DIAG \DynL  \DynE_{\!\!\E} &=&
-\frac{S_D}{d} \int_0^\infty \rmd t\, C(1,t)^{-a/2} \nn\\
&=& - \frac{1}{d} I(D)  \ .
\eea
Like in the directed case, we have
\be \label{rel2}
\DIAG \DynL  \DynF_{\!\!\E} = -d
\DIAG \DynL  \DynE_{\!\!\E}
\ . \label{fric}
\ee
This relation can  be used to simplify
the  RG-calculations. As in the directed case, it is a reflection of
the FDT, valid in  perturbation theory, and in the longitudinal case
in the full theory.

In contrast to the directed case,  the stiffness of the membrane is
renormalized through
\bea
\DIAG \DynL  \DynA_{\!\! L}&=& \int_0^L \rmd^D\!x  
\int_0^{L^2} \rmd t\,
\MOPE \DynL  \DynA  \nn\\
&=&\frac 1 {2dD} \int_0^L \rmd^D\!x \int_0^{L^2} \rmd t \, x^2\,
\frac{\partial}{\partial t}
C(x,t)^{-a/2} \nn\\
&=&-\frac 1 {2dD} S_D \int_0^L \frac{\rmd x}x   x^{D+2}\,
C(x,0)^{-a/2} +\mbox{finite}\nn\\
&=&-\frac 1 {2dD} S_D ((2-D)S_D)^{a/2} \frac{L^\E}{\E}+\mbox{finite}\ .
\label{stiff1}
\eea
The residue is therefore given by
\be
-\frac{\diagC^\ind{iso}}d:=\DIAG \DynL  \DynA_{\!\!\E}= -\frac{1}{2dD}\times S_D  
\left( (2-D)
 S_D\right)^{a/2}
\ . \label{stiff2}
\ee
Finally, the  diagram correcting the disorder vertex is
(for $a=d$ only) evaluated as follows:
\bea
\lefteqn{\DIAG \DynK  \DynJ _{\!\! L}=} \nn\\
&=&\int_0^L \rmd^{D}\! x \int_0^ L \rmd^{D}\! y \int_0^{L^2} \rmd  
\tau \int_0^{L^2} \rmd \sigma\,
\frac 4{d-2} \left(1-\frac1d\right)
\frac{\partial}{\partial \tau}\frac{\partial}{\partial \sigma}
\left(C(x,\tau)+C(y,\sigma)\right)^{-d/2+1}\nn\\
&=& \frac 4{d-2} \left(1-\frac1d\right)
S_D^2 \int_0^L \frac{\rmd x}x x^D \int_0^L \frac{\rmd y}y  
y^D\,\left(C(x,0)+C(y,0)\right)^{-d/2+1} + \mbox{finite terms}\nn\\
&=& \frac 4{d-2} \left(1-\frac1d\right)
S_D^2 \int_0^\infty \frac{\rmd x}x x^D
\,\left(C(x,0)+C(1,0)\right)^{-d/2+1} \frac{L^\E}{\E}+ \mbox{finite terms}\nn\\
&=& \frac 4{d-2} \left(1-\frac1d\right)
S_D^2 \left((2-D)S_D\right)^{d/2-1}\int_0^\infty \frac{\rmd x}x x^D
\,\left(x^{2-D}+1\right)^{-d/2+1} \frac{L^\E}{\E}+ \mbox{finite terms}\nn\\
&=& \frac 4{d-2} \left(1-\frac1d\right)
S_D^2 \left((2-D)S_D\right)^{d/2-1}
\frac{1}{2-D}\frac{\Gamma^2\left( \frac{D}{2-D}\right)}{\Gamma\left(  
\frac{2D}{2-D}\right)}\frac{L^\E}{\E}
+ \mbox{finite terms} \ .
\label{vert1}
\eea
The residue is given by the expression
\be
\diagB^\ind{iso}:=\!
\DIAG \DynK  \DynJ_{\!\E} =
\frac 4{d-2} \left(1-\frac1d\right)  \frac 1 {(2-D)^2}
\frac{\Gamma^2\left( \frac D {2-D} \right)}
{\Gamma\left( \frac {2D} {2-D} \right)}
\times S_D\left( (2-D)S_D \right)^{d/2}
\ . \label{vert2}
\ee
The factor involving the $\Gamma$-functions
is familiar from the renormalization of
the interaction-vertex for self-avoiding membranes.

\subsection{The particle limit}\label{particle limit}
In this section, we demonstrate that our model reproduces, 
both in the directed and in the isotropic case, the known results 
for the  particle upon taking the limit $D\to0$. Some subtleties
 have to be taken into account.
Study first the directed case. Then unambiguously
\be
\lim_{D\to0}\, \DIAG\DynL\DynF_{\!\E}^\ind{dir}
  =2 \ .
\ee
The other independent and non-vanishing digram is in the 
limit of small $D$ and $\E$
\be \label{particle1}
\DIAG \DynK  \DynJ_{\!\E}^\ind{dir} 
\approx \frac{2D}{d-2}  \ .
\ee
Note that it was calculated under the tacit assumption 
that $\E\to 0$ is taken {\em first}. In order to perform the limit $D\to0$, we insert
$d-2=2\frac{D-\E}{2-D}$ into \Eq{particle1} and then  perform the 
limit $\E\to0$. We obtain   
\be \label{particle2}
 \DIAG \DynK  \DynJ_{\!\E}
 \approx \frac{D(2-D)}{D-\E}
\,\  \stackrel{\E\to0}{\longrightarrow}\  2-D \ \stackrel{D\to0}{\longrightarrow} 2\ .
\ee
One easily verifies that these results coincide with a direct 
calculation for the particle, obtained by using the single 
particle propagator and response functions and by omitting the 
space integration from the very beginning.
The isotropic case reduces to the particle case along the 
same lines. First 
\bea
\lim_{D\to0}\, \DIAG\DynL\DynF_{\!\E}^\ind{iso}
	 &=& 2 \nn\\
\lim_{D\to0} \DIAG \DynL  \DynA_{\!\E}^\ind{iso}
	 &=& 0
\eea
The last diagram is 
\be
\DIAG \DynK  \DynJ_{\!\E}^\ind{iso}
\approx \frac{4D}{d-2}
\ee
Note the apparent difference in factor of 2 to \Eq{particle1}. 
Inserting now $d-2=2\frac{2D-\E}{2-D}$ yields
\be
 \DIAG \DynK  \DynJ_{\!\E}^\ind{iso}
  \approx \,  -\frac{2D(2-D)}{2D-\E} 
\,\  \stackrel{\E\to0}{\longrightarrow}\  2-D \ \stackrel{D\to0}{\longrightarrow} 2 
\ee
and thus the same final result, cmp.\ \Eq{particle2}.

Our calculations, at least at this order,
 thus always include the particle case. 

\subsection{The flow-equations}
\label{The flow-equations}
We now introduce renormalized quantities. Going back to
the original model, \Eq{model}, we  take care to reinsert
the factors of $\lambda$, which were dropped  in intermediate
expressions.  Define
\bea
r_0 &=& \sqrt{Z}\, r\nn\\
\tilde r_0 &=& \sqrt{\tilde Z}\, \tilde r\\
\lambda_0 &=& Z_\lambda \,\lambda\nn
\ .
\eea
Explicitly,  the counter-terms are as follows. First, introducing a counter-term
for $\DynE$ yields
\bea
\sqrt{Z\tilde Z} &=& 1+\DIAG\DynLT\DynE_{\!\!\E} \frac{g_\ind T}{\E}
					  +\DIAG {\DynLL } \DynE  _{\!\!\E}
						\frac{g_\ind L}{\E} \nn \\
	&=& 1+ \DIAG \DynL  \DynE_{\!\!\E} \frac{g_\ind L}{\E}
\ .
\eea
The counter-term for $\lambda \DynF$ is given by
\bea
\tilde Z Z_\lambda &=& 1- \DIAG \DynLT  \DynF_{\!\!\E} 
						\frac{g_\ind T}{\E}
					- \DIAG \DynLL  \DynF_{\!\!\E}
						\frac{g_\ind L}{\E} \nn \\
	&=& 1- \DIAG \DynL  \DynF_{\!\!\E}
\left( \frac1 d \frac{g_\ind L}{\E} + \left( 1-\frac1d \right) \frac{g_\ind T}{\E}
\right)
\ .
\eea
In the isotropic (non-directed) case, there is also a counter-term for  
$\lambda \DynA$ namely
\bea
\sqrt{Z\tilde Z} Z_\lambda &=& 1+ \DIAG \DynLT\DynA_{\!\!\E}
						\frac{g_\ind T}{\E}
					+ \DIAG \DynLL  \DynA _{\!\!\E}
						\frac{g_\ind L}{\E} \nn \\
	&=& 1+ \DIAG \DynL  \DynA_{\!\!\E} \frac{g_\ind L}{\E}
\ .
\eea
These equations are solved for $Z$, $\tilde Z$ and $Z_\lambda$:
\bea
Z_\lambda=1&+&\left( \DIAG \DynL  \DynA_{\!\!\E}-
\DIAG \DynL  \DynE_{\!\!\E}\right) \frac{g_\ind L}{\E} \\
\tilde Z = 1&+&\left( \DIAG \DynL  \DynE_{\!\!\E}-
\DIAG \DynL  \DynA_{\!\!\E}\right) \frac{g_\ind L}{\E} \nn\\
&-&\DIAG \DynL  \DynF_{\!\!\E}
\left( \frac1d \frac{g_\ind L}\E +\left( 1-\frac1d\right) \frac{g_\ind T}\E  
\right)\\
Z = 1&+&\left( \DIAG \DynL  \DynA_{\!\!\E}+
\DIAG \DynL  \DynE_{\!\!\E}\right) \frac{g_\ind L}{\E}\nn\\
&+&\DIAG \DynL  \DynF_{\!\!\E}
\left( \frac1d \frac{g_\ind L}\E +\left( 1-\frac1d\right) \frac{g_\ind T}\E \right)
\ .
\eea
It is more complicated to introduce renormalized couplings $g_\ind L$ and
$g_\ind T$. Set
\be \label{ren noise}
_\A \DynJ_\B\left( P^\ind L_{\A\B} g_\ind L + P^\ind T_{\A\B} g_\ind T  
\right) \lambda^2 \mu^\E
=_\A \DynJ_\B^0 \left( P^\ind L_{\A\B} g^0_\ind L + P^\ind T_{\A\B}  
g^0_\ind T \right) \lambda_0^2
\ .
\ee
This equation is to be understood such that quantities  on the l.h.s.\ are  
renormalized, and those  on the r.h.s.\ are bare.  Solving the
system of equations yields
\bea
g_\ind L^0 &=& \mu^\E Z_\lambda^{-2} Z^{a/2} \tilde Z^{-1} \left(g_\ind L -
\half \DIAG \DynK  \DynJ_{\!\!\E} \frac{g_\ind L g_\ind  
T}{\E} \right)
+{\cal O}(g^3)\\
g_\ind T^0 &=& \mu^\E Z_\lambda^{-2} Z^{a/2} \tilde Z^{-1} \left(g_\ind T -
\half \DIAG \DynK  \DynJ_{\!\!\E} \frac{g_\ind L g_\ind  
T}{\E} \right)
+{\cal O}(g^3)
\ .
\eea
We are now in a position, to define the so-called $\beta$-functions,
quantifying the flow of the renormalized theory upon a variation of
the renormalization scale, through
\bea
\beta_\ind L(g_\ind L,g_\ind T) &:=& \mu \frac{\partial}{\partial \mu  
}\lts_0 g_\ind L\\
\beta_\ind T(g_\ind L,g_\ind T) &:=& \mu \frac{\partial}{\partial \mu  
}\lts_0 g_\ind T
\ .
\eea
The $\beta$-function for the longitudinal disorder coupling is
\bea \label{betaL}
\beta_\ind L(g_\ind L,g_\ind T) &=& \mu \frac{\partial}{\partial \mu }\lts_0  
g_\ind L \\
  &=& \mu \frac{\partial}{\partial \mu }
\left( g^0_\ind L \mu^{-\E} Z_\lambda^2 Z^{-a/2} \tilde Z
+ \half \DIAG \DynK  \DynJ_{\!\!\E}
\frac{g_\ind L^0 g_\ind T^0}\E \mu^{-2\E} \right) +{\cal O}(g^3) \nn\\
&=& -\E g_\ind L + \left(\frac{a+2}2 \DIAG \DynL  \DynE_{\!\!\E}
+\frac{a+2}{2d} \DIAG \DynL  \DynF_{\!\!\E}\right.\nn\\
&&\left. \hspace{3cm}+\frac{a-2}2\DIAG \DynL  \DynA_{\!\!\E}
  \right) g_\ind L^2 \nn\\
 && + \left(\left(1-\frac1d\right) \frac{a+2}2
   \DIAG \DynL  \DynF_{\!\!\E}
-\half \DIAG \DynK  \DynJ_{\!\!\E} \right) g_\ind L g_\ind T \nn\\
&&+{\cal O}(g^3)
\ .
\eea
Using the relation
\be \label{rel}
\DIAG \DynL  \DynF_{\!\!\E}
=-d \DIAG \DynL  \DynE_{\!\!\E} \ ,
\ee
valid both in the directed, see \Eq{rel1}, and isotropic,
see \Eq{rel2}, case, \Eq{betaL}, results in the simplification
\bea
\beta_\ind L(g_\ind L,g_\ind T) = -\E g_\ind L
&+&\frac{a-2}2\DIAG \DynL  \DynA_{\!\!\E}
   g_\ind L^2 \nn\\
  \qquad &+& \left(\left(1-\frac1d\right) \frac{a+2}2
   \DIAG \DynL  \DynF_{\!\!\E}
-\half \DIAG \DynK  \DynJ_{\!\!\E} \right) g_\ind L g_\ind T \nn\\
\label{beta L}
& +&{\cal O}(g^3)\ .
\eea
By an analogous calculation, we obtain for the $\beta$-function
of the transversal disorder
\bea
\beta_\ind T(g_\ind L,g_\ind T) &=& \mu \frac{\partial}{\partial \mu }\lts_0  
g_\ind T \\
  &=& \mu \frac{\partial}{\partial \mu }
\left( g^0_\ind T \mu^{-\E} Z_\lambda^2 Z^{-a/2} \tilde Z
+ \half \DIAG \DynK  \DynJ_{\!\!\E}
\frac{g_\ind L^0 g_\ind T^0}\E \mu^{-2\E} \right) +{\cal O}(g^3) \nn\\
&=& -\E g_\ind T + \Bigg(\frac{a+2}2 \DIAG \DynL  \DynE_{\!\!\E}
+\frac{a+2}{2d} \DIAG \DynL  \DynF_{\!\!\E}\nn\\
 & & \qquad \qquad\quad
+\frac{a-2}2\DIAG \DynL  \DynA_{\!\!\E}  -\half \DIAG  
\DynK  \DynJ_{\!\!\E}  \Bigg) g_\ind L g_\ind T  \nn\\
&&+ \left(\left(1-\frac1d\right) \frac{a+2}2
   \DIAG \DynL  \DynF_{\!\!\E}
 \right)  g_\ind T^2 +{\cal O}(g^3)
\ .
\eea
With the simplification given by \Eq{rel} the final result
is
\bea
\beta_\ind T(g_\ind L,g_\ind T)
&=& -\E g_\ind T + \Bigg(
\frac{a-2}2\DIAG \DynL  \DynA_{\!\!\E}  -\half \DIAG  
\DynK  \DynJ_{\!\!\E}  \Bigg)
g_\ind L g_\ind T  \nn\\ \label{beta T}
&&\qquad\ \,+ \left(1-\frac1d\right) \frac{a+2}2
   \DIAG \DynL  \DynF_{\!\!\E}  g_\ind T^2  +{\cal O}(g^3)
\ .
\eea
In view of a possible stable fixed point for $g_\ind L=g_\ind T$, it is  
interesting also to study the  flow-equation
for the difference of the both couplings $g_\ind L-g_\ind T$:
\bea \label{beta diff}
\hspace{-1cm}(\beta_\ind L - \beta_\ind T)(g_\ind L,g_\ind T)
&=& -\E (g_\ind L- g_\ind T) +
\frac{a-2}2\DIAG \DynL  \DynA_{\!\!\E}
g_\ind L (g_\ind L-g_\ind T)  \nn\\
& & +\left(1-\frac1d\right) \frac{a+2}2
   \DIAG \DynL  \DynF_{\!\!\E}  g_\ind T (g_\ind L-g_\ind  
T) +{\cal O}(g^3)
\ .
\eea

\subsection{Other RG-functions and critical exponents}
\label{Other RG-functions and critical exponents}
Several critical exponents can be identified. Taking into
account scaling relations among them, there are three independent
exponents in the directed case, and four in the isotropic case.
The first is the full scaling-dimension of the membrane,
given by the fixed point value $\zeta^*$ of
\bea
	\zeta(g_\ind L,g_\ind T)&=&\frac{2-D}2 -\half \mu  
\frac{\partial}{\partial \mu}\lts_0 \ln Z \nn\\
&=& \frac{2-D}2 - \half \left( \beta_\ind L(g_\ind L,g_\ind T)
	\frac{\partial}{\partial g_\ind L} \ln Z +
	\beta_\ind T(g_\ind L,g_\ind T) \frac{\partial}{\partial g_\ind T}  
\ln Z \right)\nn\\
&=& \frac{2-D}2 +\frac{ g_\ind L}2
	 \DIAG \DynL  \DynA_{\!\!\E}
	+ \frac{g_\ind T}2 \left(1-\frac1d\right) \DIAG \DynL  \DynF  
_{\!\!\E}
+{\cal O}(g^2) \ . \nn\\&&
\eea
We again used \Eq{rel} to simplify the result.
The   dynamical exponent $z$ is
\bea
z(g_\ind L,g_\ind T)&=& 2- \mu \frac{\partial}{\partial \mu}\lts_0 \ln  
Z_\lambda \nn\\
&=& 2 - \left( \beta_\ind L(g_\ind L,g_\ind T)
	\frac{\partial}{\partial g_\ind L} \ln Z_\lambda +
	\beta_\ind T(g_\ind L,g_\ind T) \frac{\partial}{\partial g_\ind T}  
\ln Z_\lambda \right)\nn\\
&=& 2 + g_\ind L \left( \DIAG \DynL  \DynA_{\!\!\E}-
\DIAG \DynL  \DynE_{\!\!\E}\right) +{\cal O}(g^2)
\ .
\eea
The  (full) scaling dimension of the response field is
\bea
\tilde \zeta(g_\ind L,g_\ind T)&=&-\frac{2+D}2 -\half \mu  
\frac{\partial}{\partial \mu}\lts_0  \ln \tilde  Z \nn\\
&=& -\frac{2+D}2 - \half \left( \beta_\ind L(g_\ind L,g_\ind T)
	\frac{\partial}{\partial g_\ind L} \ln \tilde Z +
	\beta_\ind T(g_\ind L,g_\ind T) \frac{\partial}{\partial g_\ind T}   
\ln \tilde Z \right)\nn\\
&=& - \frac{2+D}2 - \frac{ g_\ind L}2 \left(
	 \DIAG \DynL  \DynA_{\!\!\E} + \frac2d  \DIAG  
\DynL  \DynF_{\!\!\E} \right) \nn\\
&&	\qquad\qquad- \frac{g_\ind T}2 \left(1-\frac1d\right) \DIAG \DynL  
 \DynF_{\!\!\E}+{\cal O}(g^2)
\ .
\eea
We used \Eq{rel} to simplify the result.

In the isotropic (non-directed) case, there is still
another independent exponent, which is called $\beta$ (not to be
confounded with the $\beta$-functions describing the flow
of the coupling-constants) and which measures
the anomalous dimension of the stiffness of the membrane.
\bea
	\beta(g_\ind L,g_\ind T) &=& -\mu \frac{\partial}{\partial \mu}  
\ln\left( \sqrt{Z \tilde Z}Z_\lambda \right)\nn\\
  &=& g_\ind L \DIAG \DynL  \DynA_{\!\!\E}+{\cal O}(g^2)
\label{exp beta def}
\ .
\eea

With the help of the above introduced exponents, we can quantify
the non-linearity in the response of the drift velocity to a uniform
applied force. Adding a small constant term $f$ to the random
force in \Eq{depart} results into a change in the dynamical functional
of the form
\be
\lambda \int_{x,t} \tilde r(x,t) \, f\ .
\ee
Since this term does not require a renormalization, we
deduce the exact exponent identity for the full scaling dimension  
$\left[f\right]$ of $f$
\be
\left[f\right] + \tilde \zeta + z + D = 0\ .
\ee
From the definition of $\beta$ in \Eq{exp beta def}, we have
a second identity, namely
\be
\beta = \zeta + \tilde \zeta +z +D -2\ .
\ee
Combining these two relations, we obtain
\be
\left[f\right] = \zeta -\beta - 2 \ .
\ee
Since the drift velocity has dimension
\be
\left[v\right]=\left[ \frac r t\right] \ ,
\ee
we deduce the scaling relation
\be
v \sim f^\phi \qquad \mbox{with} \qquad \phi=\frac{z-\zeta}{2+\beta-\zeta}\ .
\ee
Since, as we will see,
always $z>2$ and $\beta\le0$, $\phi$ is larger than 1 indicating
trapping of the membrane by the random flow.

Let us also mention another possible parameterization of our model,
where one does not introduce $\lambda$ and renormalization factors
for $\lambda$ and the field $r$, but associates a friction coefficient
$\eta$ with $\dot r$, entering into the dynamical functional as
\be
\eta \int_{x,t} \tilde r(x,t) \dot r(x,t)
\ee
and a temperature $T$ associated to disorder,  entering into
the dynamical functional as
\be
\eta T \int_{x,t} \tilde r^2(x,t) \ .
\ee
The corresponding renormalization factors then read
\bea
Z_{\eta}&\equiv&\sqrt{Z\tilde Z} \\
Z_{T}&\equiv& Z_\lambda \sqrt{\frac{\tilde Z}{Z}} \ .
\eea
It is especially enlightening to consider the renormalization
of temperature, which in  conserved systems (i.e.\ those
with longitudinal disorder) is preserved
by the fluctuation-dissipation theorem. Stated differently, temperature
appears as a parameter in its static partition function, and
dynamics is constructed such that via detailed balance the static
partition function is reproduced. Only non-conserved forces can give rise
to a change in temperature. This is read off from the explicit form
of $Z_\ind T$ at 1-loop order:
\bea
Z_\ind T= 1&-& \left( \frac1d \DIAG \DynL  \DynF_{\!\!\E}
+ \DIAG \DynL  \DynE_{\!\!\E} \right) \frac{g_\ind L}\E \nn\\
&-&\left(1-\frac1d \right) \DIAG \DynL  \DynF_{\!\!\E}
\frac{g_\ind T}{\E}\ .
\eea
Since the term proportional to $g_\ind L$ vanishes as found in \Eqs{rel1}  
and \eq{fric},
anomalous corrections to the temperature are proportional to $g_\ind T$ and
present whenever  $g_\ind T\neq 0$. Stated differently: In the presence
of non-potential forces, the fluctuation-dissipation theorem (FDT)
is violated.

\section{Results and discussion}
\label{Results and discussion}

In this section we analyze the general renormalization group flow
given by the two $\beta$-functions in \Eqs{beta L} and \eq{beta T}
where the coefficients have been computed in section \ref{The residues in  
the directed case}. We identify the fixed points and
compute the critical exponents at these fixed points. Before doing
so it is useful to recapitulate the
results obtained so far in the previous sections.

\subsection{Preliminaries}
To present conveniently the analysis below we introduce the 
following notation for the three independent coefficients
(two in the directed case) as
\bea
\diagA &:=&  \DIAG \DynL  \DynF_{\!\!\E}\\
\diagB &:=&\DIAG \DynK  \DynJ_{\!\E}
\\
\diagC &:=& - d \DIAG \DynL  \DynA_{\!\!\E} 
\eea
The explicit expressions for these coefficients
are given above: for $\diagA$ see \Eqs{P'} and
\eq{def I(D)}, for $\diagB$ see \Eqs{Q} and \eq{vert2} and for $\diagC$
(only non-zero for isotropic manifolds), see \Eq{stiff2}.
They are non negative. In terms of these three coefficients,
the $\beta$-functions of the two coupling constants
read:
\bea
\beta^g_\ind L(g_\ind L,g_\ind T) &=& -\E g_\ind L
+ \frac{1}{2} \left( \left(1-\frac1d\right) (a+2)
   \diagA - \diagB \right) g_\ind L g_\ind T 
-\frac{a-2}{2d}  \diagC  g_\ind L^2 
\label{new beta L g}
 \\
\beta^g_\ind T(g_\ind L,g_\ind T)
&=& -\E g_\ind T - \frac{1}{2} \left( \diagB +
 \frac{a - 2}{d} \diagC  \right)
g_\ind T g_\ind L 
+ \left(1-\frac1d\right) \frac{a+2}{2}
   \diagA  g^2_\ind T  
\label{new beta T g} 
\eea
where $a$ must be replaced by $d$ for short-range disorder (the crossover
between SR and LR disorder is also studied below).
Similarly, the expressions of the exponents are to lowest
order (from section \ref{Other RG-functions and critical exponents} 
above):
\bea
 \zeta(g_\ind L,g_\ind T) &=& \frac{2-D}2 + \frac{g_\ind T}2 \left(1-\frac1d\right) \diagA
- \frac{ g_\ind L}{2d} \diagC  \\
 z(g_\ind L,g_\ind T) &=& 2 + \frac{1}d (\diagA - \diagC) g_\ind L \\
  \beta(g_\ind L,g_\ind T) &=& - g_\ind L \frac{1}{d} \diagC 
\label{newexponents}
\eea
where we have used (\ref{rel2}). Note that in the end the exponents will
depend on the amplitudes only through their 
ratios $\diagB/\diagA$ and $\diagC/\diagA$ which are universal,
in contrast to the amplitudes themselves, which depend on
the specific normalizations of the model.

The coefficients of the RG equations 
(\ref{new beta L g})-(\ref{newexponents}),
including $\diagA$, $\diagB$ and $\diagC$,
depend a priori on $a$, $d$ and $D$, and their full dependent expressions are
given in the previous sections (where SR and LR disorder are treated in
a single calculation). However, since we are expanding to first order around
the line of critical dimension $d_c(D)$ (see \Eq{criticaldim}),
we can simplify  all these coefficients as follows: For short-range disorder
one can substitute $a=d=d_c(D)$ in all coefficients. For long-range
disorder we can substitute $a = a_c(D)$ everywhere, but one must
keep $d$ as an independent variable. This is the reason, why
 we defined $\diagA$ and $\diagC$
as $d$ independent, while $\diagB =0$ for LR disorder.
The final result for the coefficients used below, expressed 
only as a function of $D$ is, for directed manifolds:
\bea  
 \diagA^\ind{dir} &=&  2 \left[\frac{(2-D) (4 \pi)^{D/2}}{2}\right]^{\frac{2}{2-D}} \nn\\
 \diagB^\ind{dir} &=&  \frac{2+D}{D} \left[\frac{(2-D) \pi^{D/2}}{\Gamma(\frac D2)}\right]^{\frac{2}{2-D}} \label{coeffdir} \\
 \diagC^\ind{dir} &=& 0 \ . \nn
\eea
 For polymers ($D=1$) one finds
$\diagA = 2 \pi$, $\diagB = 3$ and the universal ratio 
$\diagA/\diagB = 2 \pi/3$.

For isotropic manifolds one has:
\bea 
 \diagA^\ind{iso} &=& \frac{1}{(2-D) 2^{2 + D}} \left(\frac{2}{D}\right)^{\frac{2-D}{2}} 
\left[\Gamma\!\left({\tx\frac{D}{2}}\right)\right]^{\frac{2+D}{2-D}} G_D \tilde I(D) 
\nn \\
\diagB^\ind{iso} &=&  \frac{2 + 3 D}{2 D (2-D)^2 (2 + D)} 
\frac{\Gamma\!\left(\frac{D}{2-D}\right)^2}{\Gamma\!\left({\frac{2 D}{2-D}}\right)} G_D \label{coeffiso} \\
\diagC^\ind{iso} &=&  \frac{1}{2 D (2 - D)} G_D \nn \ ,
\eea
where we denote by 
\be
	G_D = \left[ \frac{ 2 \pi^{D/2} (2-D)}{\Gamma(\frac D2)}\right]^{4/(2-D)}
\ee
the factor common to the three amplitudes (which therefore drops
in the final results). $\tilde I(D)$ is
plotted in Fig. (\ref{tilde I(D)}) and defined as
\be
\tilde I(D) := \left(\frac{D}{2}\right)^\frac{2-D}{2} 2^{2+D}
\int_0^{+\infty} \rmd t \left[{\tx \Gamma(\frac{D}{2}) + (4 t)^{\frac{2-D}{2}} 
\rme^{- \frac{1}{4 t}} -
\Gamma(\frac{D}{2},\frac{1}{4 t})} \right]^{- \frac{2+D}{2-D}} \ .
\ee
It takes the values $\tilde I(D=0) = 1$, $\tilde I(D=1) = 1.82202$
and $\tilde I(D=2) = 3$. For the coefficients for
$D=1$ we find $\diagA =1.79351 G_1$, $\diagB =\frac{5}{6} G_1$
and $\diagC = \frac{1}{2} G_1$.

We now specify to each of the four cases of interest and
compute the critical exponents.

\subsection{Directed manifolds with short-range disorder}%
\begin{figure}[tb]
\vspace{-5mm}
\centerline{ \fig{0.5\textwidth}{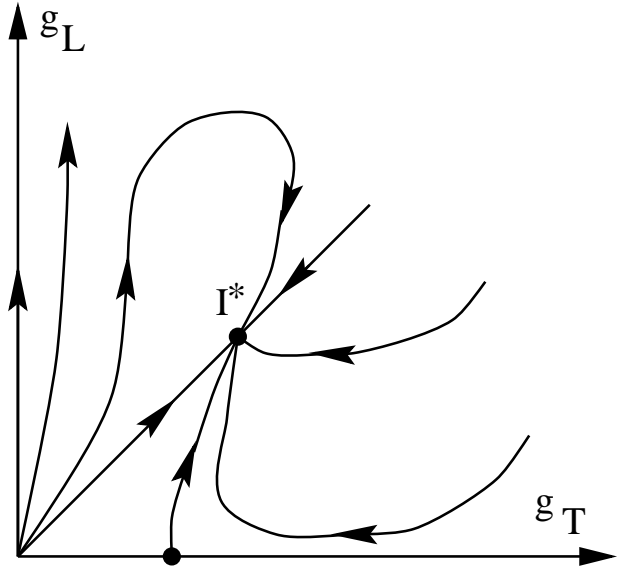} }
\vspace{1mm}
\caption{{
RG flow diagram for SR disorder.
The physics is controlled
by the fixed point $I^*$ at $g_\ind T=g_\ind L$.}}
\label{flowrg1}
\end{figure}

\label{dir SR}
The flow for this case is represented on \Fig{flowrg1}.
The expansion parameter is $\E = 2 - (2-D)d/2$.
The fixed points are found to be:

\medskip
\leftline{\bf (1) Gaussian fixed point}
The Gaussian fixed point at $g_\ind L=g_\ind T=0$ is
completely unstable for $d< d^{\ind{dir}}_c(D) = \frac4{2-D}$.

\medskip
\leftline{\bf (2) Potential disorder}
The line $g_\ind T=0$ is found to be preserved by RG
and we find a flow towards strong coupling. This case
corresponds to the dynamics of a directed manifold
in a long-range correlated random potential (short
range correlated force). It has been well studied and
it is indeed known that for $d \le d_c(D)$ (defined above)
and $D<2$, or in any $d$ for $2< D < 4$,
the physics is controlled by a
strong disorder (zero temperature) fixed point
\cite{Kardar97,BalentsFisher1993}, inaccessible
by the present method (it is accessible only via a
$D=4 -\E$ calculation). 

Further results
are available for the particle ($D=0$, $d_c=2$).
There it is known rigorously that for 
$d=1$ (Sinai model) ultra-slow logarithmic 
diffusion $r \sim (\ln t)^2$ occurs ($\nu=0$)
and it is believed (but not proven) that similar
ultra-slow diffusion also occurs for manifolds
$D>0$ (see e.g.\ \cite{vortex_review}).
At the critical dimension, for the particle
$d_c(D=0)=2$, 1- and 2-loop calculations
\cite{FisherFriedanQiuShenker1985,BouchaudComtetGeorgesLeDoussal87,%
HonkonenKarjalainen1988}
have shown that diffusion is anomalous with a continuously varying
exponent $\nu(g_\ind L)$. It was further argued that
the $\beta$ function vanishes to all orders,
and that the one loop result for $\nu(g_\ind L)$ holds exactly,
for SR disorder as well as for LR disorder
\cite{BouchaudComtetGeorgesLeDoussal87}.
While unchallenged for SR disorder (i.e.\ at $d=2$),
the exactness of the one loop exponent was recently
questioned for LR disorder in \cite{DerkachovHonkonenPismak}.
To our knowledge such issues have not been addressed
for manifolds.

\medskip
\leftline{\bf (3) Isotropic disorder fixed point}
\nopagebreak
Noting from \Eq{beta diff} that the line $g_\ind T=g_\ind L$ is preserved
by the flow at 1-loop order, we find
an isotropic fixed point at
\be
	g_\ind L=g_\ind T= g^*:=\frac{2\E d}{\left(d-1\right) (d+2)
   \diagA - d\diagB}
+{\cal O}(\E^2)
\ ,
\ee
where the coefficients are given in \Eq{coeffdir}.
Since at the critical dimension the denominator is positive for
all values of $0 \le D < 2$, $g^*$ is positive and thus  physical.
This fixed point is completely attractive, and its domain of
attraction covers all perturbative situations except the potential
case with $g_\ind T=0$. As we show below, it also controls the line
$g_\ind L=0$ (except for $D=0$).
%The
% large scale behavior of a manifold in a random short-range flow is
%determined by this isotropic fixed point.

Let us give the values of the roughness exponent $\zeta^*$ and
of the dynamical exponent $z^*$, defined in \Eqs{def zeta} and \eq{def z},
 at the fixed point:
\bea
\zeta^*&=&\zeta(g^*, g^*)= \frac{2-D}2+\frac{\E}{d+2 -\tilde R} +{\cal O}(\E^2) \\
\tilde{R}&=& \frac{\diagB}{\left(1-\frac1d\right) \diagA}
= \frac d {d-2} \left( 2^{D-1} \Gamma(D/2)\right)^{-d/2} \ ,
\label{zeta dir SR iso}
\eea
where $\E=2 - \frac{2-D}2 d$. Again one can substitute $d=d_c(D)$
in the coefficient of $\E$. Expressed as a function of
$D$ one has $\tilde{R}= \frac{2}{D} (2^{D-1} \Gamma(D/2))^{-2/(2-D)}$.
The dynamical exponent is:
\bea
z^* &=& 2+\frac{2\E}{(d-1)(d+2-\tilde{R})} +{\cal O}(\E^2)
\label{z dir SR iso}
\ .
\eea
The exponent $\beta$ equals 0 as a
consequence of the statistical tilt symmetry (see \Eq{tilt-cons}).
The exponents $\nu$ and $\phi$ can then be found
using
\be
\nu= \zeta/z \ , \qquad \phi = \frac{z-\zeta}{2-\zeta} \ .
\ee
In particular, for the polymer $D=1$, we find that
below $d_c=4$ the random flow is relevant and
 roughness and dynamical
exponents take the anomalous dimension
\bea
\zeta^*&=& \frac{1}2 \left( 1 + \frac{\E}{3 - \pi^{-1}} \right) \\
z^* &=& 2 \left(1 + \frac{\E}{6(3 - \pi^{-1})}\right)
\eea
where $\E=\frac{1}2(4-d)$. Interestingly we find that
to first order in $\E$, $\zeta > \zeta_\ind F$.

The most naive fixed $D$ expansion gives in $d=3$
(setting $\E=1$ in the above) that $\zeta=0.686$, 
$z=2.12$, $\nu=0.323$, $\phi=1.094$. In $d=2$ it gives
(setting $\E=2$) that $\zeta=0.873$, 
$z=2.249$, $\nu=0.388$, $\phi=1.22$. These estimates
can be improved since the expansion can be carried
from any point of the line $d=d_c(D)$. It is 
possible to optimize the numerical estimation of the
above critical exponents over the
expansion point $D_0$. Details of the general optimization
method are described in
\cite{WieseDavid96b} and applied to the present case in
Appendix \ref{Extrapolations}.
The results are listed in  table \ref{tab1}
and are purely indicative: we have even refrained from
giving an error bar.

{\tabcolsep2mm\begin{figure}[t]\centerline{\renewcommand{\arraystretch}{1.20}
\begin{tabular}[t]{|l|c|c|c|c|c|c|} \hline
 & $d$ &  $\zeta$& $z$ &$\nu$ & $\phi$
\\ \hline\hline
%Particle & $1$ & --- & --- & $0.76 \ldots 0.9$ & $>1$  \\  \hline
%
%  &1 & $>1$ & 2.3 & 0.9 &$>1$ \\ \cline{2-6}
Polymer & $2$ & 0.87 & 2.2 & 0.5 &$>1$ \\ \cline{2-6}
 & 3 & 0.63 & 2.1 & 0.3 &$>1$\\ \hline
%
%			& 1 &  $>1$ & 2.2  &1&$>1$\\ \cline{2-6}		 	
 			& 2 & 0.8 & 2.2  & 0.5&$>1$\\ \cline{2-6}
			& 3 & 0.5 & 2.1 & 0.33 &$>1$\\ \cline{2-6}
Membrane	& 4 & 0.4 & 2.1   &0.25&$>1$\\ \cline{2-6}
			& 6 & 0.25 & 2.07  &0.17&$>1$\\ \cline{2-6}
			& 8 & 0.2 & 2.05 &0.13&$>1$ \\ \cline{2-6}
			& 20 & 0.08 & 2.01  &0.05&$>1$ \\ \hline
\end{tabular}\renewcommand{\arraystretch}{1.0}}\vspace{2mm}
\caption{Results for directed polymers and membranes, isotropic
fixed point, SR-disorder.
Values given for the exponent $\zeta$ (obtained by extrapolating
$\zeta d$) are rather faithful.
Corrections to $z$ are small, but no plateau could be found to
optimize the extrapolations. The given values are therefore only
intended to give a rough idea. The same is true for extrapolations
for $\nu$. We have reported the values of $\nu$ obtained by setting
$D_0=D$.}%
\label{tab1}%
\end{figure}}

\medskip
\leftline{\bf (4) Transversal disorder}\nopagebreak
There is an apparent fixed point at
\be
	g_\ind L=0 \ , \qquad g_\ind T=\frac{2\E}{\left(1-\frac1d\right) (d+2) 
   \diagA } +{\cal O}(\E^2)
\ .
\ee
where the coefficients are given in \Eq{coeffdir}.
Interestingly, at this fixed point
we recover exactly the Flory-result at 1-loop order:
\bea
\zeta^* &=&\frac{2-D}2 +\frac{\E}{d+2} +{\cal O}(\E^2)
\label{zeta dir SR t}\\
z&=& 2 +{\cal O}(\E^2) \label{z dir SR t}
\ .
\eea
In the case of the particle ($D=0$) the result $\nu = \nu_\ind{F}$ was found to
be true to all orders (see e.g.\ 
\cite{HonkonenKarjalainen1988,BouchaudComtetGeorgesLeDoussal87} 
and discussion below).

However, for $D>0$ we show (see Appendix \ref{finite cor to DO}) that
finite non-transversal terms are generated in perturbation
theory. Since the above fixed point is unstable to adding
a small longitudinal disorder,  $g_\ind L>0$, these terms will
%at large scale
drive the system  towards the isotropic fixed point discussed
above.

Let us close this section by noting that we recover the known results
for the particle $D=0$, obtained via different RG techniques
in \cite{Fisher1984,FisherFriedanQiuShenker1985,%
BouchaudComtetGeorgesLeDoussal87,HonkonenKarjalainen1988}.
That this should be so was discussed in section
\ref{particle limit} but here we discuss it explicitly 
for the exponents. First we 
note that the flow diagram of Fig. 
\ref{flowrg1} has the same structure for all $D$ as the 
one found in \cite{FisherFriedanQiuShenker1985}
for the particle with an isotropic disorder fixed point and the 
transversal fixed point.
Second, expressing in all our results the coefficient 
of $\E$ as a function of
$D$ alone (using $d=d_c(D)$) yields in the limit $D \to 0$
with $\E=2-d$,
that at the isotropic fixed point
$\nu = \zeta/\E = 1/2 + O(\E^2)$ 
and $\phi = 1 + \E$ (note that one also
finds $\zeta=1 + \E/2$, $z=2 + \E$, but only
$\nu$ and $\phi$ have a physical meaning for $D=0$).
Similarly at the transversal disorder fixed point (which has a 
meaning only for the particle),
one recovers the known result $\nu=\nu_\ind{F}$ and $\phi=1$.

\subsection{Directed manifold with long-range disorder}
\label{resultsdirectedlongrange}
\begin{figure}[htb]
\vspace{-5mm}
\centerline{ \fig{0.5\textwidth}{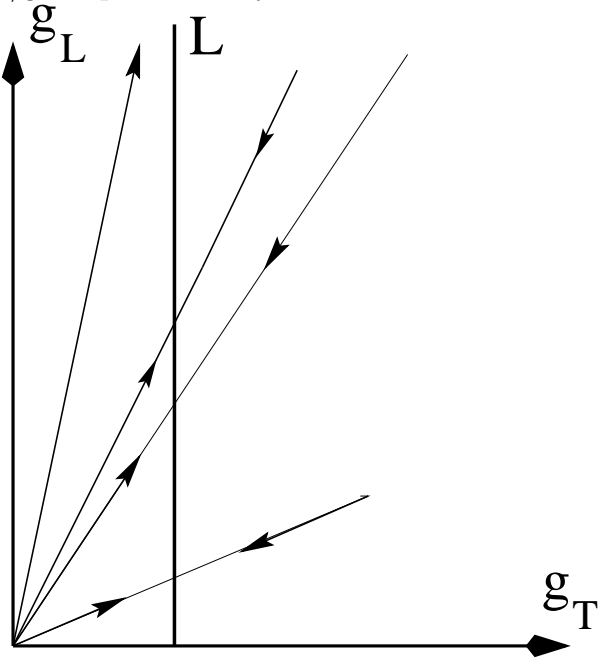} }
\vspace{1mm}
\caption{RG flow for directed manifolds in LR disorder.
 The flow is
along straight lines and  preserves the  line L
of fixed points.}
\label{flowrg3}
\end{figure}

Long range correlated disorder ($a<d$) can be studied
in an expansion near $a_c=4/(2-D)$, in the small parameter
$\delta = 2 - (2-D)\frac{a}{2}$ (with $d$ and $D$ fixed).
We have shown in section \ref{DO ren} that in this case
the vertex renormalization vanishes
\be
\diagB=\DIAG \DynK  \DynJ_{\!\E} = 0 \ .
\ee
Thus the
renormalization is driven by the corrections to
temperature and friction. The $\beta$-functions in
\Eqs{beta L} and \eq{beta T} lead to the flow diagram presented on  
\Fig{flowrg3}. Since there is no vertex renormalization,
both couplings have the same renormalization factor,
 the ratio $\kappa = g_\ind L/g_\ind T$ is preserved by RG, and the flow is along straight 
lines.%
One finds  a line L of fixed points, which
is vertical to first order in $\E$. This feature, already
noted in the particle case, $D=0$ 
\cite{HonkonenKarjalainen1988,BouchaudComtetGeorgesLeDoussal87}, thus extends
to manifolds with $D>0$. Here the fixed line
is given by:
\bea
 g^*_\ind T = \delta \frac{2 d}{(d-1)(a+2) \diagA}
\eea

For LR correlations the result for $\zeta^*$ at any
point on the fixed line can be derived to all
orders in perturbation theory. First,
there is no
counter-term to $\tilde r \Delta r$ to any order in perturbation
theory due to the fact that the interaction vertex is a difference
of the $r$-fields at the same point (``statistical tilt symmetry''),
resulting in the scaling relation
\be
\zeta^* + \tilde \zeta^* + z^* = 2-D \ ,
\ee
where $\tilde \zeta$ is the full scaling dimension of
the response field.
Second, as long as there is no renormalization of the vertex,
\be
D+2\tilde \zeta^* +2 z^* - a \zeta^* =0
\ee
is valid {\em at} the fixed point. Combining these two relations gives%
\footnote{This identity also holds at 1-loop order at the apparent
SR transversal fixed point discussed in \Eq{zeta dir SR t}, since
vertex-corrections are proportional to $g_\ind L g_\ind T$, see \Eq{DO diagram}.}
\be
 \zeta^*%_{\mbox{\scr \scriptsize F}}
= \zeta_\ind F = \frac{4-D}{2+a} \ .
\ee
Thus the Flory estimate for $\zeta$ is exact in the LR case.
Note also that $\zeta$ in that case does not depend on the type
of disorder (transversal or potential).
For a polymer $D=1$ we obtain that $\zeta=3/(2 + a)$
for $a< a_c=4$. This leads to the interesting prediction that
the polymer will become {\it overstretched} ($\zeta \ge 1$) 
when the range of the force correlation increases beyond 
$a \leq 1$.

The dynamical exponent $z$, however, and consequently $\phi$ and $\nu$,
are non-trivial, and depend continuously on the ratio
$\kappa = g_\ind T/g_\ind L$. They are given by:
\be\label{label}
z^* =  2 \left( 1 + \kappa\, \frac{\delta}{(2+a)(d-1)} \right)
\ ,
\ee
where $\delta= 2 - (2-D)\frac{a}2$. This result is 
valid to first order in $\delta$ and one may thus replace
$a$ by $a_c=4/(2-D)$ in the above formula.
It shows that the more potential the dynamics
the more glassy it is (the higher $z$).

We can now compare these results with the 
results obtained in \cite{LeDoussalCugliandoloPeliti97}.
There the exponents were obtained exactly for
$d \to \infty$ and  arbitrary $a$. A
Hartree approximation was obtained for finite $d$. The notations
of \cite{LeDoussalCugliandoloPeliti97} were different.
To translate to the present notations one must set in
\cite{LeDoussalCugliandoloPeliti97}
$\gamma \to a/2$, $g_2 \to \frac{g_\ind L}{d}$,
$g_1-g_2 \to (1-\frac{1}{d}) g_\ind T$
and $\E=2/(2-D) - \gamma \to \frac{\delta}{2-D}$.
With this correspondence the results of 
\cite{LeDoussalCugliandoloPeliti97} are as follows.
First $\zeta=\zeta_\ind{F}$, both for $d=\infty$ and for finite
$d$ 
within the Hartree approximation. This is
in agreement with the present result.
$z$ in \cite{LeDoussalCugliandoloPeliti97} however is
a highly non trivial
function of $a$, $D$ (and of $d$ in the Hartree approximation) 
implicitly given by equation (9) in 
\cite{LeDoussalCugliandoloPeliti97}. 
Remarkably, for $D<2$, expanding $z$
near $a=a_c$, which yields formula (12)
of \cite{LeDoussalCugliandoloPeliti97},
one finds to first order in $\varepsilon$ {\it exactly the same result}
as here in \Eq{label}.
Of course the two also coincide, as they should,
for $d=\infty$. Thus we find that in the long-range directed case,
the Hartree
approximation in \cite{LeDoussalCugliandoloPeliti97} 
gives the correct result for $z$ to first order, and is exact 
for $\zeta$.

Let us also give the results for the 
exponents $\nu$ and $\phi$. One finds:
\bea
&& \nu = \frac{2-D}{4} \left[1 + \delta \frac{ (\kappa^* - \kappa)(2-D)}{
2 (4-D)(d-1)} \right]  \\
&& \phi = 1 + \delta \frac{ 2(2-D) \kappa }{
(2+D) (4-D)(d-1)} 
\label{exponents directed longrange}
\eea
and thus we find that the single monomer dynamics is either subdiffusive
if $\kappa=g_\ind L/g_\ind T > \kappa^* = 2 (d-1)/(2-D)$ or 
superdiffusive if $\kappa < \kappa^*$, while
$\phi$ is always larger than $1$ (except when the
flow is purely transversal). One can also
check that our formulae agree, in the limit $D \to 0$
with the known formulae 
\cite{BouchaudComtetGeorgesLeDoussal87,BouchaudComtetGeorgesLeDoussal88,%
HonkonenKarjalainen1988} for the particle (with
$a_c=2$ and $\delta = 2 - a$).

To conclude note that in the case $\kappa=0$, 
i.e.\ in presence of LR transversal disorder alone, the
mechanism of generation of potential disorder and
barriers is different from the SR-case. LR non transversal
disorder is not generated, only SR non transversal
disorder is. (See also  the section below, where the
crossover from LR to SR is discussed). The final result
is that in  that case the exponents are given
correctly too by the above formulae setting $\kappa=0$.

\subsection{Isotropic (non-directed) manifold with short-range disorder}
\label{iso SR}
For an isotropic manifold, we find from the RG equations
\eq{beta L} and \eq{beta T} and the expressions given in section
\ref{residues isotropic} for the coefficients
that the RG flow is qualitatively
similar to the one in  the directed case depicted on \Fig{flowrg1},
with the following fixed points:

\medskip
\leftline{\bf (1) Gaussian fixed point}\nopagebreak
The Gaussian fixed point at $g_\ind L=g_\ind T=0$ is completely unstable
for $\E = 2 + D - \frac{d}2(2-D) >0$.

\medskip
\leftline{\bf (2) Potential disorder}\nopagebreak
The line $g_\ind T=0$ is preserved by RG
and again, we find a flow towards strong coupling,
which is faster than in
the directed case, as there is an additional term.
This problem describes the dynamics of an isotropic manifold
in a long-range correlated random potential (short
range correlated force). The statics of this problem 
has been much studied 
and it is indeed expected that
this problem is described by strong disorder.

In the absence of self-avoidance some features of
the problem are relatively simple to analyze.
For instance can
a polymer ($D=1$) in a $d$-dimensional random potential also be seen 
as a {\em directed} polymer in a $d+1$ dimensional random potential
(the additional coordinate being the internal dimension)
which is constant (i.e.\ correlated) in the internal dimension,
a problem extensively reinvestigated recently in relation
with flux lines in superconductors in presence of columnar
disorder 
\cite{GiamarchiLeDoussalCorrelated,BalentsKardar,Balents,vortex_review}.
It is also related to the problem of quantum localization
of a particle in a $d$ dimensional random potential (the internal
dimension being the quantum imaginary time). It is
known, and rather obvious in relation to these problems,
that if the polymer is free to move,
 it is localized in the deepest
localized state and has a finite extent.
Properties  then strongly 
depend on details, such as 
the system size, the tail of the disorder
distribution (Lifschitz tails), and the short scale cutoff \cite{CatesBall,NattermannCorrelated}.

On the other hand, the problem of many interacting directed
polymers is well defined and identical to
the problem of localization of bosons in a 
random potential. 
These properties should remain true for
membranes as well and in general we expect that in
the absence of self-avoidance,
the membrane collapses ($\zeta = 0$).

For a self-avoiding
polymer the problem is better defined. Self-avoidance is
then relevant (see section \ref{Inclusion of SA}) and properties will
depend  on whether one end of the polymer is held
fixed or is allowed to move freely. It was studied previously
with short-range potential disorder \cite{Ebert96}.
There, if the polymer is free to move one
expects again \cite{MachtaKirkpatrick1990}
a collapsed state%with $\zeta \approx D/d$
, whereas if one end is fixed there may be
another strong disorder fixed point \cite{LeDoussalMachta91} (see
section \ref{Inclusion of SA}).
In all cases it is natural to expect non-trivial
glassy dynamics.

We will not say more here about the potential case ($g_\ind T=0$)
which is beyond the reach of our perturbative method,
but let us stress that it is all the more interesting to 
investigate what happens to  localization
when transverse disorder is added. We  therefore now turn to the case
$g_\ind T>0$.

\medskip
\leftline{\bf (3) Isotropic disorder fixed point}\nopagebreak
Noting from \Eq{beta diff} that the line $g_\ind T=g_\ind L$ is preserved
by the flow (to this order), we find
an isotropic fixed point at
\be
%\lefteqn
{g_\ind L=g_\ind T= g^*:=} \frac{2\E d}{
\left(d-1\right) (d+2)
   \diagA - d \diagB -(d-2)\diagC} +{\cal O}(\E^2)
\label{iso fixed point}
\ ,
\ee
where the coefficients  are given
in \Eq{coeffiso}.
We have checked numerically, that the denominator of \Eq{iso fixed point}
is always positive, which is necessary for this fixed point to be  stable
and to be in the physical domain. We have also checked numerically that
this fixed point is completely attractive, and its domain of
attraction covers all perturbative situations except the potential
case $g_\ind T=0$. As in the directed case, it also controls the line
$g_\ind L=0$ (except for $D=0$) and describes
the large scale behavior of an isotropic manifold in a random short-range  
force flow, see appendix \ref{finite cor to DO}.

The critical exponents at this fixed point, $\zeta^*$ and
$z^*$, defined in \Eqs{def zeta} and \eq{def z}, are with the 
same diagrams, given in \Eq{coeffiso}
\bea
%\lefteqn
{\zeta^*=\zeta(g^*,g^*)=\frac{2-D}2}
+\frac{\left( 
\left(d-1\right)
   \diagA- \diagC \right)\E  }
{\left(d-1\right) (d+2)\diagA - d\diagB-(d-2)\diagC} 
+{\cal O}(\E^2)
\ .
\label{zeta iso SR iso}
\eea
Note the difference with the directed case. The first
coefficient in the numerator, $\diagA$,
is positive as before, since it arises from the upward corrections
to the temperature. However, since there is no tilt symmetry any more, the
elasticity is also renormalized upwards (the polymer tends to
shrink to take advantage of favorable regions).
This produces the second coefficient
$-\diagC$, which is negative.
The competition between the two opposite effects however gives
a positive sum and the membrane is stretched.
Also note that the  $\E$-correction vanishes like $1/d$ for
$D\to2$.

A similar formula is valid for the dynamic exponent $z$:
\bea
{z^*=z(g^*,g^*)=}&2&+\frac{2 \left( \diagA-\diagC \right)\E  }
{\left(d-1\right) (d+2)\diagA - d\diagB-(d-2)\diagC}
+{\cal O}(\E^2)
\label{z iso SR iso}
\eea
with the same coefficients. The $\E$-correction is always
positive, but vanishes like $1/d^2$ for $D\to 2$.

{\tabcolsep3mm\begin{figure}[t]\centerline{\renewcommand{\arraystretch}{1.20}
\begin{tabular}[t]{|l|c|c|c|c|c|c|c|c|} \hline
 & $d$ &  $\zeta$& $z$ & $\nu$ & $\beta$ & $\phi$
\\ \hline\hline
% Particle & $1$ &--- & --- & $0.58\ldots 0.66$  & --- & $>1$\\ \hline
%
  & $2$ & $\approx 1$  & $ >2$ & 0.5 & $-0.08$  & $>1$ \\ \cline{2-7}
Polymer & 3 & $0.8$ & $ >2$ &0.40 &  $-0.06$ &$>1$ \\ \cline{2-7}
  & 4 & 0.67  &$ >2$ & 0.33 &$-0.03$  & $>1$
\\ \hline
			& 3 & $0.8$ &$ >2$ & 0.40 & $-0.2 \ldots 0$ & $>1$  
\\ \cline{2-7}		 	
 			& 4 & 0.68 & $ >2$ & 0.33  & $-0.2 \ldots 0$ &$>1$\\  
\cline{2-7}
Membrane	& 6 & 0.5 & $ >2$ & 0.25 &  $-0.2 \ldots 0$& $>1$\\ \cline{2-7}
			& 8 & 0.4 & $ >2$ & 0.20 &  $-0.2 \ldots 0$&$>1$ \\  
\cline{2-7}
			& 20 & 0.2 & $ >2$ & 0.09 & $-0.2 \ldots 0$ &$>1$ \\ \hline
\end{tabular}\renewcommand{\arraystretch}{1.0}}\vspace{2mm}
\caption{Results for isotropic  polymers and membranes at the isotropic fixed point, SR-disorder.
Results for $\zeta$ are obtained from the extrapolation of
$\zeta d$, and are the most faithful. One observes a nice
plateau in extrapolations for $\nu$, but no corrections to
the Flory approximation can be deduced. The exponents $z$ and $\phi$
do not allow for direct extrapolations, but are always corrected upwards.
Results for $\beta$ are significant for polymers. For membranes
only a bound seems to emerge from extrapolations.
}%
\label{tab2}%
\end{figure}}%
Since there is no  tilt symmetry any more, the
elasticity is also renormalized upwards and gives rise to a
 non-trivial exponent $\beta^*$ (with the coefficients again given in 
\Eq{coeffiso}):
\bea
%\lefteqn{\beta^*=}\nn\\
\beta^*&=&\frac{- 2 \, \diagC \E  }
{\left(d-1\right) (d+2)\diagA - d\diagB-(d-2)\diagC}
+{\cal O}(\E^2)
\ . \label{beta iso SR iso}
\eea

The other exponents can be obtained as $\nu = \zeta/z$ and
$\phi = (z-\zeta)/(2- \zeta + \beta)$. One notes that in the 
limit $D \to 0$ one recovers again correctly the results
for the particle for $\nu$ and $\phi$.

For a polymer, $D=1$, the disorder becomes relevant
below $d=6$ and setting $D=1$, we find that the above results yield:
\bea
\zeta^*=0.5 + 0.130792 \E\ ,  \qquad
z^*=2 + 0.03996 \E\ , \qquad \beta^*=- 0.015446 \E
\eea
with $\E=\frac{1}2(6-d)$, as well as
$\nu = 1/4 + 0.060401 \E$ and 
$\phi = 1 + 0.0369373 \E $.
The most naive extrapolation 
to $d=3$ would be (setting $\E = 1.5$)
that $\zeta=0.70$, $z=2.06$ and $\beta=-0.023$
and in $d=2$ that $\zeta=0.76$, $z=2.08$ and
$\beta=-0.031$.

One can try to obtain more reliable estimates for
these critical exponents from expressions
\eq{zeta iso SR iso}, \eq{z iso SR iso} and \eq{beta iso SR iso}
for polymers ($D=1$) and membranes ($D=2$) in three or two dimensions 
by optimizing on the expansion point. This is a tedious task 
since $\E$ is rather big.
The numerical values obtained by the methods of \cite{WieseDavid96b} are not
very precise, as we could not
find a combination of the exponents and $D$ or $d$, which in
suitable extrapolation variables build up a nice plateau.
Different extrapolation schemes yielded strongly varying
results. Some indicative values obtained by this
methods are summarized in \Fig{tab2}.

Since $\zeta$ seems to increase rapidly as $d$
decreases, an interesting question is whether
there is a dimension $d_{l}$ below which the
polymer will be fully stretched ($\zeta=1$).
The result of \Fig{tab2} seems to indicate that
$d_l$ could be around $d_l=2$. Our calculations are not precise 
enough to decide on whether the polymer is already 
over-stretched or not in $d=2$ but that would be
an interesting point to try to answer by other 
methods or by numerical simulations (in $d=2$).

\medskip
\leftline{\bf (4) Transversal disorder fixed point}\nopagebreak
The transversal fixed point ($g_\ind L=0$) is at
\be
	g_\ind L=0 \ , \qquad g_\ind T=\frac{2\E}{\left(1-\frac1d\right) (d+2) 
   \diagA } +{\cal O}(\E^2)
\ ,
\ee
where the diagram is given in \Eq{coeffiso}.
It is unstable towards perturbations of $g_\ind L$ see appendix
\ref{finite cor to DO}. For the
critical exponents $\zeta^*$ and $z^*$, we
 recover exactly the Flory-result at 1-loop order:
\bea
\zeta^* &=&\frac{2-D}2 +\frac{\E}{d+2} +{\cal O}(\E^2)
\label{zeta iso SR t}\\
z^*&=& 2 +{\cal O}(\E^2) \label{z iso SR t}
\ .
\eea

As was discussed for the directed case,
this result was also found to be true
for the particle ($D=0$) to all orders, but
for $D>0$ as we show (see Appendix \ref{finite cor to DO})
finite non-transversal terms are generated in perturbation
theory which will drive the system  towards
the isotropic fixed point discussed
above.

\subsection{Isotropic (non-directed) manifold with long-range disorder}
\label{Iso LR}

\begin{figure}[tb]
\vspace{-5mm}
\centerline{ \fig{0.5\textwidth}{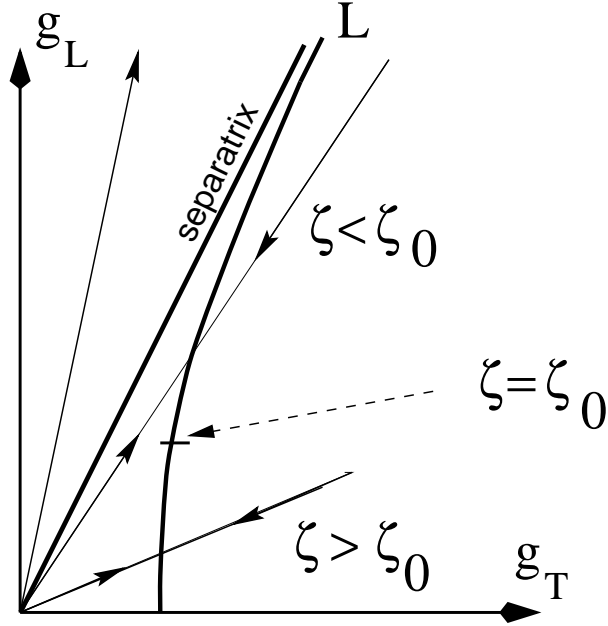} }
\vspace{1mm}
\caption{RG flow diagram for isotropic manifolds in
LR disorder. The flow is
along straight lines. There is
a line of fixed points $L$ and an apparent
separatrix. $\zeta-\zeta_0$ changes sign
upon increasing $g_\ind L/g_\ind T$, suggesting progressive
localization.  }\label{flowrg2}
\end{figure}

For isotropic manifolds, LR correlated disorder becomes
relevant for $a < a_c(D) = 2 (2+D)/(2-D)$. For fixed dimension
$d > a_c(D)$, it can be studied in an expansion 
in $\delta =2 + D - \frac{(2-D)}2 a$.
As for directed manifolds, the vertex renormalization
vanishes,
\be
\diagB=\DIAG \DynK  \DynJ_{\!\E}  =0 \ .
\ee
Thus the
renormalization is driven by the corrections to
temperature, friction and elasticity.
The $\beta$-functions \eq{new beta L g} and \eq{new beta T g} with the coefficients given in \Eq{coeffiso},
lead to the flow diagram represented on
\Fig{flowrg2}. Since there is no vertex renormalization
both couplings have the same renormalization factor
and thus the ratio $\kappa = g_\ind L/g_\ind T$ is again preserved by
RG.
As in the case of directed manifolds, one finds that the
flow is along straight lines and that there is 
a line L of fixed points (see \Fig{flowrg2}). Its
equation parameterized by the bare ratio
of coupling constants $\kappa = g_\ind L(0)/g_\ind T(0)$
is:
\bea
 g_\ind T^* = \frac{2 d \delta }{(d-1) (a+2) \diagA - \kappa (a-2) \diagC}
\eea
and $g^*_\ind L = \kappa g^*_\ind T$, where $a$ should be replaced by 
$a=a_c(D)=2(2+D)/(2-D)$.

\medskip

A novel and interesting feature is the existence of
a separatrix, as depicted on
\Fig{flowrg2}:
When starting with mostly potential
disorder, i.e.\ a large enough ratio:
\be
\kappa = \frac {g_\ind L}{g_\ind T} > 
\kappa_c := (d-1) \frac{a+2}{a-2} \frac{\diagA}{\diagC} \ ,
\ee
there is no perturbative fixed point and the
system, within the present expansion, 
always flows to strong coupling
(for the coefficients see \Eq{coeffiso}). For polymers $D=1$, since we
expand around $a=a_c=6$ this value is rather
large $\kappa_c(d) = 7.17404 (d-1)$.
This property suggests the existence of two
phases, or two regimes, one described by the
above fixed line, the other by a fixed point, which is non-perturbative
in $\delta$ (at fixed $d$). It could also be that
while the perturbative one has a continuously varying 
exponent $\zeta$ the non-perturbative corresponds to 
complete localization $\zeta=0$. It  would be interesting
to check this scenario by non-perturbative methods as e.g.\ in
\cite{LeDoussalCugliandoloPeliti97}. A study of
the crossover between SR and LR in the next section will
shed some more light on the nature of the change of
regimes observed here.

In the perturbative (non-localized) phase $\kappa < \kappa_c$
the exponents as a function of
$\kappa = \frac {g_\ind L}{g_\ind T} $ read:
\bea
\zeta^*&=&\frac{2-D}2+\delta \frac{
(d - 1) \diagA - \kappa \diagC }{
(d - 1)(a+2)  \diagA  - (a-2) \kappa \diagC}
\\
z^*&=&2+2\delta \kappa
\frac{\diagA -  \diagC }{
(d-1)(a+2)  \diagA  - (a-2) \kappa \diagC}
\eea
where $a$ should be replaced by $a=a_c(D)=2(2+D)/(2-D)$
and $\delta = 2 + D - \frac{2-D}{2} a$.
Again the diagrams are given in \Eq{coeffiso}.

These results contain interesting novel features compared to the case
of directed manifolds and can be studied as follows. Let us denote
the universal ratio $R=\diagA/\diagC$. We can express all exponents as follows
to first order in $\delta = 2 + D - (2-D)\frac{a}{2}$:
\bea
 \zeta &=& \frac{2-D}2+ \frac{\delta}{a-2}
\frac{ \kappa_0 - \kappa }{\kappa_c - \kappa} \nn\\
z &=& 2+ \frac{ 2 \delta}{a-2} 
\frac{\kappa (R-1)}{ \kappa_c - \kappa} \nn \\
 \beta &=& - \frac{2 \delta}{a-2} \frac{\kappa}{\kappa_c - \kappa} \\
\nu &=& \frac{2-D}{4} + \frac{\delta}{4 (a-2)} 
(D + (2-D) R) \frac{\kappa^* - \kappa}{\kappa_c - \kappa} \nn\\
 \phi  &=& 1 + \frac{4 \delta}{(a-2)} 
\frac{R \kappa}{(2+D) (\kappa_c - \kappa)}
\label{exponentsIso LR}
\eea
with $\kappa_0 = (d-1) R$ and $\kappa^* =
2 (d-1) R/(D + (2-D) R)$ and one should set 
$a=a_c(D)=2(2+D)/(2-D)$, i.e.\ $a=a_c=6$ for the polymer.

In the case of the polymer ($D=1$) we find that 
$R = 3.58702$, and $\kappa_c = 2 R (d-1) = 7.17404 (d-1)$
varies in the range $35.8702 < \kappa_c < +\infty$
for the range of dimensions $a_c=6 < d < +\infty$
where the result holds. $\kappa_0 = R (d-1) = \kappa_c/2 = 3.58702 (d-1)$ and 
$\kappa^* = 1.56399 (d-1)$ which varies
between $7.81993 < \kappa^* < +\infty$ as $d$
increases from $d=6$ to $d= +\infty$.
Note that one has $\kappa^* < \kappa_0 < \kappa_c$.

The above given lowest order results for the exponents in the expansion
in $\delta = 2+D-\frac{2-D}2 a$ show some new features as compared to
directed manifolds which result from the competition
between the effect of temperature increase (which
tends to stretch) and the effect of increase in elastic stiffness
(which tends to localize).
First we find that for 
$\kappa < \kappa_0$ (with $\kappa_0 = \kappa_c/2$ for polymers)
the manifold
is stretched $\zeta > \zeta_0$ compared to manifolds in the
absence of disorder (as was always the case 
for directed manifolds) while for large potential
disorder $\kappa > \kappa_0$ the manifold
is more localized $\zeta < \zeta_0$. The effect of
potential disorder, which acts to localize isotropic polymers
as was discussed above
thus becomes predominant for $\kappa > \kappa_0$
(to lowest order in $\delta$). The single monomer diffusion
exponent $r \sim t^{\nu}$ also
exibits a change from superdiffusive ($\nu > 1/4$ for polymers)
to subdiffusive ($\nu < 1/4$) behavior when $\kappa$ becomes
larger than $\kappa^*$. Finally note that in all
cases $\kappa >0$ one has $z>2$ and $\phi >1$
which indicates glassy dynamics and trapping by the flow.

\subsection{Crossover from short-range to long-range correlated disorder}
\label{Crossover}

\begin{figure}[t]
%\vspace{-3mm}
\centerline{ \fig{\textwidth}{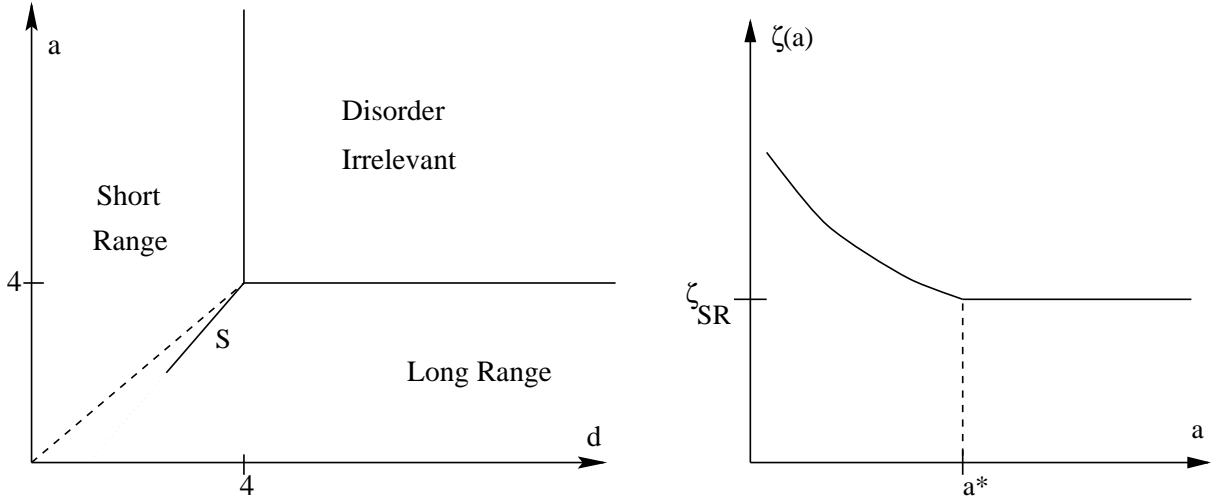} }
\caption{Schematic sketch  of the crossover in the $d,a$ plane
between LR and SR regimes for a directed polymer (left). The expansion
point is $d=a=4$ and the SR exponents hold until the line S. The exponent
$\zeta$ is for $d$ fixed $a$-independent in the SR-regime and $a$-dependent
in the LR-regime, see the sketch on the right.}
\label{flow}
\end{figure}

It is interesting to discuss the crossover from short to long-range 
disorder. It can be studied analytically around the point
$a=d=d_c(D)$ in a double expansion. 
In the case of the particle, this has been done 
by  several authors (see 
\cite{HonkonenKarjalainen1988} and references
therein) in a double expansion near $a=d=2$.
A complicated scenario was found with several sectors.
We will recover their result and find novel ones
for manifolds. For polymers we will
expand near $d=a=4$ for directed polymers and
$d=a=6$ for isotropic ones. Typically one expects a line $a_c(d)$
where the SR exponents cease to hold, as represented
on figure \ref{flow}. It is thus important to study the
crossover in detail in view of numerical simulations 
since in the physical dimensions $d=2,3$ there will
be a wide range of correlation decay
exponents $a$ accessible in practice, where the
competition between long-range and short-range disorder
obtained here can be tested.

\subsubsection{General structure of the RG flow}

Here, the situation is analyzed as 
follows. Denote  by $g_\ind L$ and $g_\ind T$ the couplings for SR-disorder, and 
by $b_\ind L$ and $b_\ind T$ the couplings for LR-disorder. Their canonical dimensions
are in the case of isotropic manifolds 
\be
\left[ g_\ind L \right] =\left[ g_\ind T \right]= \E=2+D-{2-D\over 2}d \ , \qquad 
\left[ b_\ind L \right] =\left[ b_\ind T \right] = \D=2+D-{2-D\over 2}a
\ee
and for directed manifolds
\be
\left[ g_\ind L \right] =\left[ g_\ind T \right]= \E=2-{2-D\over 2}d \ , \qquad 
\left[ b_\ind L \right] =\left[ b_\ind T \right] = \D=2-{2-D\over 2}a\ .
\ee
Since $a<d$, we always have $\delta>\E$.
 In Sec.~\ref{The flow-equations} we have  calculated the renormalization of  
elasticity, friction and thermal disorder. To generalize this to the 
mixed case discussed here, one simply has to replace $g_\ind{L}$ by $g_\ind L+b_\ind L$ and $g_\ind T$ by 
$g_\ind T+b_\ind T$ in the corresponding
renormalization functions.  

In addition, we 
now have to take care of the  mixture of  both types of disorder
under renormalization. As discussed in Sec.~\ref{DO ren}, no 
LR-disorder 
is generated. If we denote  SR-disorder by $\DynJ$, and LR-disorder
by $\DynJcross$ (for isotropic disorder for simplicity,
generalization to transversal and longitudinal disorder are straightforward), then we obtain the following egality 
for the residues of the pole-terms
\bea
\DIAG \DynK  \DynJ_{\E}&=&\DIAG \DynKc  \DynJ_{\delta^{-1}} \nn\\
&=&
\DIAG \DynKcc  \DynJ_{(2\delta-\E)^{-1}} =\diagB\ .\label{resid egality}
\eea
Note that this is a statement about the leading term, i.e.\ the 
residue, but that the finite terms are different and will 
lead to a less trivial result at the 2-loop level. A similar statement
holds for the particle case, as one can see by a procedure analogously
to that presented in section \ref{particle limit}.
The generalized 1-loop $\beta$-functions then read
\bea
\beta^g_\ind L(g_\ind L,g_\ind T,b_\ind L,b_\ind T) &=& -\E g_\ind L
-\frac{d-2}{2d} \diagC 
   g_\ind L(g_\ind L+b_\ind L) %\nn\\&&
+ \left(1-\frac1d\right) \frac{d+2}2
   \diagA
     g_\ind L( g_\ind T +b_\ind T) \label{beta L g}
\nn\\
&&-\half \diagB  
(g_\ind L+b_\ind L)( g_\ind T +b_\ind T) %+\,\ldots 
\\
\beta^g_\ind T(g_\ind L,g_\ind T,b_\ind L,b_\ind T)
&=& -\E g_\ind T - 
\frac{d - 2}{2d} \diagC  
g_\ind T(g_\ind L + b_\ind{L}) %\nn\\&&
+ \left(1-\frac1d\right) \frac{d+2}2
   \diagA  g_\ind T(g_\ind T+b_\ind T) \nn\\
&& -\half \diagB 
(g_\ind L\!+\!b_\ind{L}) (g_\ind T+b_\ind T)
% +\, \ldots 
\label{beta T g} \\
\beta^b_\ind L(g_\ind L,g_\ind T,b_\ind L,b_\ind T) &=& -\delta\, b_\ind L
-\frac{d-2}{2d} \diagC 
   b_\ind L (g_\ind L+b_\ind L) %\nn\\&&
+ \left(1-\frac1d\right) \frac{d+2}2
   \diagA b_\ind L 
	(g_\ind T + b_\ind T)  %+ \, \ldots
\label{beta L b} \\
\beta^b_\ind T(g_\ind L,g_\ind T,b_\ind L,b_\ind T)
&=& -\delta\, b_\ind T - 
\frac{d-2}{2d} \diagC   
b_\ind T(g_\ind L+b_\ind L)   %\nn\\ &&
+ \left(1-\frac1d\right) \frac{d+2}2
   \diagA  b_\ind T(g_\ind T+b_\ind T) \qquad\qquad
%  +\, \ldots
\label{beta T b}
\eea
Once fixed points have been found, the exponents can
then be obtained, since one has simply:
\bea
\zeta(g_\ind L,g_\ind T,b_\ind L,b_\ind T) =
\zeta(g_\ind L + b_\ind L, g_\ind T + b_\ind T)
\eea
and similarly for the exponents $z$ and $\beta$, 
given by formula (\ref{newexponents}).

Let us now discuss the structure of these flow-equations. 
This is not a trivial task since the whole depends on the values
of the diagrams $\diagA$, $\diagB$ and $\diagC$, all of them 
functions of the expansion point, and on $\delta$ and $\E$. 
We can furthermore make the observation that
\be
\kappa := \frac{b_\ind L}{b_\ind T}
\ee
is again unchanged under renormalization, thus entering as an 
additional parameter. 
The situations can then be analyzed by the following observations:

\medskip
\leftline{(1) Relevance of the disorder}\nopagebreak
If both $\E$ and $\delta$ are negative, then disorder is 
indeed irrelevant, i.e.\ when starting with any combination 
of the couplings, the flow is always towards 0.

\medskip
\leftline{(2) Attractivity of the isotropic fixed point $g:=g_\ind L=g_\ind T$}
\nopagebreak
The best way to study this is to write down the flow equation for 
the combination $(g_\ind L-g_\ind T)/(g_\ind L+g_\ind T)$, which 
using \Eqs{beta L g} and \eq{beta T g} evaluates to 
\be \label{converge isotropic}
\mu \frac \p {\p \mu}\lts_0 \frac{g_\ind L-g_\ind T}{g_\ind L+g_\ind T}
	= \diagB \,\frac{(g_\ind L-g_\ind T)(g_\ind L+b_\ind L)(g_\ind T+b_\ind T)}
{(g_\ind L+g_\ind T)^2}
\ee 
Since the r.h.s.\ is always positive, the ratio $(g_\ind L-g_\ind T)/(g_\ind L+g_\ind T)$ will flow to 0. We can therefore study the simplified 
flow-equations for which we set 
\bea
 g&:=&g_\ind L=g_\ind T\nn \\
b&:=& b_\ind T\\
 b_\ind L&=&\kappa b \nn\ .
\eea
These equations read
\bea
\beta^g(g,b) &=& -\E g
-\frac{d-2}{2d} \diagC 
   g(g+\kappa b) %\nn\\&&
+ \left(1-\frac1d\right) \frac{d+2}2
   \diagA
     g( g +b) \label{beta g}-\half \diagB (g+\kappa b)(g+b)\qquad\quad
\\
\beta^b(g,b) &=& -\delta\, b
-\frac{d-2}{2d} \diagC 
   b (g+\kappa b) + \left(1-\frac1d\right) \frac{d+2}2
   \diagA b 
	(g + b)  
\label{beta b}
\eea

\medskip
\leftline{(3) Stability of the SR-fixed point}\nopagebreak
Next discuss the stability of the SR-fixed point with $b=0$, present
for $\E>0$ and consequently also $\delta>0$.  Its stability
on the axis $b=0$ was established before for all cases of interest. 
To analyze its stability towards perturbations of LR-disorder, we
linearize \Eq{beta b} for small $b$. 
Two cases can be distinguished. If (with the diagrams 
explicitly given in \Eqs{coeffdir} and \eq{coeffiso})
\be \label{SR cond}
{\delta} <\E\, \frac{
\left(d-1\right) (d+2)
   \diagA -(d-2) \diagC }
{\left(d-1\right) (d+2)
   \diagA - d\diagB-(d-2) \diagC } \ ,
\ee
 then LR-disorder  is irrelevant
and the SR-fixed point alone is stable.
The exponents are then given by the results for
the SR disorder problem.
Note that the coefficient
on the r.h.s.\ of \Eq{SR cond} is always larger than 1, such that
this condition is satisfied for weakly LR-disorder.
$\delta < \E$ would be the result of a naive
dimensional estimate, and the above results gives the correct
result to lowest order, i.e.\ the tangent to the curve
in the $a,d$ plane above which SR exponents are correct.
For a directed polymer $D=1$ one finds the condition
\be \label{condition dir}
4-a < (4 - d)/(1-\frac{1}{3 \pi}) = 1.1187 (4-d)
\ee  
and for an isotropic polymer 
\be \label{condition iso}
6-a < 1.07723 (6 - d)
\ee
for the exponents to be given by their SR expressions
(see sections \ref{dir SR} and \ref{iso SR}).

\medskip
\leftline{(4)  Repulsiveness of the line $g=0$}\nopagebreak
By analyzing the flow-equation \Eq{beta g} for vanishing
coupling $g$, we see that the flow on the axes $g=0$ is always pointing 
upwards (increase of $b$) at least for $\kappa >0$. 

\medskip
\leftline{(5) The complete flow-diagram}\nopagebreak
Let us now draw the complete flow-diagram for both couplings $b$ and 
$g$ in the situation where the SR-fixed point is stable and $\kappa$
large. (This domain includes all values of $\kappa>1$, 
but is significantly larger. We will detail on the exact phase
boundary below.)%
\begin{figure}[t]
{\fboxsep0mm%\fbox%
{\unitlength0.033333333\textwidth
\begin{picture}(30,20)
\put(0,11){\epsfxsize=0.32\textwidth\epsfbox{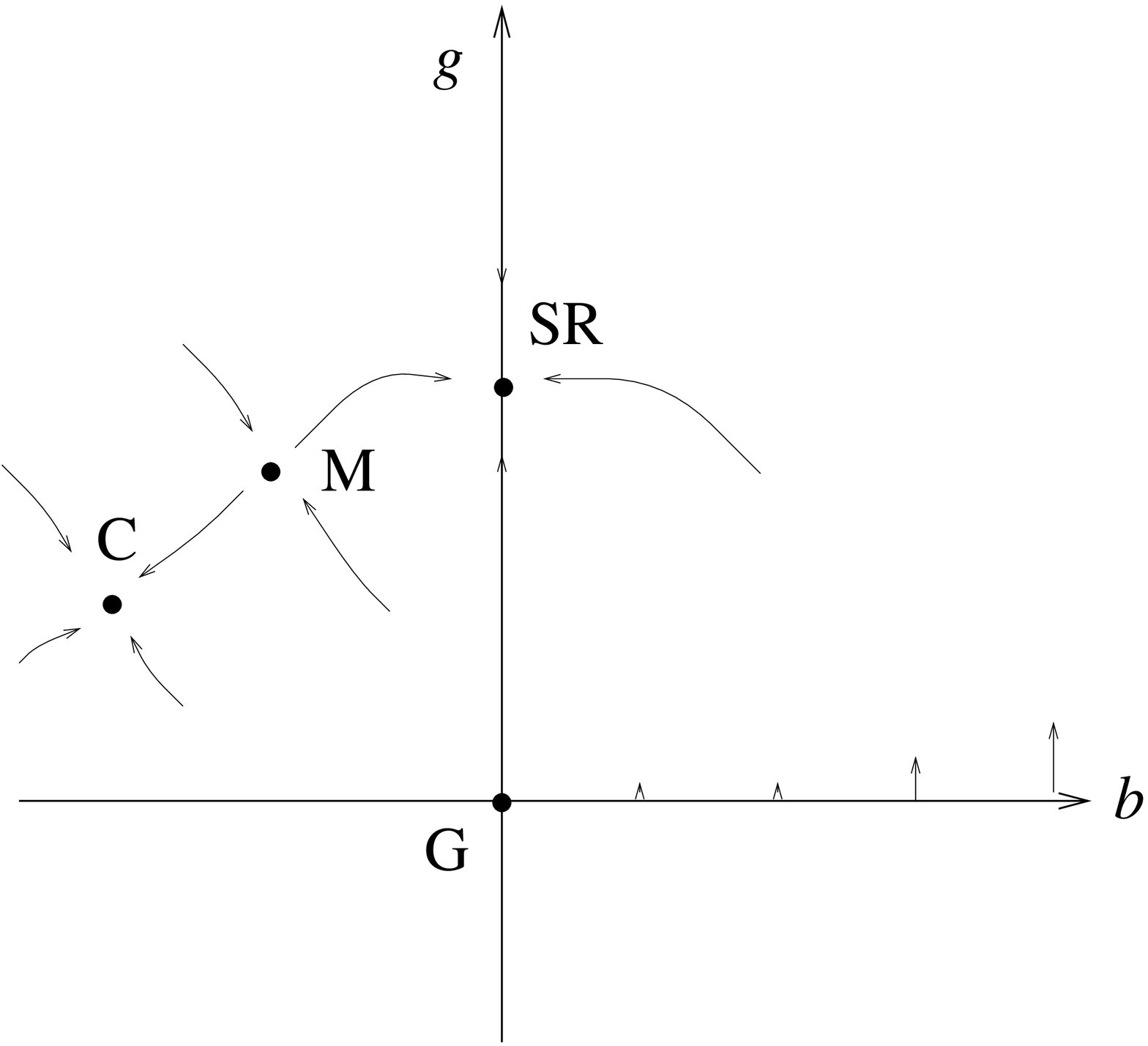}}
\put(10,11){\epsfxsize=0.32\textwidth\epsfbox{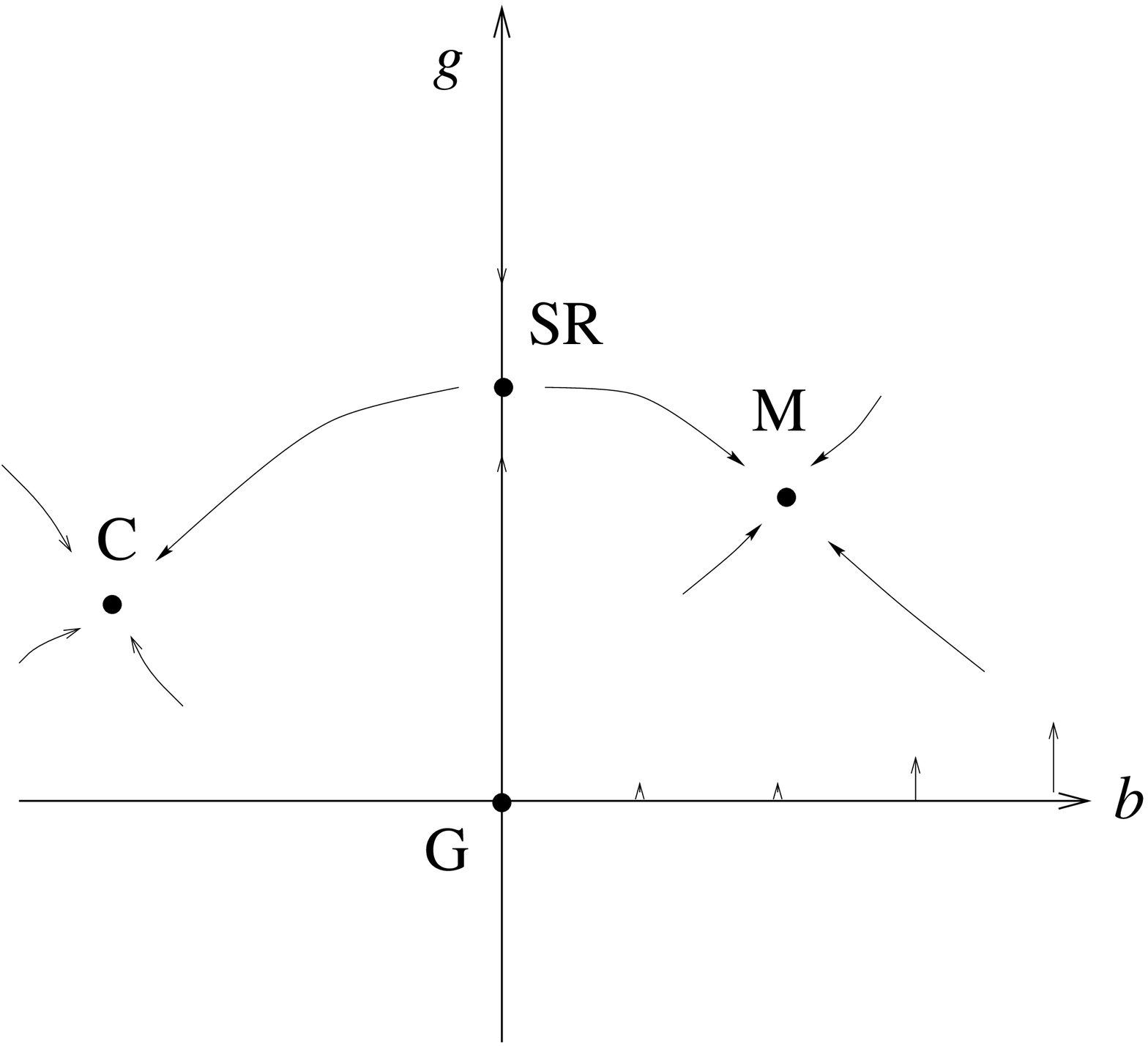}}
\put(20,11){\epsfxsize=0.32\textwidth\epsfbox{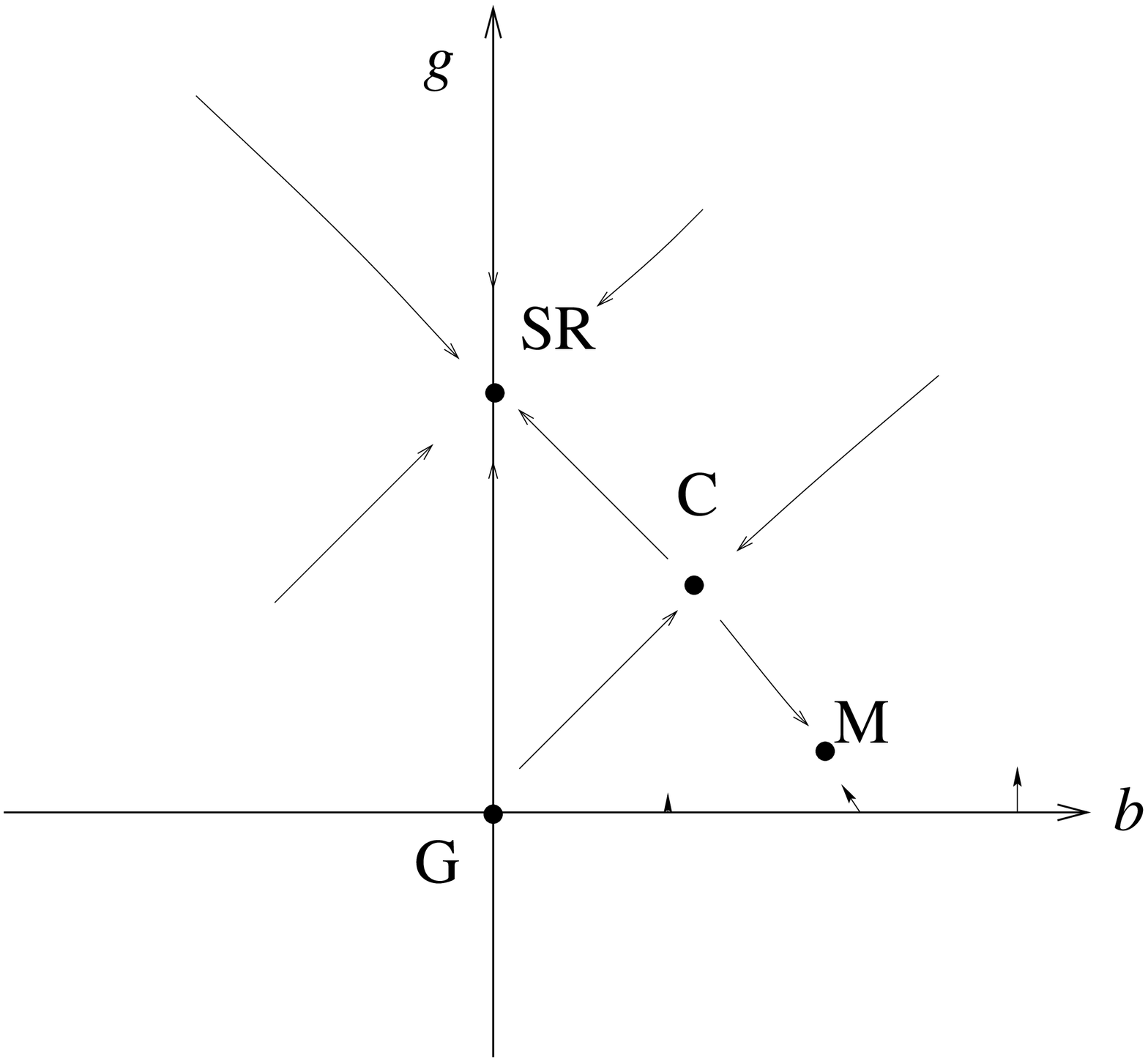}}
\put(0,1){\epsfxsize=0.32\textwidth\epsfbox{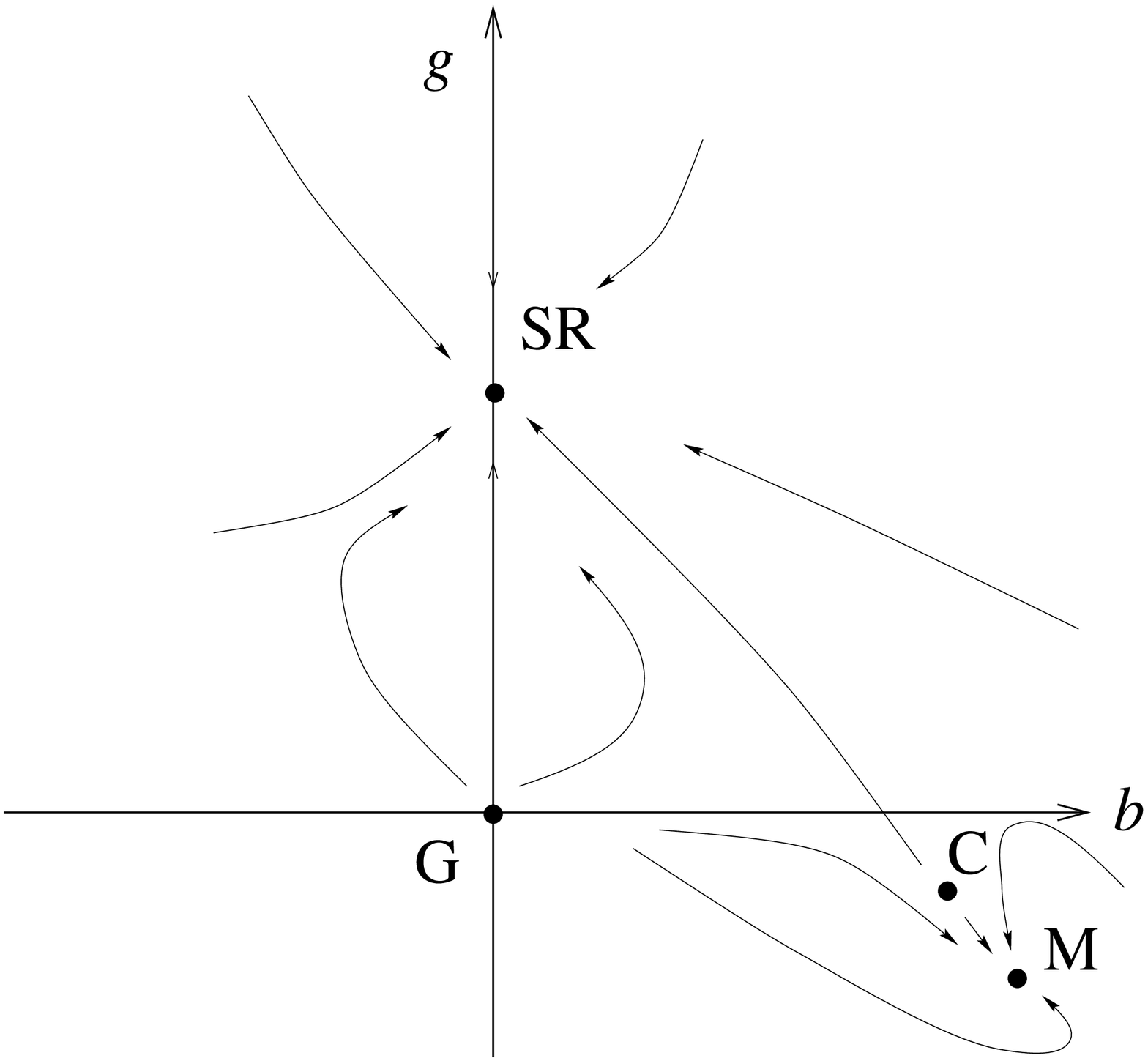}}
\put(10,1){\epsfxsize=0.32\textwidth\epsfbox{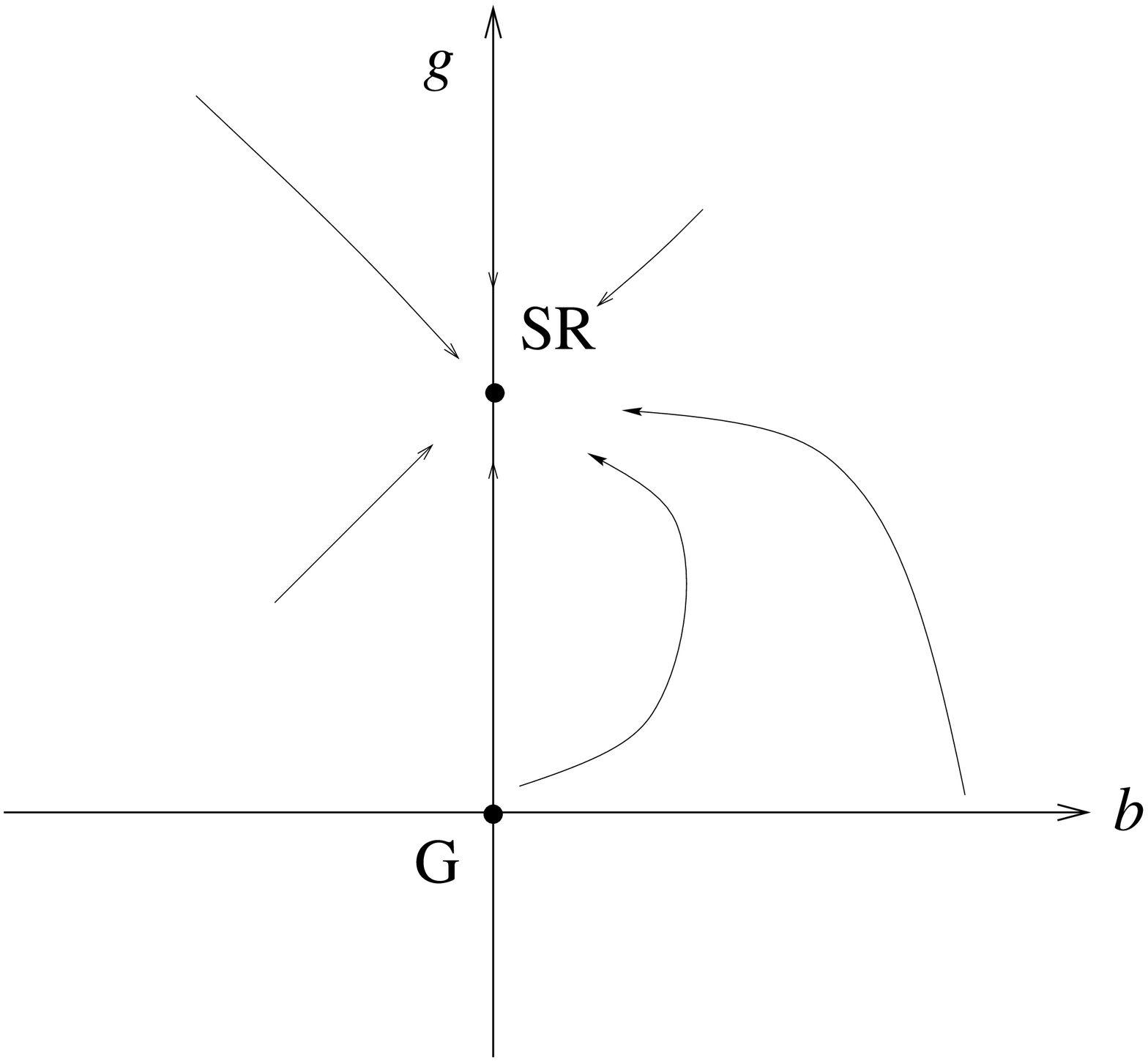}}
\put(20,1){\epsfxsize=0.32\textwidth\epsfbox{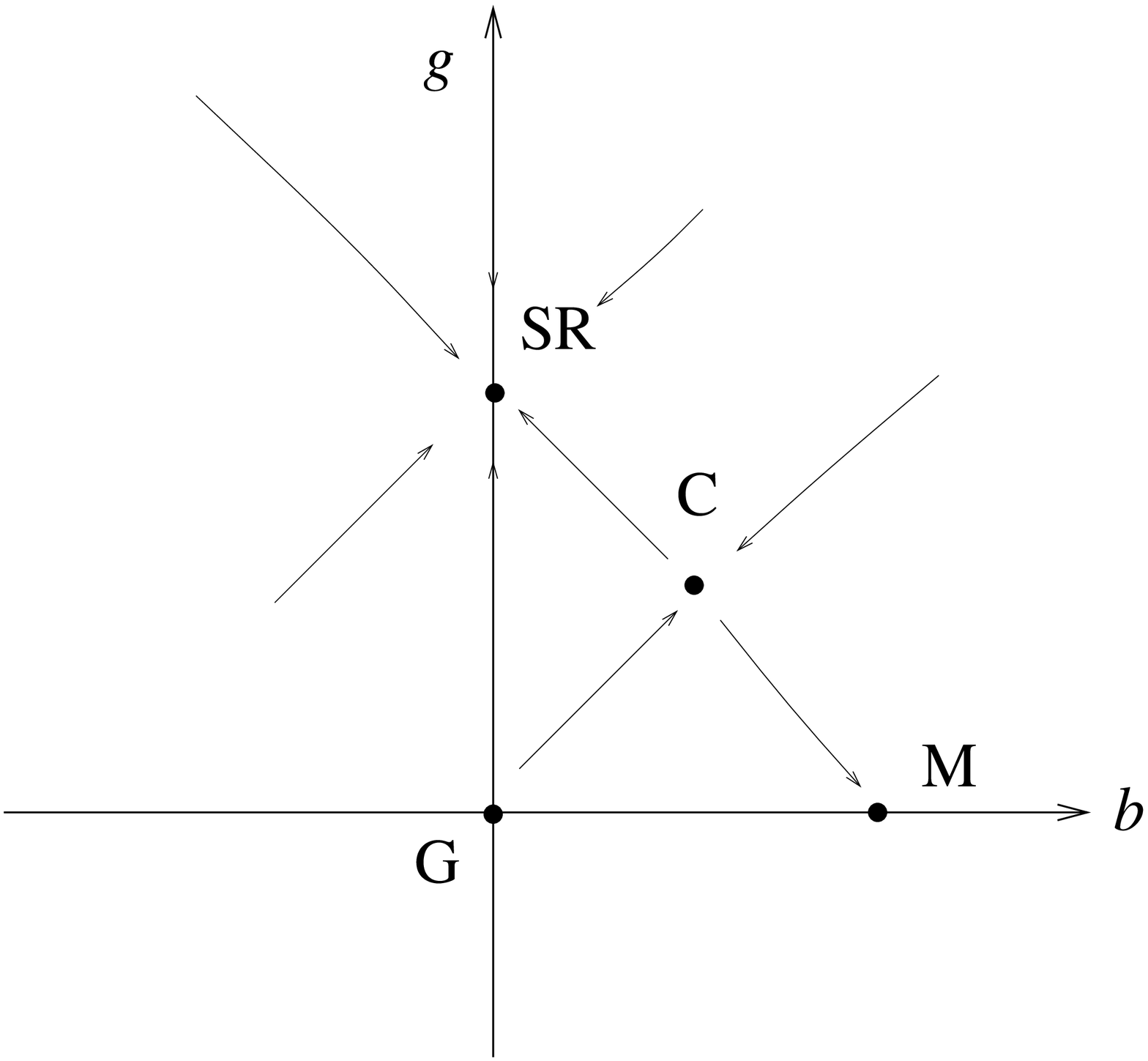}}
\put(0.3,19.3){${\cal M}$}
\put(10.3,19.3){${\cal N}$}
\put(20.3,19.3){${\cal O}$}
\put(0.3,9.3){${\cal P}$}
\put(10.3,9.3){${\cal Q}$}
\put(20.3,9.3){${\cal R}$}
\end{picture}}}
\caption{Schematic sketch of the possible flow-diagrams for $\E>0$.
For directed manifolds, $\cal M$ and $\cal N$ are sufficient and the
fixed point C is rejected to infinity. $\cal M$ is relevant 
when SR is attractive, and $N$ else. 
The other situations become important for the isotropic case, see
the discussion below. }
\label{schematic flow}
\end{figure}
By direct integration of the flow-equations, one finds the situation  $\cal M$
sketched on  \Fig{schematic flow}. 
The figure contains 4 fixed points, namely G, M, C and SR, as expected on 
general grounds. (For directed manifolds, C however is rejected to $\infty$.)
How does then the SR-fixed point become unstable?

\medskip
\leftline{(6) Crossover from the SR-fixed point to the mixed fixed point M}
\nopagebreak
Above, we have given the conditions when the SR-fixed point is stable. 
If one now increases $\delta$ such that \Eq{SR cond} is no longer
satisfied, then the flow-diagram changes to $\cal N$ on
 \Fig{schematic flow}. Note that in this case, the mixed fixed point M
takes over the stability of the SR-fixed point. 

\medskip
\noindent{(7) Completeness of the scenarios when there is 
a repulsive SR-fixed point ($\E>0$ and \Eq{SR cond} not satisfied).}\nopagebreak

We claim that under these conditions, the scenario on the 
mid of \Fig{schematic flow} is the only possible one, i.e.\ that 
no new stable fixed point can appear in the
 physically relevant sector $b,g>0$, nor can the fixed point
M disappear other than in a reverse of what we discussed in point
(6), making the SR-fixed point stable. 
 Let us  focus on the fixed point M. It can neither pass
through the axes $b=0$ other than through the SR-fixed point, 
as else there were 3 fixed points on the axes. Nor can it pass  
through the axes $g=0$, since this axes is always repulsive 
and thus does not admit a completely stable fixed point. M can 
not pass through the Gaussian fixed point G since $\delta>\E>0$. 
The last way for M to escape is via infinity. There are situations
were this happens (see e.g.\ \cite{WieseKardar98a}), but not here: 
At the fixed point, both \Eq{beta g} and \Eq{beta b} have to vanish.
Combining them, one obtains for $b>0$ and $g>0$
\be \label{critical eps delta}
\delta-\E=\half \diagB \,\frac{(g+\kappa b)(g+b)}{g} > \half \diagB\, g
\ee
bounding $g$. 

\medskip
\leftline{(8) The case of a completely stable SR fixed point}\nopagebreak
The same scenario is valid, if $\E>0$ and the SR fixed
point is attractive (see left of \Fig{schematic flow}), with 
two possible exceptions: First, the fixed points C and M can coalesce 
and then annihilate, leaving only two fixed points, G and SR, (flow
$\cal Q$ on \Fig{schematic flow}), 
which always have to exist as discussed above.  Formally, C and M 
have become complex. This is only possible 
in the isotropic case, see the discussion below. 

Further changing the 
parameters, the pair of them can re-appear in the physical sector
$b,g>0$. One then obtains two completely stable fixed points, 
SR and M, with their domains of attraction divided by a 
phase separatrix, passing through C. The most obvious such situation 
is when $\kappa=0$, since then there has to be a fixed point on the
axes $g=0$, see $\cal R$. This is the limiting case of the situation $\cal O$ 
on \Fig{schematic flow} 
for $\kappa\to 0$. (The axes $g=0$ is no longer repulsive.)

\begin{figure}{}
{\fboxsep0mm%\fbox%
{\unitlength0.05\textwidth
\begin{picture}(20,10)
\put(0.2,0.5){%\fbox
{\epsfxsize=0.48\textwidth\epsfbox{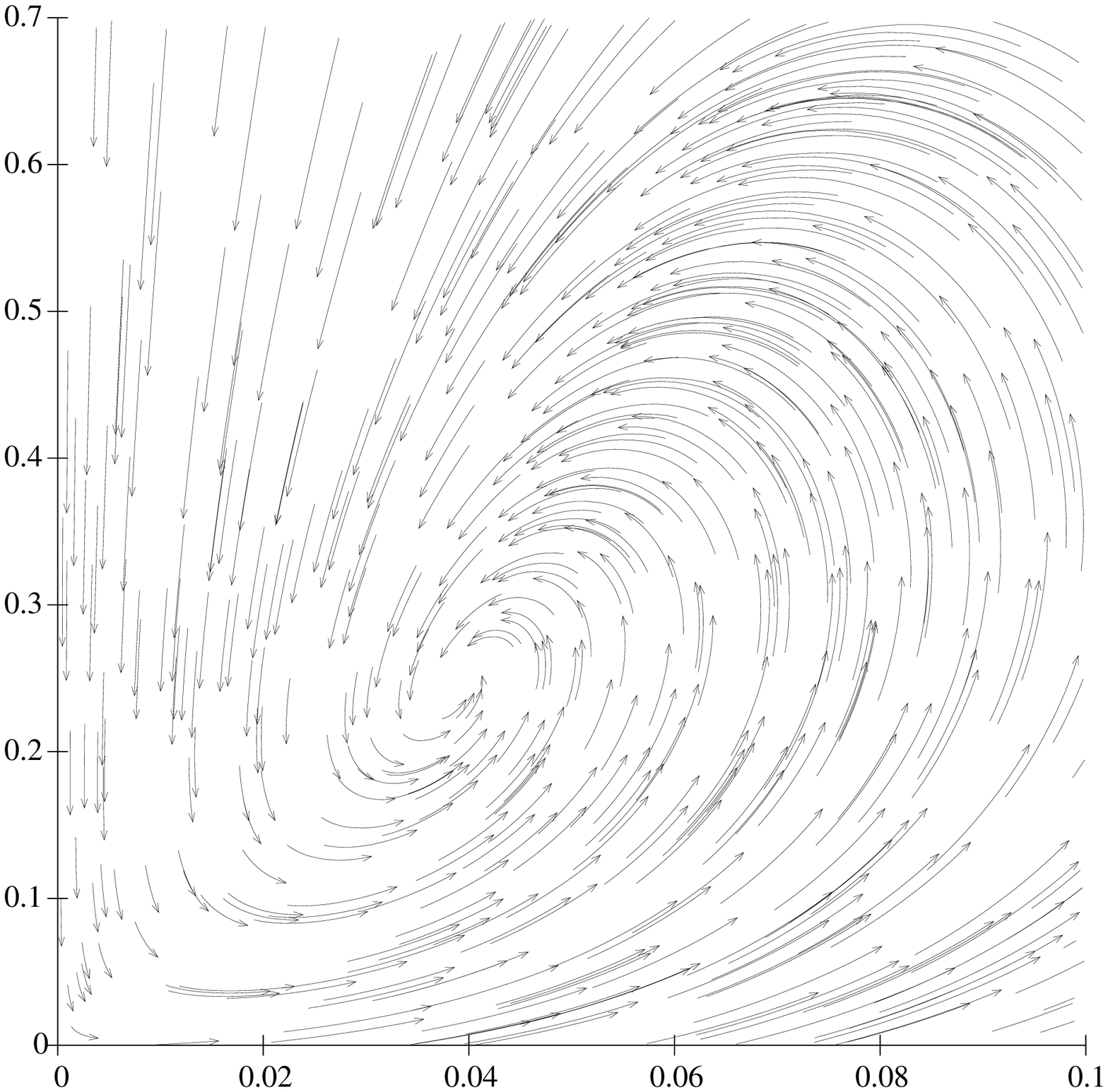}}}
\put(10.2,0.5){%\fbox
{\epsfxsize=0.48\textwidth\epsfbox{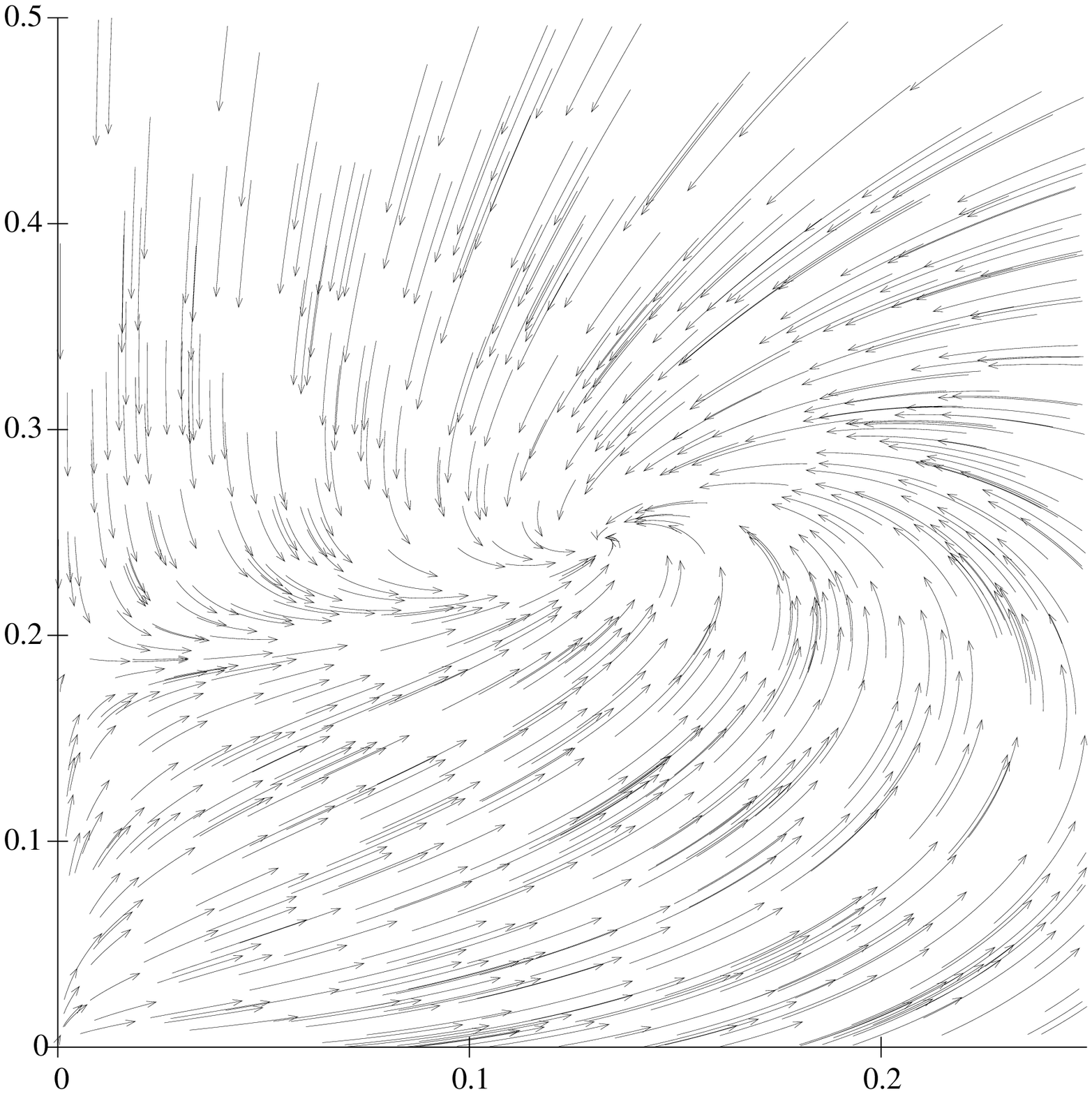}}}
\put(0.3,9){$g$}
\put(10.3,9){$g$}
\put(8.7,0.3){$b$}
\put(18.7,0.3){$b$}
\end{picture}}}
\caption{Some flow-diagrams in the case of 
isotropic disorder. The left one is for $D_0=1$, $\E=-1$, 
$\delta=1$, $\kappa=100$, the right one for
$D_0=1$, $\E=1$, $\delta=2$, $\kappa=10$. Note also the repulsive SR-fixed
point on the axis $b=0$.}
\label{example flow}
\end{figure}

\medskip
\leftline{(9) The case $\E<0$, $\delta>0$}\nopagebreak
The last case of interest is, when by say keeping $\delta$ fixed 
$\E$ becomes negative. Then the SR-fixed point disappears and 
only the mixed fixed point stays in the physically relevant sector
$b \ge 0$ and $g \ge 0$. With the same reasoning as above, one 
concludes that it cannot escape, as long as $\delta>0$. If 
however $\delta$ changes sign, it will disappear through the
Gaussian fixed point at the origin, which then will become 
stable, as expected. We have plotted
some of the flow-diagrams in the case $\E<0$, $\delta>0$
on \Fig{example flow}, which show amusing spiral behavior.
We note, as will be discussed below in details that
in that case, as $\E/\delta$ becomes very negative,
the value of the couplings can become rather large.
This terminates the discussion of the topology 
of the flow-diagrams.

Note also that in the directed case, the fixed point C is rejected to 
$\infty$, and thus the flows $\cal M$ and $\cal N$ are sufficient to 
descrive the crossover.

We will now give detailed results for the critical
exponents at $M$ and discuss how they 
differ from their values at the LR fixed point 
discussed earlier. In general, if $\E$ and
$\delta$ are comparable, the critical exponents are continuous 
functions of both arguments, modifying the pure
LR-behavior. On the other hand, when
$\delta$ is small compared to the modulus of $\E$ (which then 
necessarily is negative) then the exponents tend to
their values given by the LR-fixed points, except
in one situation discussed below.

\subsubsection{Results for directed manifolds}
Let us introduce the ratio $S = (1-\frac{1}{d})(d+2) \diagA/\diagB$.
For polymers one expands around $a=d=4$ and $S=3 \pi$. Then 
the fixed point $M$ has coordinates:
\bea
&& b^*_\ind T + g^* = \frac{ 2 d \delta}{(d-1)(d+2) \diagA} \\
&& b^*_\ind L + g^* = \kappa b^*_\ind T + g^* = \tilde{\kappa} (b^*_\ind T + g^*) \\
&& \tilde{\kappa} = \kappa \frac{1 - \frac{\E}{\delta}}{
1 + \frac{\kappa -1}{S} - \frac{\E}{\delta} }
\eea
and is physical, i.e.\ has a 
positive $b^*_\ind T = (b^*_\ind T + g^*) (1 - \frac{1}{S} - \frac{\E}{\delta})
/(1 + \frac{\kappa -1}{S} - \frac{\E}{\delta})$ as long as the above condition 
$\E/\delta < 1 - \frac{1}{S}$ is satisfied (it collides with
the SR fixed point at this value, except for $\kappa=0$). The fixed point $C$ in
the present case is rejected to $b_\ind T = - \infty$.

For $\E/\delta < 1 - \frac{1}{S}$ one thus recovers the
previous results for the LR fixed line of directed manifolds,
section \ref{resultsdirectedlongrange},
except that $\kappa$ has to be replaced in all
expressions (e.g.\ in \Eqs{label} - \eq{exponents directed longrange})
by the value $\tilde{\kappa}$ which
depends on the ratio $\E/\delta$. Thus we have that
$z_M(\kappa) = z_{LR}(\tilde{\kappa})$ etc..
When $\E/\delta=1-\frac{1}{S}$
one has that all initial $\kappa$ yield $\tilde{\kappa} = 1$ (isotropic
disorder) and one recovers the SR exponents,
while when $\E/\delta$ decreases towards $- \infty$ one
recovers progressively the LR exponents since $\tilde{\kappa} \to \kappa$
when $\E/\delta \to - \infty$. An interesting property is 
that $\tilde{\kappa}(\kappa)$ is an increasing function which
is bounded $\tilde{\kappa}(\kappa = \infty) 
= S(1 - \frac{\E}{\delta})$. Thus only a finite interval
of the vertical fixed line $L$
studied in the LR expansion can really be reached for a fixed
finite $\E/\delta$.

In particular the Flory exponent for $\zeta=\zeta_\ind{F}=(4-D)/(2 +a)$ is
still exact at $M$. This yields the exact relation for 
$a^*(d)$ at which the SR exponent will start being corrected
as $a^*(d) = (4-D)/\zeta_\ind{SR}(d) - 2$, a relation which
could be checked in numerical simulations. For polymers 
using the improved estimates it gives $a^*(d=3) \approx 2.76$
(while the naive extrapolation in \Eq{condition dir} gives $a^*(d=3) = 2.88$)
and $a^*(d=2) \approx 1.44$
(while the naive extrapolation in \Eq{condition dir} gives $a^*(d=3) = 1.76$).

The case $\kappa=0$ must be discussed separately.
For concreteness, start with both $b_\ind L=0$ and 
$g_\ind L=0$. We know that only SR potential disorder is generated,
yielding a finite $g_\ind L$,
but that $b_\ind L$ remains zero.
The short-range part  still converges
to the isotropic line $g_\ind L=g_\ind T=g$, as shown in 
\Eq{converge isotropic} and the
above flow equations are still valid setting $\kappa=0$.
There will be the SR fixed point as before,
and the  fixed point M is on the axis $g=0$. 
At the critical value of $\E/\delta$ (see \Eq{critical eps delta}) the
 fixed points SR and M exchange stability, but without colliding
(right at the critical value there is a fixed line extending
from one to the other). In any case,  the values of
the exponents will be as given above setting  
$\kappa=0$.

We have also checked  that the results for the crossover from LR to SR
for the particle given e.g.\ in Eq.\ (62) of \cite{HonkonenKarjalainen1988}
are recovered from our formulae.

\subsubsection{Results for isotropic manifolds}
With the same notations as previously (but different coefficients
$\diagA$, $\diagB$ and $\diagC$) we find the
coordinates of the fixed points $M$ and $C$ as:
\bea
&& g^* + b_\ind T^* = (g^* + b_\ind T^*)^\ind{dir}\, X \\
&& g^* + b_\ind L^* = g^* + \kappa b_\ind T^* = (g^* + b_\ind T^*)^\ind{dir}
\, \kappa_c (X - 1) \ .
\eea
$X$ satisfies an equation with the two roots
\bea \label{xsolu}
X_{\pm}= \frac{1}{2} (1 + \rho G \pm \sqrt{ (1 + \rho G)^2 - 4 G } ) \ ,
\eea
where we have defined $\rho = 1 - \frac{\kappa}{\kappa_c}$
and $G = S (1- \frac{\E}{\delta})/(1-\kappa)$.
We shall concentrate on the case of the polymer $D=1$,
where one has $\kappa_c = 35.8702$ and $S=\frac{2}{5} \kappa_c$.
We found that in the range of interest the fixed point $M$
corresponds to $X_{+}$ for $\kappa >1$ and
to $X_{-}$ for $\kappa < 1$ and is in the physical
sector as long as the condition $\E/\delta < 1 - 
\frac{1}{S} \frac{\kappa_c}{\kappa_c -1}$ holds.
The  fixed point $C$ is mostly in the unphysical
region. (For the exception see below.)

\begin{figure}[t]
{\fboxsep0mm%\fbox%
{\unitlength0.05\textwidth
\begin{picture}(20,9)
\put(1,1){\epsfxsize=0.45\textwidth\epsfbox{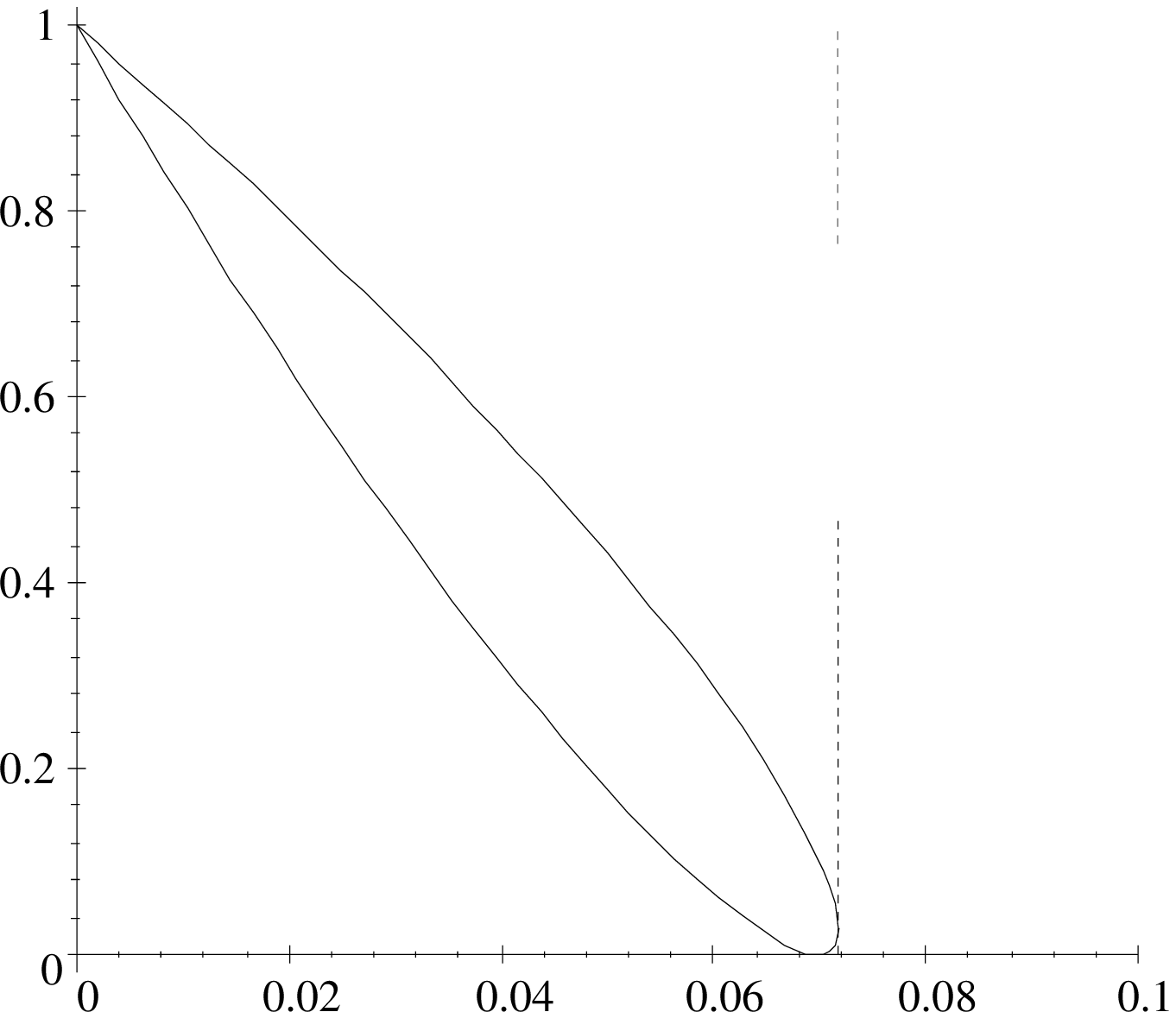}}
\put(11,1.05){\epsfxsize=0.45\textwidth\epsfbox{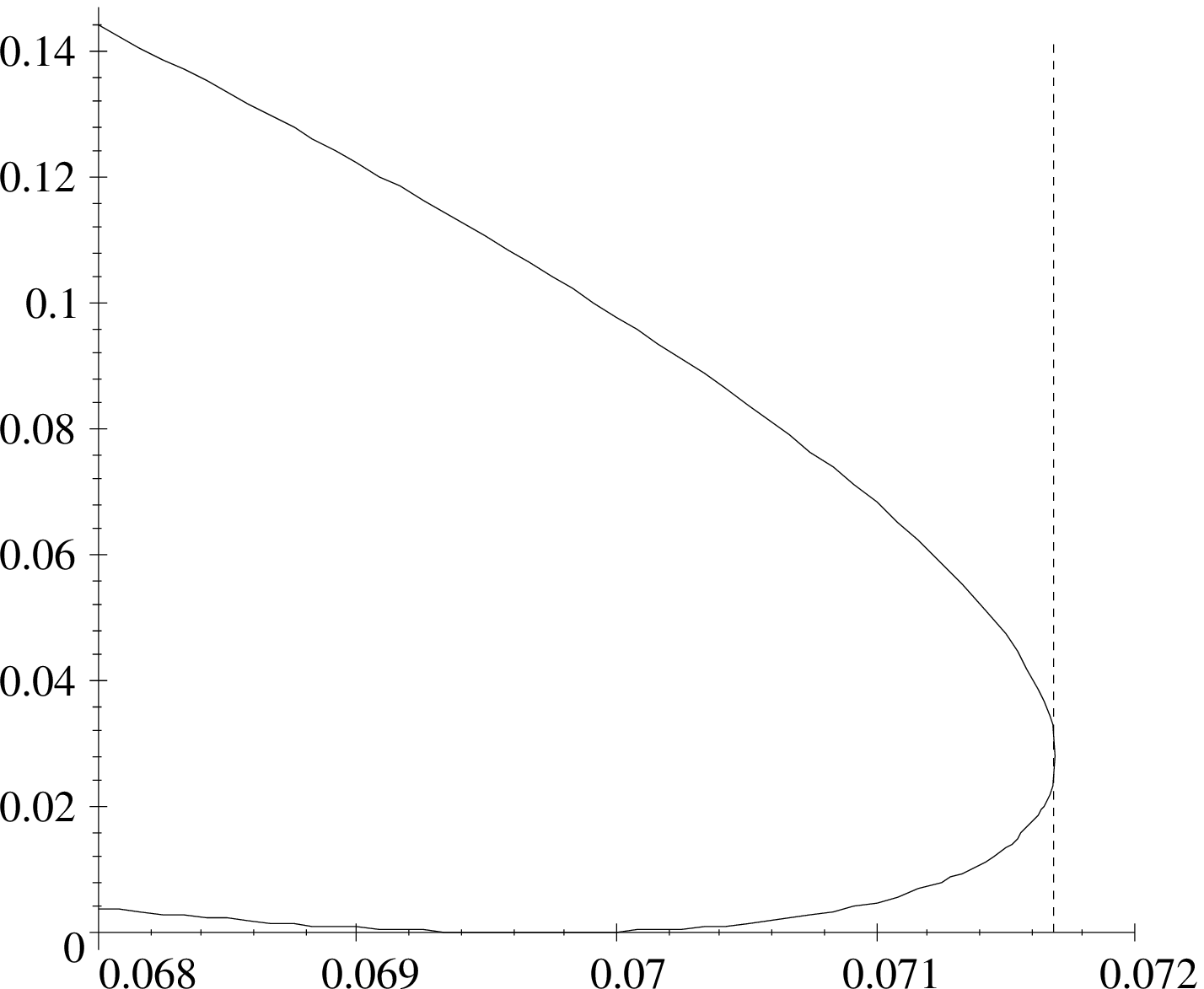}}
\put(0.3,8){$\kappa$}
\put(10.3,8){$\kappa$}
\put(8.7,0.3){$ 1-\frac\E\D$}
\put(18.7,0.3){$ 1-\frac\E\D$}
\put(6,3){\epsfxsize=0.38\textwidth\epsfbox{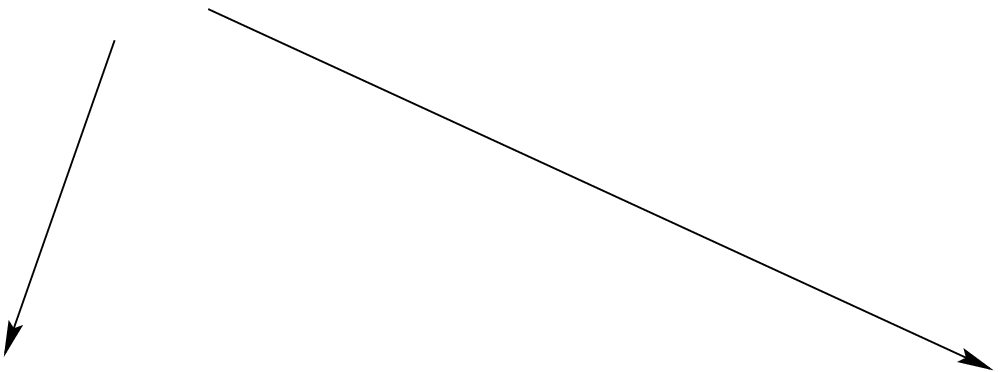}}
\put(5,6){\begin{minipage}{3.5cm}\small
fixed points M and C do not exist\end{minipage}}
\put(18.2,1.5){ $\cal O$}
\put(6,8){${\cal M}$}\put(17,7){${\cal M}$}
\put(8,3){${\cal N}$}\put(19.4,4){${\cal N}$}
\put(3,3){${\cal P}$}
\put(4.5,4.5){${\cal Q}$}\put(14.5,4.5){${\cal Q}$}
%\put(18.3,1.0){ $\cal R$}
\end{picture}}}
\caption{Domains for which the argument of the root in \protect\Eq{xsolu} is negative for polymers ($D=1$), 
and therefore the fixed points M and C do not exist. The right figure 
is just a magnification of the left one, on which e.g.\ domain $\cal 
O$ cannot be identified. The dotted line
is the boundary of the domain of stability for the  SR fixed point
as given by \protect\Eq{SR cond}, such that to the left 
of it, the SR fixed point is stable. Note that it is tangential
to the domain, where the  fixed points M and C do not exist.
We have also labeled the regions according to the flow-diagrams in
\Fig{schematic flow}. Situation $\cal R$, not drawn here, is obtained by 
starting from the domain $\cal O$ and taking the limit $\kappa \to 0$.}
\label{roots plot}
\end{figure}
There are several interesting properties. First,
one can now study in more detail the surprising 
property of the separatrix found before. One finds that
as $\E/\delta$ is decreased to
$- \infty$ at fixed $\kappa$, two radically different behaviours
occur with a transition at $\kappa=\kappa_c$. 
Either $\kappa < \kappa_c$ and then one finds
(for $\kappa >1$) that as $\E/\delta  \to 
- \infty$, $X_{+}$ saturates at the value:
\bea \label{limitX1}
X_{+} \to X_{+}^{+\infty} = \frac{\kappa_c}{\kappa_c - \kappa} 
\eea
while for $\kappa > \kappa_c$, $X_{+}$
keeps on growing indefinitely as  $ \frac{\E}{\delta} \to - \infty$ as
\bea \label{limitX2}
X_{+} \sim \frac{- \E}{\delta} \frac{\kappa - \kappa_c}{\kappa_c (\kappa -1)} 
S
\eea

In the first case $\kappa < \kappa_c$ the exponents at the 
fixed point $M$ will evolve 
smoothly, as $\E/\delta \to - \infty$ to their values
on the fixed line obtained in section (\ref{Iso LR}) in a 
first order expansion in $\delta$ at fixed $d$.
However for $\kappa > \kappa_c$ there is no
well defined expansion in $\delta$ at fixed $d$,
as the fixed point $M$ becomes non-perturbative in
$\delta$ at fixed $\E<0$. The fixed point
can still be reached and exponents be computed
but only in a perturbative expansion
in {\it negative} $\E$ (for $\delta>0$).

For a given $\kappa$ and a fixed ratio $\frac{\E}{\delta}$
one can compute $X$ using (\ref{xsolu}) and obtain the critical
exponents at $M$ to first order in 
$\delta = 2 + D - (2-D)\frac{a}{2}$ as:
\bea
 \zeta &=& \frac{2-D}2 + \frac{\delta}{(d-2) \kappa_c} (\kappa_0 - \tilde{\kappa}) X \\
 z &=& 2+ \frac{2 \delta}{d-2} (R-1)(X-1) \\
  \beta &=& - \frac{2 \delta}{d-2} (X-1)
\eea
with $\tilde{\kappa} = \kappa_c (1 - 1/X)$,
$\kappa_0 = (d-1) R$, $R = \diagA/\diagC$,
 and one should set 
$d=2(2+D)/(2-D)$, i.e.\ $d=6$ for the polymer.
Using the above limit (\ref{limitX1}) one checks that
indeed the formulae (\ref{exponentsIso LR}) are recovered when $\E/\delta
\to - \infty$ for $\kappa < \kappa_c$.

We can now obtain the exponents of the regime 
$\kappa > \kappa_c$ (which was non-perturbative
in $\delta$) in an expansion in negative $\E$.
We find, using the above limits (\ref{limitX2}):
\bea   
 \zeta &=& \frac{2-D}2 - 
\frac{4(-\E)}{(d+2)(d-2)(\kappa -1)}  \left(\frac{\kappa}{\kappa_c} -1\right) S \\
 z &=& 2+ \frac{2 (-\E)}{(d-2)(\kappa -1)} (R-1) \left(\frac{\kappa}{\kappa_c} -1\right) S \\ 
\beta &=& \frac{2\E (\kappa-\kappa_c)}{d (\kappa-1)} \frac{\diagC}{\diagB}
\eea

The general analysis of the RG flow is complicated
and we have therefore restricted ourselves to the case of 
the polymer ($D=1$, $d\approx d_c=6$). 
Interestingly, there is a domain of values of the parameters
$\epsilon/\delta$ and $\kappa$ where the two roots given
above become complex (no physical fixed point $M$ or $C$).
This happens in general for $\kappa <1$ when:
\bea   
\frac{1}{S} 
\frac{1-\kappa}{(1 + \sqrt{\frac{\kappa}{\kappa_c}})^2} <
1 - \frac{\epsilon}{\delta} < \frac{1}{S} 
\frac{1-\kappa}{(1- \sqrt{\frac{\kappa}{\kappa_c}})^2}
\label{defq}
\eea
with $S=2 \kappa_c/5$. For clarity the 
different regions in the plane of parameters
$\kappa$ and $1-\frac\E\D$ are drawn on \Fig{roots plot}.
The corresponding flow-diagrams are given in \Fig{schematic flow}.
We invite the reader to start with flow $\cal M$ and then to continue
to $\cal N$, $\cal O$, $\cal P$, and back to $\cal M$, and thus to 
observe how the flow is modified. The domain $\cal Q$ is defined
by (\ref{defq}) and corresponds to the case where
the two 
fixed points M and C have annihilated and only the fixed points G and
SR remain. Note also that the domain $\cal O$, where both the SR and
the mixed fixed points are attractive, and where the both of them are 
separated by a separatrix as shown on $\cal O$ of \Fig{schematic flow},
is very small. Although it is a rather narrow region which may thus be
difficult to observe numerically
(in addition the flow towards both fixed points 
along the line joining them and passing through $C$ is very slow)
it is intriguing physically
since in that region there will be two different phases and possible
values of the exponent depending on the respective 
bare values of LR and
SR disorder. 

\subsection{Inclusion of self-avoidance}
\label{Inclusion of SA}
Up to now we have not taken  into
account self-avoidance. While this is justified for directed
polymers (and membranes) by construction, 
physically, self-avoidance is always present for isotropic
polymers (and membranes). Although it is difficult to 
study in a controlled way for the present problem, we
give here some arguments which indicate that
self-avoidance does  probably not change the
behavior in the new glassy phase. We will
only address the case of short-range disorder.

Let us examine the operator for self-avoidance.
In the dynamical theory of section \ref{FFTRG} it reads:
\be \label{SA}
\int_{x,x',t}\DynB=2\int_{x,x',t} \tilde{r}(x,t) \frac\p{\p r(x,t)} \delta^{d}(r(x,t) - r(x',t))\ .
\ee
Power counting at the Gaussian fixed point shows that
in the absence of disorder, self-avoidance is relevant for
$d< d^0_{\mbox{\scr SA}}(D) = 4 D/(2-D)$. On the other hand we have found that
short-range disorder alone becomes relevant, for
isotropic manifolds, for $d< d_c(D)=(4 + 2 D)/(2-D)$.
Since we study $D<2$, self-avoidance is always
irrelevant near the upper critical dimension $d_c(D)$ around
which we expand. Thus the calculations of the previous
sections are consistent.

There is always the possibility that self-avoidance
becomes relevant as the dimension $d$ is lowered further.
One can  give some
qualitative estimate. We first consider the naive dimension
of the self-avoiding operator at the disorder dominated fixed point
studied in the previous sections. The power-counting used here
and in our previous work \cite{LeDoussalWiese97a} gives for the eigenvalue
of the above self-avoiding operator:
\begin{eqnarray} \label{SA-PC}
\lambda_{\mbox{\scr SA}}^{\ind{PC}} &=& D + 2 + \beta - (d+2) \zeta \ \nn\\
&=& 2D - d \zeta -\delta^{\ind{FDT-violation}}\ ,
\end{eqnarray}
where 
$\delta^{\ind{FDT-violation}}$ measures the violation of the FDT,
\be
\delta^{\ind{FDT-violation}}=\zeta-z-\tilde \zeta = D-2-\beta+2\zeta \ .
\ee
Of course, $\delta^{\ind{FDT-violation}}$ is zero in the purely
potential case, and one recovers the power-counting of self-avoiding
membranes without disorder. At the non-trivial fixed point, 
$\beta$ is very small and $\zeta$ much larger than its Gaussian
value $\frac{2-D}2$, such that $\delta^{\ind{FDT-violation}}$ is
positive and renders self-avoidance less relevant.

Let us now try to estimate \eq{SA-PC}.
If one first inserts the Flory values $\zeta=4/(d+2)$ and $\beta=0$,
one finds that $\lambda_{\mbox{\scr SA}}^{\ind{PC}} = D - 2 + \beta<0$
(since always $\beta\le 0$) and thus that self-avoidance is
never relevant. More generally, neglecting $\beta$ (which  is always small)
one finds that self-avoidance becomes relevant if $\zeta_{\mbox{\scr dis}} < \frac{2 + D}{2 + d}$.
The upper bound coincides with the Flory value for the roughness exponent in
presence of self-avoidance, which is known to be
a good approximation of the true value. Thus we obtain
the rough estimate that as long as  $\zeta_{\mbox{\scr dis}} > \zeta_{\mbox{\scr SA}}$ where 
$\zeta_{\mbox{\scr SA}}$ is the roughness exponent with self-avoidance only, 
 self-avoidance can safely be neglected. This is in agreement
with the naive expectation, that always the operator which results into
the larger exponent $\zeta$ will be dominant.
A look at the estimates
for $\zeta_{\mbox{\scr dis}}$
obtained in section \ref{iso SR} shows that for 
polymers the neglect of self-avoidance should be justified.

The above approximation is rather naive when applied to 
the {\em non-trivial} fixed point, and there are situations where
this kind of arguments goes wrong. Since a complete analysis of the 
situation away from $\E=0$ is technically impossible, one can only
study the stability of the fixed point with respect to perturbations
by self-avoidance; at first order in $\E$ this stability-analysis is then 
 {\em exact}\,\footnote{It is worth 
mentioning that in the
case of a 2-body self-avoiding interaction only, 
 3-body self-avoidance is 
 seemingly relevant by power-counting,
  but that with the help of the above mentioned
 stability-analysis one can falsify this assertion \cite{WieseHabil}.}.
The complete expression for $\lambda_{\ind{SA}}$ is
\be \label{SA-full}
\lambda_{\ind{SA}}^\ind{full}= 
2D - d \zeta -\delta^{\ind{FDT-violation}} + \mu \frac{\partial}{\partial \mu}
\lts_{g_\ind L =g_\ind T=g^*} \ln Z_\rho \ ,
\ee
where $Z_\rho$ is the renormalization group $Z$-factor for self-avoidance.
Let us only state the result here ($\DynB$ denotes the self-avoidance 
interaction, see \Eq{SA}):
\bea
Z_\rho &=& 1-2d\left(g_\ind L -\left(1-\frac1d\right) g_\ind T \right) \DIAG\DynS\DynB_{\!\!\E} \nn\\
\MOPE\DynS\DynB&=&  \frac1d \frac{2}{a-2} \frac{\p}{\p \sigma}\frac{\p}{\p \tau}
\left[ C(x,\tau) + C(y,\sigma)\right]^{1-a/2}
\eea
In analogy to \Eq{vert2}, the diagram is evaluated as
\be
\DIAG\DynS\DynB_{\!\!\E}=\frac1d\frac2{a-2}\frac1{(2-D)^2}\frac{\Gamma^2\!\left(\frac D{2-D}\right)}{\Gamma\left(\frac {2D}{2-D}\right)}\times S_D((2-D)S_D)^{a/2}
\ee
Since 
\be
\mu \frac{\partial}{\partial \mu}
\lts_{g_\ind L =g_\ind T=g^*} \ln Z_\rho = 2 g^* \DIAG\DynS\DynB_{\!\!\E} >0 \ ,
\ee
and with the help of \Eqs{SA-PC} and \eq{SA-full},
$\lambda_{\ind{SA}}^\ind{full}>\lambda_{\ind{SA}}^\ind{PC}$, such that
vertex-corrections render self-avoidance {\em more} relevant.
The influence of these corrections for {\em large} $\E$ is difficult to estimate. Let us only remark, that for large $d$, 
$\DIAG\DynS\DynB_{\!\E}$ is exponentially suppressed.

Thus we obtain that at the isotropic fixed point self-avoidance is 
presumably not important, since the random flow has already
strongly stretched the polymer or membrane.

There is one case however, well studied previously
\cite{Ebert96,LeDoussalMachta91,Harris1983,MachtaKirkpatrick1990,SmailerMachtaRedner1993}, where we know that
self-avoidance is important. This is when
the flow is the gradient of a short-range random potential,
a case not studied here, but which we mention for completeness.
There disorder and self-avoidance become relevant
{\it simultaneously} at $d=4 D/(2-D)$. The available
RG treatments for this problem \cite{Ebert96,Harris1983,MachtaKirkpatrick1990} exhibit a runaway flow to
strong coupling, as in the present
method, and thus the problem can not completely be
understood. One open question is whether the
exponent $\zeta$ is the same as for pure
self-avoidance $\zeta_{\mbox{\scr SA}}$ \cite{Harris1983}, and
 even how universal  
 the strong coupling regime is.
The question was addressed in \cite{LeDoussalMachta91}
and it was concluded that the problem is governed by a new strong disorder
fixed point, very much like the directed polymer problem,
where {\it both} self-avoidance and disorder are relevant.
Flory arguments were presented leading to $\zeta \approx 0.8$
in $d=2$ in good agreement with later simulations
\cite{SmailerMachtaRedner1993}. This problem however deserves further studies.

\subsection{Polymer in a hydrodynamic flow: toy model for generation of barriers}
\label{A toy model }

One of the remarkable results obtained in previous 
sections is that for a manifold with $D>0$ in a
purely divergenceless flow
new terms which violate the divergence free condition
are generated in perturbation theory (this is shown explicitly
in Appendix \ref{finite cor to DO}). Although they are
finite, they eventually drive the system to the
fixed point where $g_\ind T=g_\ind L$. This is in sharp contrast
to the particle case $D=0$ where the divergenceless flow
is a physical fixed point, distinct from the
fixed point at $g_\ind T=g_\ind L$. Physically this is because barriers
are generated in the case of the manifold but not in
the case of the particle.

In the case of the particle the fact that
the line $g_\ind T=0$ is exactly preserved to all orders
can be understood by noticing that the exact
stationary distribution is known, and is
simply the spatially uniform distribution. This exact property
implies strong constraints, as it must exactly  be preserved by
renormalization. In the case of manifolds, one can
check whether the measure $\mu_{\mbox{\scr eq}} = \exp( - c/(2T) \int_x (\nabla r_x)^2)$
is also stationary. If it was, it would imply similar
properties as for the particle, and in particular no generation
of barriers (since it is translationally invariant). 
There are some simple cases of flows where $\mu_{\mbox{\scr eq}}$ is
indeed stationary. One example is the generalization to
polymers \cite{OshaninBlumen1994} of the de Marsilly Matheron model of a 
layered random flow. There one can show explicitly that
$\mu_{\mbox{\scr eq}}$ satisfies the corresponding Fokker Planck equation
and is stationary. This problem was studied analytically
in \cite{OshaninBlumen1994}. Although it can be solved, it is in a
sense misleading because it misses the physics of the generation
of barriers unveiled here: for a polymer in a {\it generic}
divergenceless random flow such a simple translationally
invariant stationary measure does not exist.

In this section we will analyze this mechanism further.
We will present some toy models
which allow to illustrate in a simple way 
how elasticity leads to dynamical generation
of barriers even in a divergenceless flow.

\begin{figure}[t]
%\vspace{-3mm}
\centerline{ \fig{0.4\textwidth}{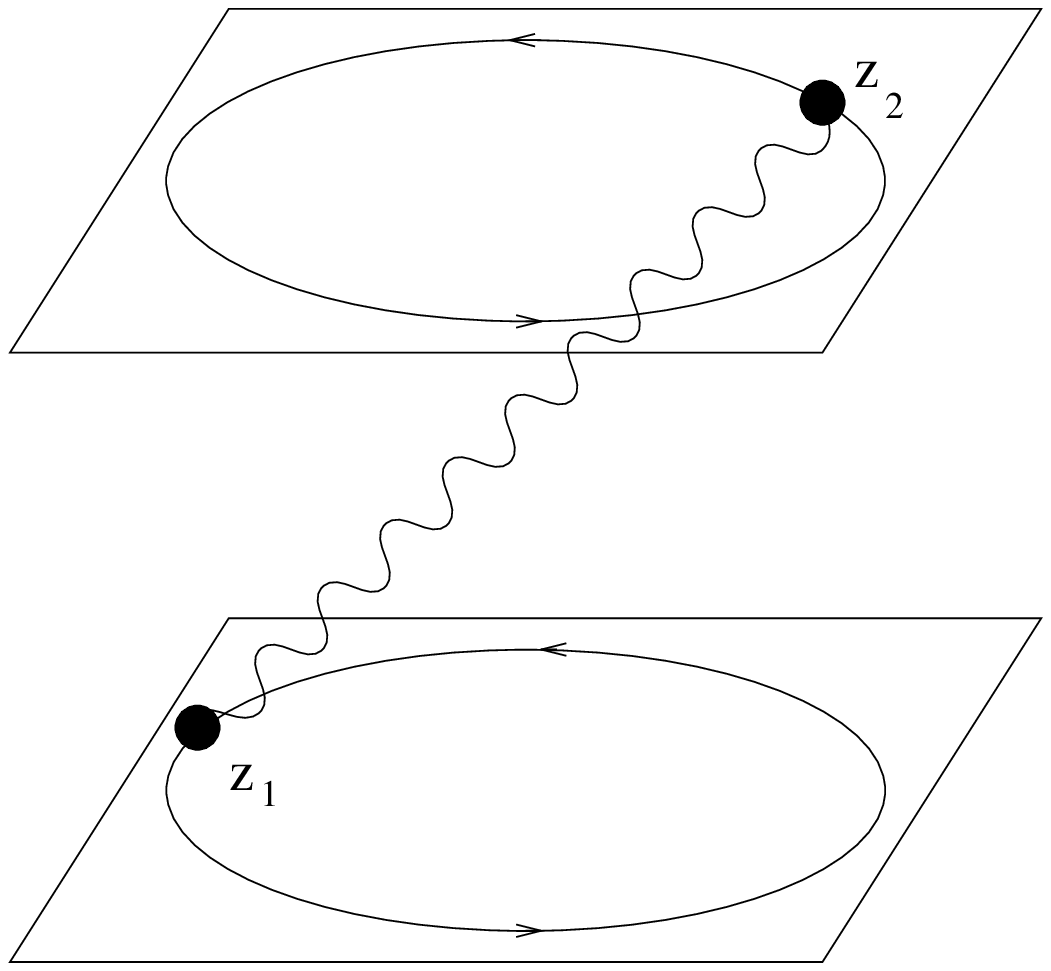} }
\caption{A dumbbell diffusing with each end on a rotating plane,
according to \Eq{2particules}.}
 \label{dumbbell}
\end{figure}%
We start with a single particle convected 
in such a flow, which we denote as
$v_\alpha(x)$. The probability distribution
$P(x,t)$ for the particle to be at position $x$ at time $t$,
satisfies the Fokker Planck equation ($T$ is the temperature):
\begin{equation}
\partial_t P(x,t) = \partial_\alpha ( T \p_\alpha P(x,t)  - v_\alpha(x) P(x,t))
\end{equation}
Since $\partial_\alpha v_\alpha =0$ one checks easily
that the spatially uniform measure $P(x)= 1/V$ 
($V$ is the volume of the system) is a stationary
solution. At large time (and $T>0$) the particle density will be
uniformly distributed over the system.
Upon applying a force $f$, the drift velocity in response to
the force will be exactly $f$. This can be checked by
methods similar to the one introduced in \cite{Derrida1983}.

Let us now consider two elastically
coupled particles (a dumbbell, see figure \ref{dumbbell}). We will start with 
two particles each located in a 2d plane.
For convenience we will denote $z_k = x_k + i y_k$ the complex
position of particle $k=1,2$ in its plane.
We will model the directed polymer situation,
and thus each particle will see a different
flow. The equation of motion is:
\bea  \label{2particules}
&& \dot{z}_1 = c (z_2 - z_1) + i \omega_1 z_1 + \eta_1  \nonumber \\
&& \dot{z}_2 = c (z_1 - z_2) + i \omega_2 z_2 + \eta_2  \ ,
\eea
where $c$ is the elastic coupling and
$\left<\eta_i \eta_j^*\right>=4 T \delta_{ij}$
the thermal noise. The flow in each plane is thus
simply $v_x= - \omega_k y$, $v_y=  \omega_k x$, i.e\ %
circulating around the center $x=y=0$
(where the stream function $\psi(x,y)= \omega_k (x^2 + y^2)$
has a maximum). The two flows have identical center,
but a slightly different rotation rate. We denote
by $2 \delta = \omega_1 - \omega_2$ the difference, and by $\omega$ the 
mean.
Although this model is linear, it already exhibits
striking behavior.

In the absence of ``disorder'', $\delta=0$,
the motion is simply a collective rotation
around the center $z=0$. The eigenvalues of the linear
system at $T=0$ are $i \omega$,
which describes the global rotation, and 
$-2 c + i \omega$ which describes the decay towards
the steady state $z_1=z_2$. At any $T$ the global rotation can
be eliminated by redefining $z_k = \exp(i \omega t)\, \tilde{z}_k$,
and similarly redefining the thermal noise. Thus the steady
state measure is simply:
\begin{eqnarray}
P(z_1,z_2) = e^{- \frac{c}{2 T} |z_1 - z_2|^2 } \ ,
\end{eqnarray}
i.e.\ uniform in space and proportional to the 
Boltzmann measure of the elastic system in the rotating
frame. Thus again, at $T>0$, one  observes free collective
diffusion of the center of mass, and a linear response $v=f$ of its
velocity $v$ to
an applied force $f$ .

Remarkably, as soon as $\delta >0$,
the zero mode disappears and the two particles converge
towards the center. At $T=0$ the
linear system eigenvalues are now
$\lambda_{\pm}=-c + i \omega \pm \sqrt{c^2-\delta^2}$. 
The global rotation can be eliminated as before, but
the differential rotation cannot. The convergence
of the dumbbell towards $(z_1=0,z_2=0)$ is exponential as
$|z_k(t)| \sim e^{- (c - \sqrt{c^2-\delta^2}) t}$,
with a characteristic relaxation time $\tau \sim 2 c/\delta^2$
at small $\delta$ (which interestingly is much larger
than the characteristic time $1/|\omega_1-\omega_2|$).
The slowest decaying eigenmode has $z_1/z_2 \sim e^{i \delta/c}$
at small $\delta$, i.e.\ it is twisted by an angle $\sim \delta/c$.

Most interestingly the effect persists to $T>0$. It is
difficult to find the full stationary measure for
the process (which seems to be non Gaussian) but it is possible 
to compute explicitly the mean squared positions,
by integration of the equations of motion. We find:
\begin{eqnarray}
\left<|z_k|^2\right> = \frac{2 T}{c \delta^2} (c^2 + \delta^2) \ ,  \qquad k=1,2\ .
\end{eqnarray}
This indicates a genuine bound state at $z=0$.
At small $\delta$, we find $|z_k|^2 \sim T \tau$ with the
same relaxation time $\tau$ introduced above. Since $T$
is the diffusion coefficient, the picture is that the 
dumbbell will diffuse away from the center at $z=0$
for a time $\sim \tau$ until it is pulled back in again.
The nature of this bound state is intriguing. One
can easily see that the stationary conformation is twisted.
Indeed a similar calculation gives:\
\be
\left<z_1 z_2^*\right> = 2 T (c + i \delta)/\delta^2 \ .
\ee
Thus, remarkably, as soon as one considers two elastically
coupled particles, preferred regions appear and
the stationary measure is 
no longer spatially uniform. Physically this
is because the two particles prefer regions
where the elastic energy is reduced. The 
center of mass motion thus becomes coupled to the
internal elastic mode. 

Let us now apply an external force. One easily sees
that this only amounts to translate the center of the bound state.
In shifted coordinates the motion remains identical.
The new center has coordinates: %Checked, Kay
\begin{eqnarray} \label{shifted}
z_{1,2} = f \frac{ \omega \mp \delta + 2 i c}{
2 \omega c - i (\omega^2 - \delta^2) } \ ,
\end{eqnarray}
where $f=f_x + i f_y$ is the force applied on
both monomers. Interestingly the new center
now depends explicitly on $\omega$ (the global rotation 
cannot be eliminated anymore), in a complicated way.
Also one has $z_1 \neq z_2$, i.e.\ the dumbbell is slightly
inclined. For small $\delta \ll \omega,c$ it reads:
\begin{eqnarray}
z_{1,2} \approx f \left( \frac{- 1}{i \omega} \mp \frac{\delta}{2 c \omega - i \omega^2} \right) \ .
\end{eqnarray}
In the limit $\delta \to 0$ the dumbbell center is close
to $z_1=z_2=-1/i \omega$ (with a weak restoring spring).
Amusingly, in that limit the 
dumbbell thus responds perpendicularly to the force, as a vortex.

So far we have described situations where the location of the
extrema of the stream functions in each plane coincide, but the
rotation rate does not. If these locations are shifted, the same
phenomena arise, i.e.\ a bound state occurs around a new center
with non trivial position.

This model can be extended in several directions. First one
may consider a  directed chain with $N$ monomers,
again with aligned stream function extrema at $z_k=0$, and 
different frequencies $\omega_k$. One then sees that the
decay towards the center at $T=0$ exactly maps onto
the problem of spin depolarization studied in 
\cite{MitraLeDoussal91}, a problem which arises in the
context of the so called ``motional narrowing'' in nuclear
magnetic resonance (NMR). Using the results of 
\cite{MitraLeDoussal91} leads to a stretched exponential
decay of $\left<|z_k(t)|^2\right>$ towards the center.

Nonlinear extensions of \Eq{2particules} can also be studied. They will 
show the generation of barriers. For instance
in a model with two uncorrelated random stream functions 
$\psi_k(z_k)$, local long living bound states will
be located where extrema are close together (note that
saddle points are rapidly escaped from). The dumbbell will eventually
escape over barriers between these metastable states.

Finally one can see that the effect persists even if
the two monomers see the same flow, though in that
case it must be nonlinear. A related example of that phenomena
was discovered by Thual and Fauve
\cite{ThualFauve1988}  in a different context.
If one identifies their complex Landau Ginsburg order
parameter $W(x) = z(x)$ as the complex position in $2d$ of
monomers in a gaussian elastic chain at $T=0$, their
complex Landau Ginsburg equation describes the dynamics
of such a chain in the complex $2d$ velocity field
$v = ( \mu + (\beta_r + i \beta_i) |z|^2 + 
(\gamma_r + i \gamma_i) |z|^4 ) z $ (which describes
a double well potential force plus a circulating force
with a rotation depending on the distance to the center).
They found that this equation can have stationary localized
solutions, which in our language correspond to a rotating 
and twisted polymer in a bound state.

To conclude, some insight can be gained by studying these
simple models,
and exploring further the physics of these problems 
would be of great interest.

\section{Conclusion}
\label{Conclusion}
In this article, we have studied a model for
polymers, and manifolds of internal dimension $D$,
diffusing and convected by static random flows.
Our results are obtained
using field theoretical dynamical RG near the
upper critical dimension $d_c(D)$, which is
$d_c=4$ and $d_c=6$ for
directed and isotropic polymers respectively, in short-range correlated
flows. In principle arbitrary $D$ can be considered, although
technically, since $d_c(D)$ diverges as $D \to 2^{-}$,
our quantitative analysis is valid for $D<2$.

We found that below the upper critical dimension $d_c(D)$
the random flow is relevant and changes the large scale
behavior. This yields
a new universal RG fixed point for polymers and manifolds in
non-potential static random flows with short-range correlations.
We have analyzed the critical theory using multilocal operator
product expansion techniques. As a result,
all divergences can be absorbed by multilocal counter-terms,
of which we explicitly give the 1-loop contribution.

The new RG fixed
points are in general characterized  by three exponents,
the roughness of the manifold $\zeta$,  the dynamical
exponent $z$, and $\phi$ related to the behavior of the
drift velocity $v(f)$ under a small applied force $f$.
They have been computed in a dimensional expansion, and
estimated numerically in physical dimensions. For
directed manifolds, a relation exists between these three
exponents. In the limit $D \to 0$ we recover previously
known results for the particle.

Our results show that in static random nonpotential flows
polymers and manifolds are in a non-trivial steady state with
glassy characteristics. In this state, the polymer or membrane is
stretched by the flow with a
roughness exponent, which is larger than in the absence
of the flow (e.g.\ $\zeta \approx 0.63$ in $d=3$ and
$\zeta \approx 0.8$ in $d=2$ for directed polymers in
SR random flows instead of $\zeta_0=1/2$ in the absence of
disorder). The internal dynamics is slower, with $z >2$.
Upon applying a small uniform external force $f$, the polymer
will move with a reduced velocity $v(f) \sim f^\phi$
with $\phi>1$. This state is characterized by a
zero {\it linear mobility} $\mu=\frac{dv}{df}|_{f=0}$.
Thus one can consider that the polymer is trapped in
preferential regions of the flow. On a simpler
toy model with only two particles, we have been able to
describe more precisely the effect of trapping,
but in the case of the polymer and
manifolds its detailed physics remains to be investigated.
Numerical simulations would be extremely useful to check our
results for the exponents and to further
characterize this glassy phase. These simulations
should be easier to perform than for
conventional potential glassy systems, since in the
present case the convergence to the steady state is expected to be
faster.

The glassy state that we found for this non-potential
system can be compared to the one arising in potential systems.
The main difference is that as soon as disorder is non-potential,
an effective temperature is generated, resulting in a
violation of the fluctuation dissipation
 theorem. The increase of temperature
competes with the relevance of disorder which tends
 to make the temperature
irrelevant. Since the important dimensionless disorder parameter
is proportional to the inverse temperature,
contrarily to most potential systems,
disorder does not flow to strong coupling.
Instead, it flows to a perturbatively accessible fixed point,
studied here. The effect
of disorder, and thus the glassy characteristics, are in
effect weaker than in potential glasses. Indeed, we find
a finite dynamical exponent $z>2$
and $v \sim f^\phi$ rather than the much
stronger non linear ``creep'' behavior $v \sim \exp(-c (f_c/f)^{\mu})$
obtained for manifolds with $D<4$ and $z$ effectively infinite,
which arises from a zero temperature
fixed point \cite{vortex_review,LemerleEtAl1998,ChauveEtAl98}.
In fact the state found here is reminiscent of the
so called ``marginal glasses'' predicted in the statics for
periodic manifolds at their lower critical dimension $D=2$
\cite{CarpentierLeDoussal1997,CardyOstlund82,HwaFisher1994,%
GoldschmidtSchaub1985,TsaiShapir1994a,TsaiShapir1994b,TsaiShapir1995},
where also thermal effects are important
and which have a similar $v \sim f^\phi$ characteristics.
This corresponds effectively to barriers which grow
logarithmically with size (and inverse applied force),
and it is remarkable that the ``randomly driven polymer''
also exhibits such divergent barriers
(in an effective way, since the notion of barriers is not
 clear-cut in nonpotential systems). It is all the
more remarkable in the case of a purely hydrodynamic
(i.e.\ divergenceless) flow, which for the single particle
leads to a trivial, spatially uniform, stationary distribution
(thus without barriers). Here, these barriers are generated 
from a non trivial interplay between elasticity, disorder 
and thermal fluctuations.

In the case of LR correlated flows we have found
via an expansion in the range of the correlation
$a$ near $a=a_c(D)=d_c(D)$ at fixed $d$ that the
behavior of the system is controlled by a line of
fixed points, and thus
depends continuously on the ratio 
$\kappa = g_\ind L/g_\ind T$ of the longitudinal
to transversal part of disorder (which
remains unchanged under renormalization).
For directed manifolds, our analytical
results were found to be in agreement with the exact large $d$
result obtained in \cite{LeDoussalCugliandoloPeliti97}
and even, to first order in $\E$
with the corresponding Hartree approximation.
In that case $\zeta$ was found to be exactly equal to
its Flory approximation $\zeta_\ind{F} = (4-D)/(2+a)$
while the other exponents are continuously varying.
Several remarkable results were obtained
in the case of the isotropic manifold.
We found that there exists a critical value $\kappa_{cr}$
of $\kappa$, beyond which no fixed point is found
in the fixed dimension expansion, as well as a smaller
characteristic value $\kappa_0$ beyond which the manifold 
becomes compressed $\zeta < \zeta_0$ rather than stretched
by the flow (as it is in all other cases).
Since this problem also corresponds to
directed polymers subjected to
non-potential ``correlated disorder'', these effects
may be understood as a result of a competition between
localization (due to the correlated nature of the force)
and driving. To further investigate this competition
we also studied in detail the intricate crossover between SR and LR 
disorder, in a dimensional expansion around the point
$a=d=6$ for polymers, important for studies in the physical dimensions
$d=2,3$. In view of the rich variety of obtained behaviors,
numerical simulations would be welcome
to study the competition between SR and LR disorder
effects. 

We also note that the glassy phase obtained here is via the RG
perturbatively controlled  around the line $d_c(D)$ (see
\Fig{fig-correl}), whereas the large $d$ approach
of \cite{LeDoussalCugliandoloPeliti97}
works for any $D$ and for LR disorder. A last and complementary method
is the functional renormalization group (FRG)
which can be used in a $D=4 - \E$ expansion
and thus allows to explore the region of larger
$D$ than the present method. Recent progresses on the
finite temperature FRG in \cite{LeDoussalGiamarchi1997} which showed
the existence of such a finite temperature fixed point, leads us to
conjecture that a similar fixed point will be accessible
by FRG in the present model and should merge smoothly into the
one found here upon decreasing $D$. Work is in progress
to verify this conjecture \cite{LeDoussalWieseProgress}.

To conclude, let us mention an interesting speculation, inviting for
further investigation \cite{HwaLeDoussalWieseProgress}:
It is well-known that a directed polymer in a
random potential maps via the Cole-Hopf transformation
to the Burgers-Kardar-Parisi-Zhang equation \cite{KPZ}.
The question  is whether the generalization of the directed
polymer diffusing in a non-potential random force field
instead of a potential one, has some effective correspondence in a
generalization of the Burgers equation
to non-potential driving and is
thus possibly related to the Navier Stokes equation.

%>>>>>

\section*{Acknowledgements}
It is a pleasure to thank
 L.\ Cugliandolo, F.\ David,  T.\ Giamarchi, J.\ Kurchan
and L.\ Sch\"afer for discussions.

\appendix

\section{Correlator and Response-function}
\label{appendixfree}
In this appendix we give a collection of formulas for the correlator
and the response function in the It\^o-discretization.
\bea
\frac1d\left< r(k,\omega) r(k',\omega')\right>_0 &=&  
\frac{2\lambda}{\omega^2 + (\lambda k^2)^2}\, (2\pi)^D \delta^D(k+k') (2 \pi)  
\delta(\omega+\omega') \\
\frac1d\left< r(k,\omega) \tilde r(k',\omega')\right>_0 &=& \frac{1}{i\omega  
+ \lambda k^2}\, (2\pi)^D \delta^D(k+k') (2 \pi) \delta(\omega+\omega') \\
\frac1d\left< \tilde r(k,\omega) \tilde r(k',\omega')\right>_0&=&0
\ .
\eea
Fourier-transformation with respect to $\omega$ yields
\bea
\frac1d\left< r(k, t) r(k', t')\right>_0 &=& \frac{1}{k^2} \rme^{-\lambda |  
t- t'|k^2}
(2\pi)^D\delta^D(k+k') \label{A1}\\
\frac1d\left< r(k, t) \tilde r(k', t')\right>_0 &=& \Theta( t> t')  
\rme^{-\lambda| t- t'| k^2}(2\pi)^D\delta^D(k+k') \\
\frac1d\left< \tilde r(k, t) \tilde r(k', t')\right>_0&=&0
\ .
\eea
We will mostly use the real space representation. Let us start with the
correlator.  The Fourier-transformation of \Eq{A1}
cannot be done explicitly but involves an error-function.
In addition, to obtain a well-defined function for $D<2$ one has
to subtract the zero-mode. We therefore define
\bea
C(x, t) &:=& \frac1{2d} \left< (r(x, t) -r(0,0) )^2\right>_0
\ .
\eea
The limiting cases are
\be
	C(x, t) = \left\{ {%\renewcommand{\arraystretch}2
\begin{array}{l@{\qquad}l}
\displaystyle \frac1{(2-D)S_D} \,|x|^{2-D} & \mbox{for}\   t\to0 \\
\displaystyle \rule{0mm}{7mm}\frac2{(2-D)(4\pi)^{D/2}} \,|\lambda   
t|^{(2-D)/2} & \mbox{for}\  x\to0
\end{array}}
\right. %}
\ ,
\ee
with 
\be
S_D=2 \frac{\pi^{\frac D 2}}{\Gamma(\frac D2)} .
\ee
Let us also note the ratio
\be
\frac{C(1,0)}{C(0,1)} = 2^{D-2}\, \Gamma\left(\frac D2\right)
\ .
\ee
The response-function
\be
R(x, t) := \frac1d\left< r(x, t) \tilde r(0,0)\right>_0
\ee
is calculated analytically
\be
	R(x,t) = \Theta( t>0) \left( 4 \pi \lambda  t \right)^{-D/2}  
\rme^{-x^2/4\lambda  t}
\ .
\ee
The (perturbative) fluctuation dissipation theorem (FDT) reads
\be
\Theta( t) \frac{\partial}{\partial \lambda  t} C(x, t) = R(x, t)
\ .
\ee
	
\section{Extrapolations}
\label{Extrapolations}
\begin{figure}[b]\centerline{
\epsfxsize=0.7\textwidth \parbox{0.7\textwidth}{\epsfbox{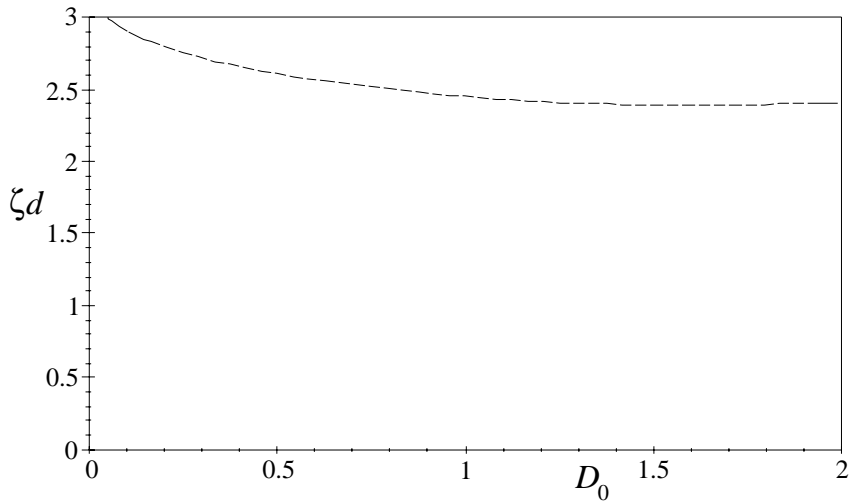}}
}
\caption{Extrapolations for $\zeta d$ to
$(D=1,d=3)$.}
\label{extr1}
\end{figure}%
The numerical values for the exponents reported in table \ref{tab1}
and \ref{tab2} were obtained by an optimization procedure using
the freedom to expand about any point on the critical curve $\E(D,d)=0$.
This procedure, first proposed in \cite{Hwa90}, was later refined in
\cite{WieseDavid96b} to obtain reliable estimates for the scaling dimension
of self-avoiding polymerized tethered membranes in the frame work of
a 2-loop calculation.
As will become apparent, various extrapolation schemes are possible,
and choosing the best one is almost an art; we shall rely heavily on the
methods  developed in Ref.~\cite{WieseDavid96b} to which the interested
reader is referred to for further details and discussion.
Our presentation is inspired by \cite{WieseKardar98a},
where  complementary material can be found.
The general idea is of course to expand about some
point  $(D_0,d_0)$ on the critical curve  $\E(D_0,d_0)=0$.
The simplest scheme is to extrapolate towards the physical theories
for $D=1,2$ and $d=2,3,\ldots$, using the expansion parameters
$D-D_0$ and $d-d_0$.
However, as shown in Ref.~\cite{WieseDavid96b}, this set of expansion
parameters is not optimal, and better results are obtained by using
$D_c(d)$, (solution of the equation $\E(D_c(d),d)=0$)
and $\E(D,d)$.
Furthermore, it is advantageous to make expansions for quantities
such as $\zeta d$  rather than  $\zeta$.

After selecting one of these schemes, the next step is to re-express
the quantity which shall be  expanded in terms of $D_c(d)$ and $\E(D,d)$.
If we are  interested in polymers ($D=1$) in
$d=3$ subject to isotropic disorder,
we have to evaluate this expression for $\E=3/2$.
However, we are still free
to choose the expansion point along the critical curve, which then fixes
$D_0$. As the expansion point is varied, different values for $\zeta d$
(i.e.\ $3\zeta$ for $d=3$) are obtained, as plotted on \Fig{extr1}.
The criterion for selecting a value for $\zeta$ from such curves
is that of minimal sensitivity to the expansion point $D_0$.
We thus evaluate $\zeta$ at the extrema of the curves.
The broadness of the extremum provides a measure of the goodness of the result, 
and the expansion scheme. The robustness of this choice
was explicitly checked in Ref.~\cite{WieseDavid96b} in the case of self-avoiding
tethered membranes, by going to the second order.
(For additional discussions of such ``plateau phenomena'' see Sec.~12.3 of
Ref.~\cite{WieseDavid96b}.)

\begin{figure}[t]\centerline{
\epsfxsize=0.5\textwidth \parbox{0.5\textwidth}{\epsfbox{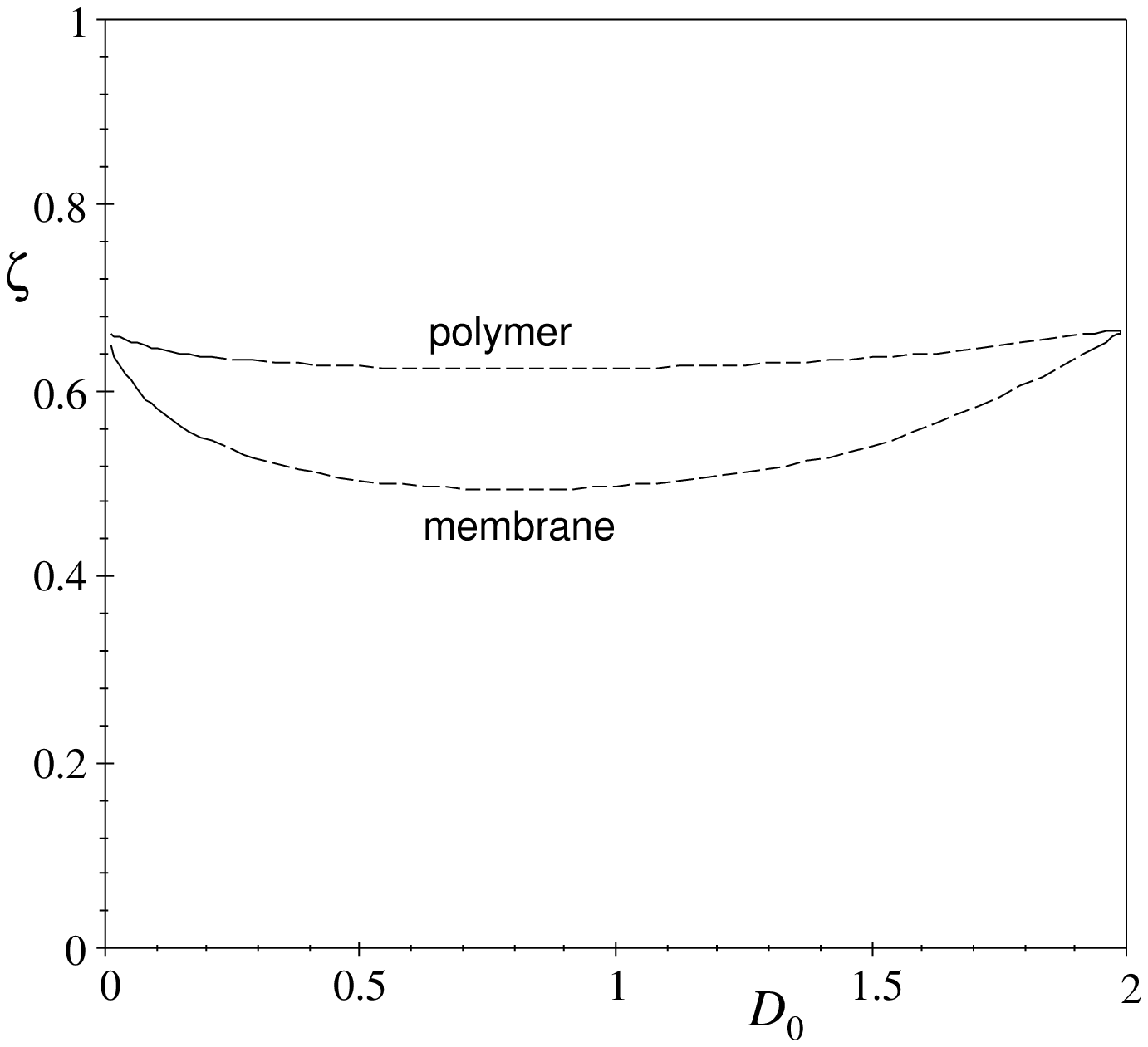}}
\epsfxsize=0.5\textwidth \parbox{0.5\textwidth}{\epsfbox{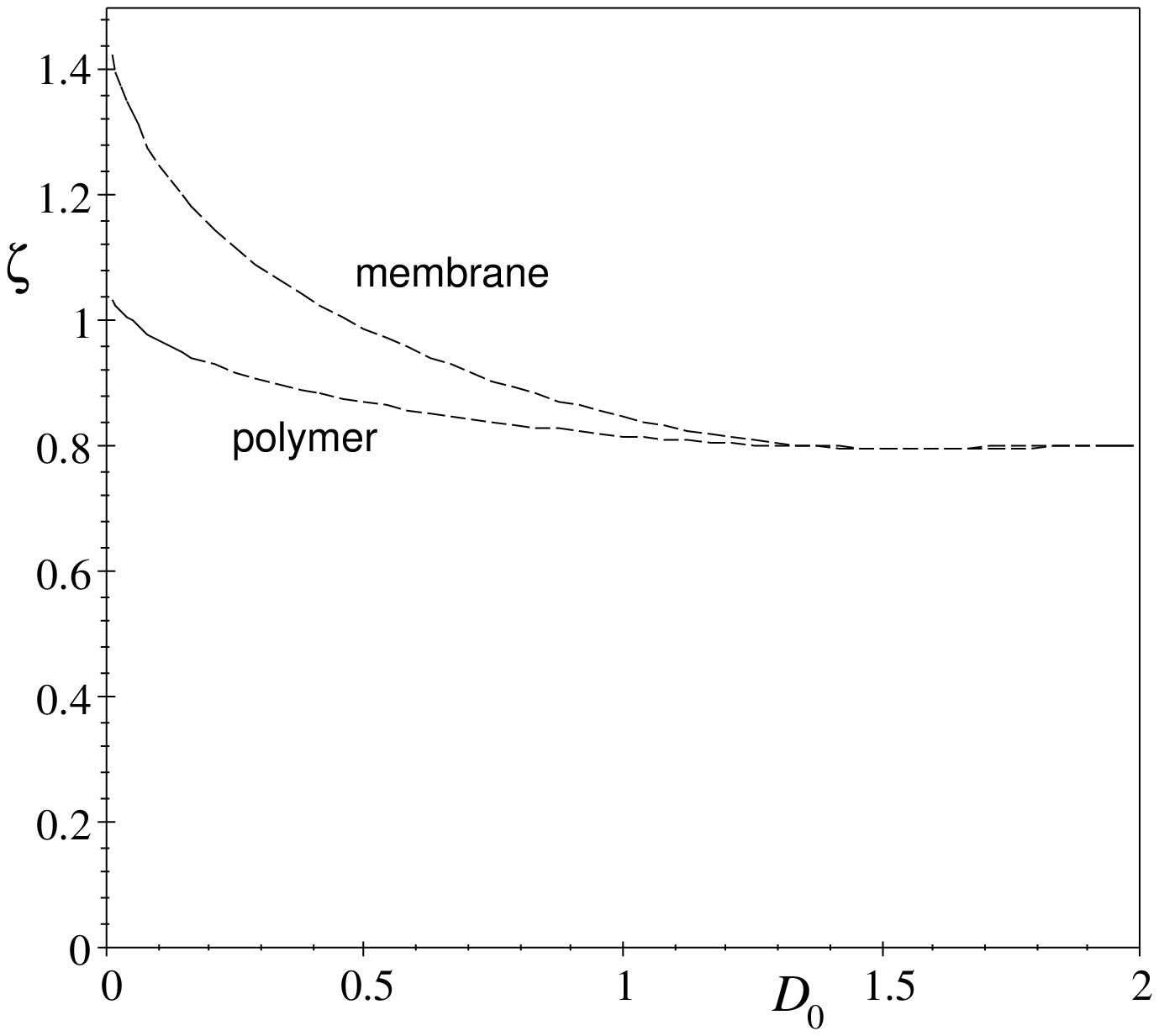}}}
\caption{Extrapolations for $\zeta$ to polymers and membranes
in 3 dimensions, using the extrapolation of $\zeta d$.
The left diagram is for the directed, the right
for the isotropic case.}
\label{zeta striking}
\end{figure}

\begin{figure}[tbh]\centerline{
\epsfxsize=0.5103\textwidth \parbox{0.5103\textwidth}{\epsfbox{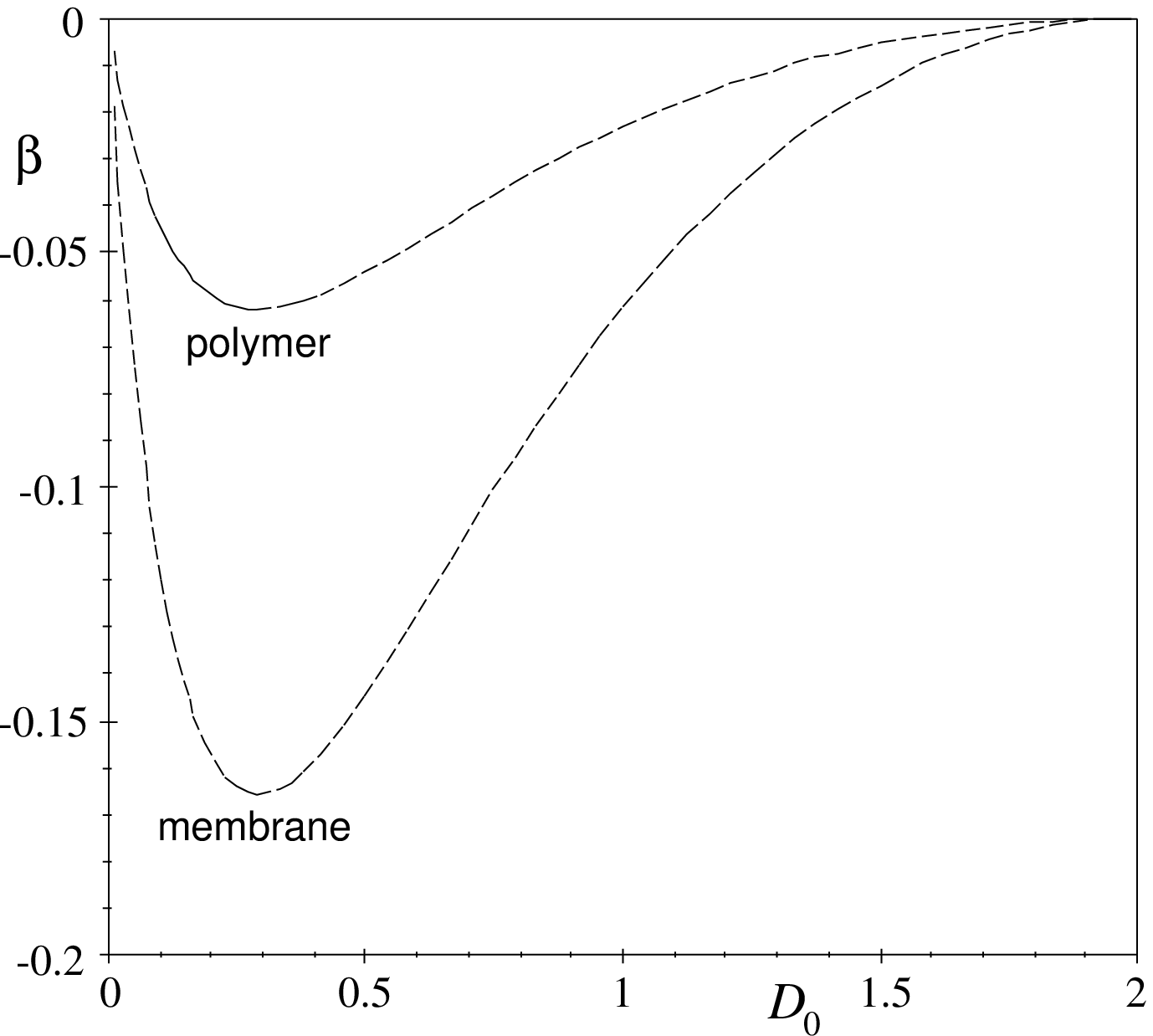}}
\epsfxsize=0.4896\textwidth  
\raisebox{-0.3mm}{\parbox{0.4896\textwidth}{\epsfbox{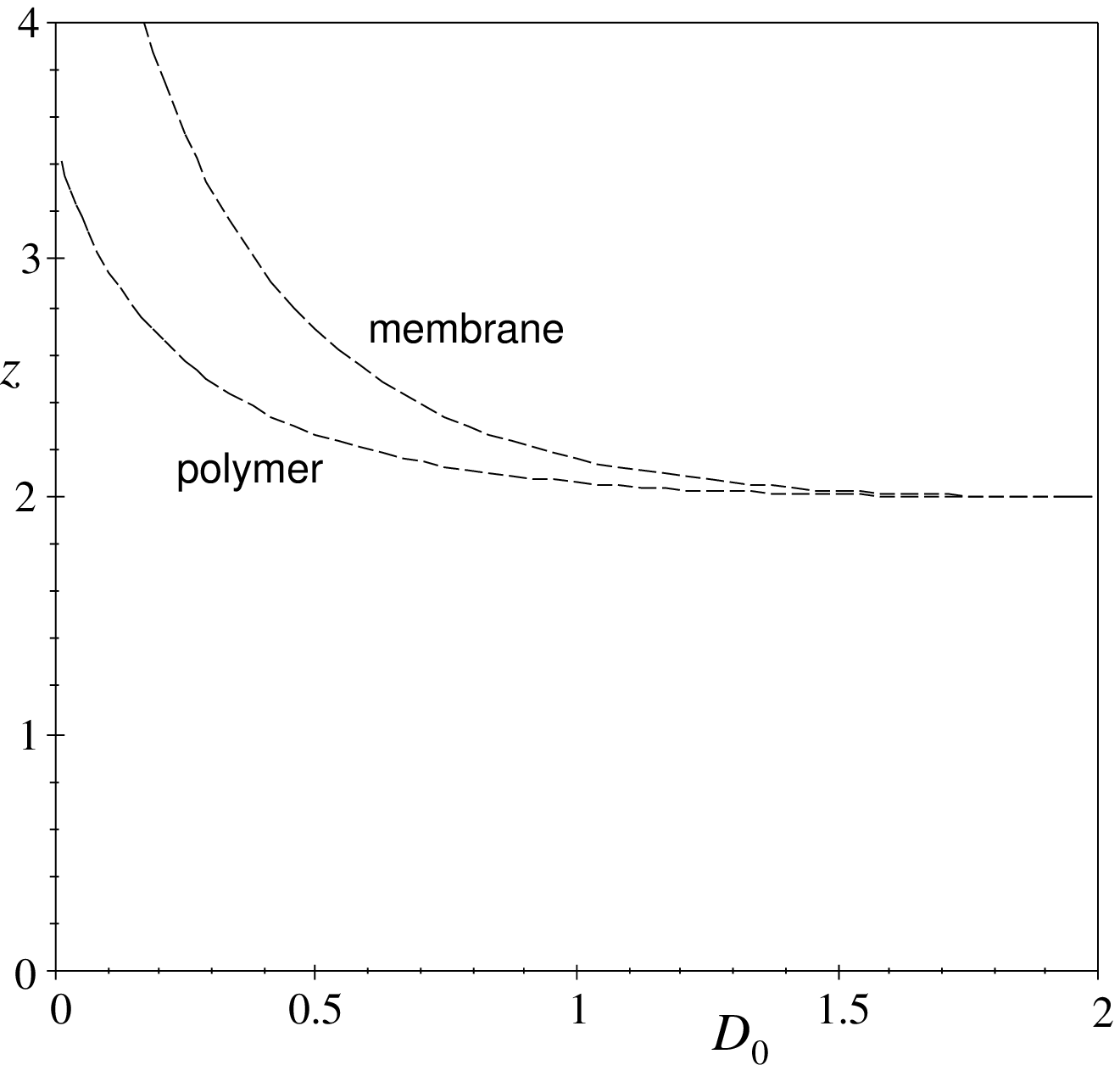}}}}
\caption{Extrapolations for $\beta$ (left) and $z$ (right)
to polymers and membranes
in 3 dimensions, isotropic disorder. Note that extrapolations are impossible
for $z$ and unreliable for $\beta$.}
\label{beta z}
\end{figure}

While we examined several such curves, only a selection is reproduced in
 Figs.~\ref{extr1}, \ref{zeta striking}, and \ref{beta z}.
While quantitative predictions were already reproduced in tables
\ref{tab1} and \ref{tab2}, let us focus on some important qualitative
observations. First, it is striking to observe that the dependence
of $\zeta$ on $D$ in the directed case and the independence on
$D$ in the isotropic case, read off from the extrapolations on
\Fig{zeta striking}, is already predicted by the Flory estimate.
Indeed, the latter is hard to improve numerically.

For the exponents $z$ and $\beta$, we did not find
a reliable extrapolation scheme: As can be deduced from \Fig{beta z},
the `minimum sensitivity point' in extrapolations for $\beta$ is
roughly independent of $D_0$, which indicates that the extrapolation
for $\beta$ only depends on $\E$, but not in a non-trivial way
on $D$ and $d$.

Extrapolations for $z$ show a plateau
when approaching $D_0\to 2$ only. Since the structure of the perturbation
expansion is such that expanding about $D_0=2$ always yields
the free-theory result $z_0=2$, extrapolations merely permit
in concluding on $z>2$. On the other hand,
it is possible to build up combinations
of $z$ and $D$ and $d$, which upon expansion build up a nice plateau.
However the answer obtained by this method strongly depends on the
selected scheme, and therefore is unreliable.

\section{Finite corrections to longitudinal disorder in the purely
transversal case}\label{finite cor to DO}
The aim of this appendix is to show that in the pure transversal 
case, longitudinal disorder is generated at the 2-loop order. 
A full 2-loop calculation is very difficult to perform
(see \cite{WieseDavid96b}). We will therefore only show, 
that at least one MOPE-coefficient, namely $\MOPE{\DynP \renewcommand{\arraystretch}{0.28}\begin{array}{c}{}\ind T\\ {}\ind T\\{} \ind T\end{array}}\DynJL$
does not vanish, as this was the case for 
$\MOPE {\DynK  {_\ind T \atop ^\ind T}} \DynJL$.
Since already the evaluation of $\MOPE{\DynP \renewcommand{\arraystretch}{0.28}\begin{array}{c}{}\ind T\\ {}\ind T\\{} \ind T\end{array}}\DynJL$
is a formidable task, we simplify the task further by 
considering only a special limit. This 
is accomplished by evaluating the leading singularity when approaching
two of its vertices as
\bea
\MOPE{\DynPsub \renewcommand{\arraystretch}{0.35}\begin{array}{c}{}\ind T\\ {}\ind T\\{} \ind T\end{array}}\DynJL 
&=&  \MOPE {\DynK  \renewcommand{\arraystretch}{0.35}\begin{array}{c}{}\ind T\\ {} \ind T\end{array}} \DynJT 
\times
\MOPE {\DynK  \renewcommand{\arraystretch}{0.35}\begin{array}{c}{}\ind T\\ {} \ind T\end{array}} \DynJL \nn \\
&+& 
\MOPE {\DynK  \renewcommand{\arraystretch}{0.35}\begin{array}{c}{}\ind T\\ {} \ind T\end{array}} \DynJL 
\times
\MOPE {\DynK  \renewcommand{\arraystretch}{0.35}\begin{array}{c}{}\ind L\\ {} \ind T\end{array}}\DynJL \nonumber \\
&+&
 \MOPE {\DynK  \renewcommand{\arraystretch}{0.35}\begin{array}{c}{}\ind T\\ {} \ind T\end{array}} \DynRT 
\times
\MOPE {\DynQ  \renewcommand{\arraystretch}{0.35}\begin{array}{c}{}\ind T\\ {} \ind T\end{array}} \DynJL \nn \\
&+&  
\MOPE {\DynK  \renewcommand{\arraystretch}{0.35}\begin{array}{c}{}\ind T\\ {} \ind T\end{array}} \DynRL 
\times
\MOPE {\DynQ  \renewcommand{\arraystretch}{0.35}\begin{array}{c}{}\ind L\\ {} \ind T\end{array}}\DynJL \nonumber \\
&+& \ldots  \label{appC1}
\eea
The symbol $\DynR$ hereby denotes the subleading operator (with
the possible transversal or longitudinal projectors unspecified)
\be
\DynR=\int_k \tilde r(x,t)\, k^2 \,\rme^{ik\left[r(x,t)-r(y,t')\right]}\tilde r(y,t')\ .
\ee
Since we know from the first-order calculations, see
section \ref{DO ren}, that  $\MOPE {\DynK  \renewcommand{\arraystretch}{0.28}\begin{array}{c}{}\ind T\\ {} \ind T\end{array}} \DynJT=\MOPE {\DynK  \renewcommand{\arraystretch}{0.28}\begin{array}{c}{}\ind T\\ {} \ind T\end{array}} \DynJL=0$, the first two terms on the r.h.s.\ in
 \Eq{appC1} vanish. From the calculations in section \ref{DO ren}
 also follows that $\MOPE {\DynQ  \renewcommand{\arraystretch}{0.28}\begin{array}{c}{}\ind T\\ {} \ind T\end{array}} \DynJL=0$, but that $\MOPE {\DynQ  \renewcommand{\arraystretch}{0.28}\begin{array}{c}{}\ind L\\ {} \ind T\end{array}}\DynJL$ does {\em not} vanish. Since moreover
 $\DynR$ is after
 $\DynJ$ the next-to-leading term in the MOPE, at least for $D\to2$, it is sufficient to show
that 
$\MOPE {\DynK  \renewcommand{\arraystretch}{0.35}\begin{array}{c}{}\ind T\\ {} \ind T\end{array}} \DynRL$ is non-zero in order to establish that 
$\MOPE{\DynP \renewcommand{\arraystretch}{0.28}\begin{array}{c}{}\ind T\\ {}\ind T\\{} \ind T\end{array}}\DynJL$ does not vanish identically.
This is the aim of the following calculation. Let us however note
that the argument is not conclusive, when $\DynR$ is not the most relevant
subleading calculation. This applies especially to the case of the
particle with $D=0$, where the canonical dimension of $\nabla r$ 
becomes 0 and there is thus an infinity of marginal operators, 
which can be constructed from $\DynJ$ by adding an arbitray power of
 $\nabla r$. We comment on this special case later, see appendix
 \ref{Non-renormalization of transversal disorder in 
the particle case}. 

We now study the MOPE
\be
\DynK \longrightarrow \DynJ + \mbox{derivatives of }\DynJ \ ,
\ee
To simplify notations,
we give all results for the isotropic case and
short-range disorder ($a=d$).
We start from the expression \eq{3.33} for the contraction $D_1$ in the
notation of figure \ref{disren} and  the analogue expression for
$D_2$, which together read
\bea
\lefteqn{2\int_k\int_p
\rme^{ik\left[r(x,0)-r(y,t)\right]}\,
\rme^{(-p^2+\frac{k^2}4) \left[  C(x-x',\tau)+C(y-y',\sigma) \right]}
 R(x-x',\tau) R(y-y',\sigma)\times
}\nn\\
&&
\left[ \tilde r_\A(x,0)  \tilde r_\B(y,t) \,{\textstyle
(p-\frac k2)_\G\, (p-\frac k2)_\D -\tilde r_\A(x,0)\,\tilde r_\D(y,t)\,
  (p-\frac k2)_\G \,(p+\frac k2)_\B}
\right] \times
\nn\\
&&\left[ {\textstyle g_\ind T P^\ind T_{\A\B}(p-\frac k2) + g_\ind L P^\ind  
L_{\A\B}(p-\frac k2) }
\right] \left[ {\textstyle g_\ind T P^\ind T_{\G\D}(p+\frac k2) + g_\ind L  
P^\ind L_{\G\D}(p+\frac k2) }
\right]\ .  \hspace{1cm}
\label{D2}
\eea
We want to write this expression in the form
\be
\int_k \tilde r_\A(x,0)  \tilde r_\B(y,t) \,\rme^{ik\left[r(x,0)-r(y,t)\right]} \,\Delta_{\A\B}(k) \ .
\label{D3}
\ee
In order to decompose $\Delta_{\A\B}(k)$ into longitudinal and
transversal part, we remark that if we denote
\be
\Delta_{\A\B}(k) = \Delta_\ind L(k) P^\ind L_{\A\B}(k) + \Delta_\ind T(k)  
P^\ind T_{\A\B}(k) \ ,
\ee
then
\bea
\Delta_\ind L(k) & =&  P^\ind L_{\A\B}(k) \Delta_{\A\B}(k) \nn\\
\Delta_\ind T(k) &=&  \frac1{d-1}P^\ind T_{\A\B}(k) \Delta_{\A\B}(k) \ .
\eea
Applying the longitudinal projector $P^\ind L_{\A\B}(k)=\frac{ k_\A k_\B}{k^2}$
onto $\Delta_{\A\B}(k)$ as obtained from \Eqs{D2} and \eq{D3} yields
\bea
\lefteqn{2\int_p \rme^{(-p^2+\frac{k^2}4)\left[C(x-x',\tau)+C(y-y',\sigma)\right]}
R(x-x',\tau)R(y-y',\sigma)}\nn\\
&& \left\{
\frac1{k^2} \left[g_\ind T k^2 + (g_\ind L-g_\ind T)\frac{(pk-\frac{k^2}2)^2}{
p^2-pk+\frac{k^2}4}\right]
\left[ g_\ind T(p^2-pk+\frac{k^2}4)+(g_\ind L-g_\ind T)\frac{(p^2-\frac{k^2}4)^2}
{p^2+pk+\frac{k^2}4}\right] \right. \nn\\
&&\ \ -\frac1{k^2}\left[
g_\ind T(pk+\frac{k^2}2)+(g_\ind L-g_\ind  
T)\frac{(pk-\frac{k^2}2)(p^2-\frac{k^2}4)}
{p^2-pk+\frac{k^2}4}
\right]  \times
\nn\\&&\qquad\qquad
\left.\times\left[ g_\ind T(pk-\frac{k^2}2)+(g_\ind L-g_\ind  
T)\frac{(p^2-\frac{k^2}4)(pk+\frac{k^2}2)}
{p^2+pk+\frac{k^2}4}\right]
\right\}
\label{complete long}
\eea
Using the FDT \eq{FDT} and expanding up to second order in $k$ gives
 for the longitudinal part
\bea
\lefteqn{\int_p
\frac{\partial}{\partial \sigma}\frac{\partial}{\partial \tau}
\rme^{-p^2\left[C(x-x',\tau)+C(y-y',\sigma)\right]}}
\nn\\
&& \left\{\left(2\,{\frac {p^2k^2\!-\!pk^2}{k^2p^4}}
\!+\!\frac{7\,pk^2p^2k^2\!-\!3p^4k^4\!-\!4{pk}^4}{p^8k^2}
\!+\!{\frac  
{(p^2k^2\!-\!pk^2)\left[C(x\!-\!x',\tau)\!+\!C(y\!-\!y',\sigma)\right]  
}{2p^4}}
\right)
\!g_\ind L g_\ind T \right. \nn\\ %%}
&&
%% {
\qquad \left. +{\frac {4\,\left (p^2k^2 -pk^2\right )^2}{p^8 k^2}}\,g_\ind T^2 
+\frac{k^2}{p^4}\,g_\ind L^2 \right\}
\label{2 ord long}
\eea
Two remarks are in order: First, integrating over $p$, we recover
from the first term proportional to $g_\ind L g_\ind T$ the leading
contribution already given in \Eq{DO diagram}. Second,
and most importantly, there is a term proportional to $g_\ind T^2$,
which  is positive for all dimension $d>1$. Thus, even when starting
with pure transversal disorder, longitudinal disorder will be generated
under renormalization. Although finite at 1-loop order, it will necessitate
a counter-term at the 2-loop level, as discussed at the beginning of 
this section.

For completeness, let us also analyze corrections to the transversal part of
disorder. Since it is calculatory more convenient, instead of
contracting $\Delta_{\A\B}(k)$ with the transversal projector, we contract
it  with the  unit-matrix $ \delta_{\A\B}$, resulting in
$(d-1) \Delta_\ind T + \Delta_\ind L$. It is then easy to
subtract $\Delta_\ind L$ already given in \Eqs{complete long} and
\eq{2 ord long}. Ergo
\bea
\lefteqn{\hspace{-0.47cm}\Delta_{\A\A}=2 \int_p
\rme^{(-p^2+\frac{k^2}4)\left[C(x-x',\tau)+C(y-y',\sigma)\right]}
R(x-x',\tau)R(y-y',\sigma)}&&\nn\\
&& \left\{
\left[ { g_\ind T}\,(d-1) +{ g_\ind L}\right]\left[{
g_\ind T} (p^2-{ pk}+{\frac {k^2}4} )+ ({ g_\ind L}-{ g_\ind T} ){\frac {\left ({
p}^2-{\frac {k^2}4}\right )^2}{{
p}^2+{ pk}+{\frac {k^2}4}}}\right] \right.\nn\\ %%\}
&& %%\{
\left.
-\left[ g_\ind T^2(p
^2-{\frac {k^2}4})+2\left ({ g_\ind L- g_\ind T}\right ){
g_\ind T}(p^2-{\frac {k^2}4})+{\frac {({ g_\ind L
}-{ g_\ind T} )^2 \left(p^2-{\frac {k^2}4}\right)^{3}}{
\left (p^2-{ pk}+{\frac {k^2}4}\right )\left (p^2+{ pk
}+{\frac {k^2}4}\right )}} \right]\right\}\qquad
\label{complete diag}
\eea
Using the FDT, \Eq{FDT}, and expanding up to order $k^2$,
this can be written as
\bea
\lefteqn{\int_p
\frac{\partial}{\partial \sigma}\frac{\partial}{\partial \tau}
\rme^{-p^2\left[C(x-x',\tau)+C(y-y',\sigma)\right]}}
\nn\\
&&
\left\{\left(\frac{2\left(d-1 \right)}{p^2}
+\frac{4d\, pk^2 -(d+3) p^2 k^2}{2p^6}
+\frac{k^2(d-1)}{2\,p^2} \left[C(x-x',\tau)+C(y-y',\sigma)\right]
\right)
{ g_\ind L g_\ind T}
\right. \nn\\
&&\qquad\left.
+{\frac {2d \left( p^2k^2 - { pk }^2\right)}{p^{6}}}{ g_\ind T^
2}
+{\frac {k^2 }{p^4}}g_\ind L^2
\right\}
\label{2 ord diag}
\eea
The transversal part can now be obtained
by subtracting \Eq{2 ord long} from \Eq{2 ord diag}.
The interesting thing to note is that there is no component
proportional to $g_\ind L^2$ as expected from the fluctuation
dissipation theorem, which states that potential disorder
is preserved under renormalization. On the other hand, there will be
all other combinations of terms.

\section{Non-renormalization of transversal disorder in 
the particle case}
\label{Non-renormalization of transversal disorder in 
the particle case}
In this appendix, we show, why in the particle case 
 purely transversal disorder is not
renormalized. An analogous result in the context of the Focker-Planck 
equation was obtained in \cite{HonkonenKarjalainen1988}.

To this aim consider $n$ interaction vertices $\DynJT$, 
and choose from its $2n$ end-points the one who
is the most retarded in time. 
Without loss of generality due to translation-invariance in time
its time-argument can be set to 0, leading to $\tilde r_\alpha(0)$. 
(Note that for a particle, there
is only a time argument, but no space argument $x$.)
To construct a new interaction with exactly two response-fields per
vertex from the $n$ interaction vertices, it is 
necessary to contract $\tilde r_\alpha(0)$. Now, any of the other 
vertices has the structure ${\cal V}:=\int_k \tilde r(t) \rme^{ik\left[r(t)-r(t')\right]}\tilde r(t')$.
Contracting $ \tilde r_\alpha(0)$ with $ {\cal V}$ yields 
$ik_\alpha R(t) -ik_\alpha R(t')$. Since by assumption $t>0$ and $t'>0$,
$R(t)=R(t')=1$ and the both terms cancel. This cancelation
is independent of the kind of disorder, and also present 
for longitudinal one. (Note, that the argument is not valid  for 
$D>0$.)
The difference between the two kinds of disorder 
comes from the last possible
contraction, namely by contracting $\tilde r_\alpha(0)$ 
with the other end of the same vertex. Denoting by
$P^{\ind{T}/\ind{L}}_{\alpha\beta}(k)$ the transversal respectively 
longitudinal projector, 
this vertex is
$\int_k P^{\ind{T}/\ind{L}}_{\alpha\beta}(k) \tilde r_\alpha(0) \rme^{ik\left[r(0)-r(t)\right]}\tilde r_\beta(t)$, and 
contracting $\tilde r_\alpha(0)$ with $\rme^{-ikr(t)}$ yields
$-ik_\alpha$ which is multiplied by $P^{\ind{T}/\ind{L}}_{\alpha\beta}(k)$.
In the transversal case, the product $k_\alpha
P^{\ind{T}}_{\alpha\beta}(k)$ vanishes identically. This is not
so for longitudinal disorder, leading to a renormalization of 
disorder.

\newcommand{\tit}[1]{#1}

\end{document}